\documentclass[12pt,a4paper]{article}
\usepackage[utf8]{inputenc}
\usepackage[T1]{fontenc}
\usepackage[english]{babel}
\usepackage[a4paper,top=2cm,bottom=2cm,left=2cm,right=2cm,marginparwidth=1.75cm]{geometry}
\usepackage{authblk}
\usepackage{balance}
\usepackage{bm}
\usepackage{siunitx}
\usepackage{graphicx}
\usepackage{array}
\usepackage[usenames,dvipsnames]{xcolor}
\usepackage{booktabs}
\usepackage{multirow}
\usepackage{caption}
\usepackage{subcaption}
\usepackage{soul}
\usepackage{url}
\usepackage{hyperref}
\usepackage{doi}
\usepackage{amsmath, amssymb}
\usepackage[export]{adjustbox}
\usepackage[numbers]{natbib}
\usepackage[prefix]{nomencl}
\usepackage{changes}
\usepackage{tikz}
\usetikzlibrary{tikzmark}
\usepackage{pdflscape}
\usepackage[nottoc]{tocbibind}
\usepackage[section]{placeins}
\newcommand{\subcaptionmarker}[1]{(#1)}
\newcommand{\subcaptionmark}{\phantomsubcaption{}\subcaptionmarker{\thesubfigure}}

\DeclareSIUnit\bar{bar}

\newcommand*\circled[1]{\tikz[baseline=(char.base)]{
            \node[shape=circle,draw,inner sep=1pt] (char) {#1};}}

\renewcommand{\nomgroup}[1]{%

  \ifthenelse{\equal{#1}{A}}{\item[\textbf{Latin symbols}]}{%
  \ifthenelse{\equal{#1}{B}}{\item[\textbf{Greek symbols}]}{%
  \ifthenelse{\equal{#1}{C}}{\item[\textbf{Subscripts}]}{}}}}
\title{WhY shape matters: Hydrodynamics of a Y-shaped membraneless electrolyzer}
\author[1]{Karl Schoppmann\footnote{Corresponding author: karl.schoppmann@tu-dresden.de}}
\author[2,3]{Hannes Rox \footnote{Karl Schoppmann and Hannes Rox contributed equally to this work.}} 
\author[1]{Erik Frense}
\author[1]{Frank Rüdiger}
\author[2]{Xuegeng Yang}
\author[2,3,4]{Kerstin Eckert} 
\author[1]{Jochen Fröhlich}

\affil[1]{Institute of Fluid Mechanics, TU Dresden, 01062 Dresden, Germany}
\affil[2]{Institute of Fluid Dynamics, Helmholtz-Zentrum Dresden-Rossendorf, 01328 Dresden, Germany}
\affil[3]{Institute of Process Engineering and Environmental Technology, Technische Universität Dresden, 01062 Dresden, Germany}
\affil[4]{Hydrogen Lab, School of Engineering, Technische Universität Dresden, 01062 Dresden, Germany}

\hypersetup{colorlinks = true,linkcolor = blue,anchorcolor =red,citecolor = blue,filecolor = red,urlcolor = red, pdfauthor=author}

\begin{document}

\date{}	
\maketitle
\begin{abstract}
A novel Y-shaped membraneless flow-through electrolyzer is introduced to achieve a homogeneous electrochemical reaction across the entire electrode in a cost-efficient cell design with effective product separation. 
Numerical simulations of the electrolyte flow and electrical current within the already known I- and T-shaped cells motivate the newly proposed Y-shape cell. 
Furthermore, a new design criterion is developed based on the balance between bubble removal and gas generation. 
As proof-of-concept experimental results using the Y-shaped electrolyzer are presented, showing homogeneous gas distributions across the electrode and efficient product separation by the electrolyte flow.
\end{abstract}

\maketitle
\section{Introduction}
\label{sec:intro}
Water electrolysis is of great importance in the generation of H\textsubscript{2}, a clean and versatile energy carrier which is needed for the transition towards net-zero-emissions industry \cite{Spek2022}. In particular, membraneless electrolyzers offer advantages in terms of simplicity and cost effectiveness \cite{Manzotti2022}. In contrast to conventional technologies like alkaline water electrolysis or proton exchange membrane, membraneless designs can diminish operating and capital expenditures associated with membrane degradation and resistance \cite{Esposito2017}.

\begin{table}[!ht]
    \centering
    \caption{Illustration of concepts for membraneless flow-through electrolyzers designs. Qualitative sketches throughout. }
    \begin{tabular}{l|c|c|c}
        \toprule
        & I-shape & T-shape & Y-shape \\
        \hline & & & \\
        
        \parbox[t]{2mm}{\multirow{1}{*}[7em]{\rotatebox[origin=c]{90}{Arrangement}}}
        & \includegraphics[width=0.25\textwidth]{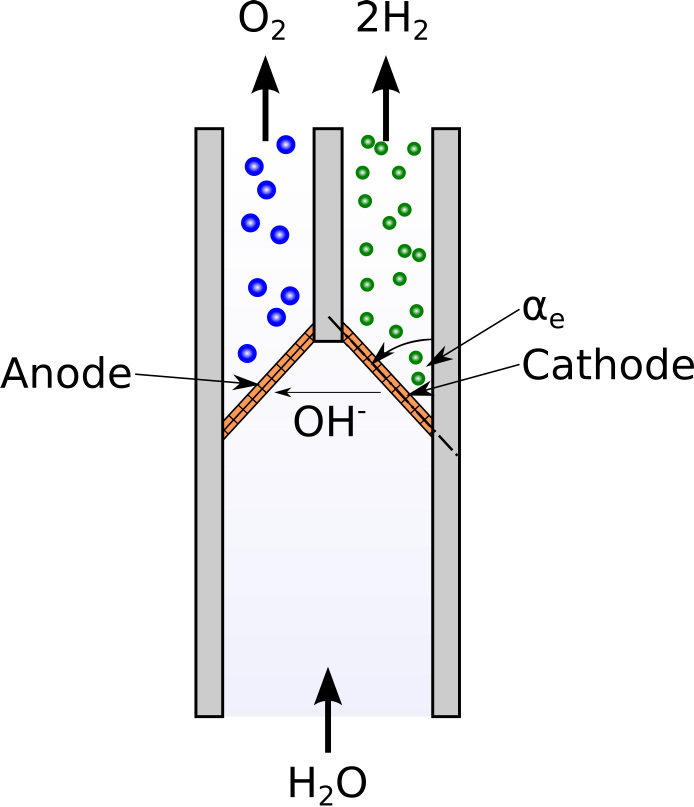} 
        & \includegraphics[width=0.30\textwidth]{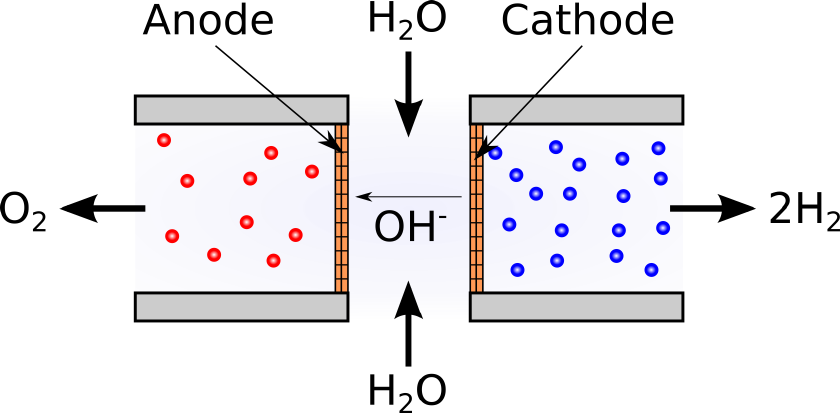} 
        & \includegraphics[width=0.25\textwidth]{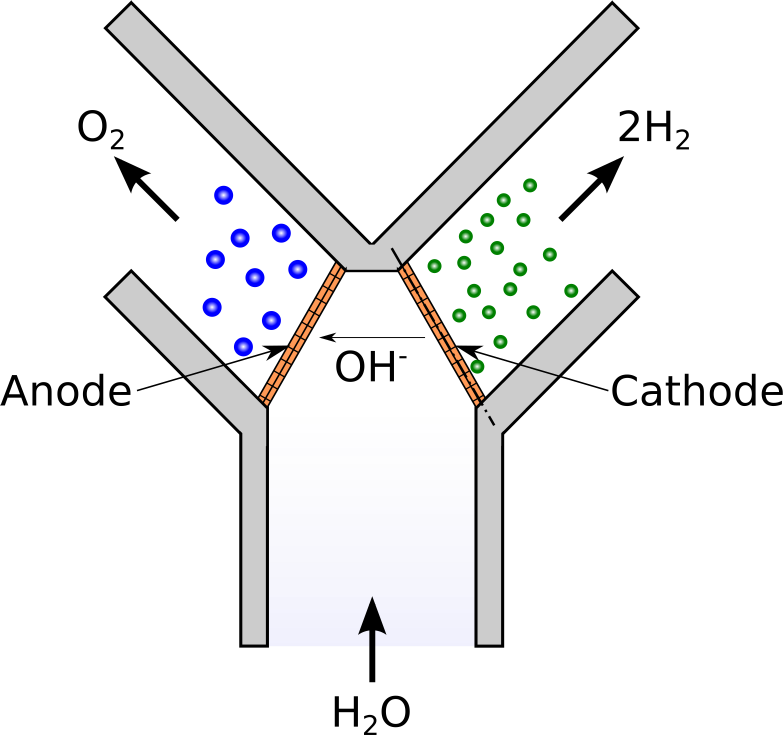}  \\ & & & \\
        
        \parbox[t]{2mm}{\multirow{2}{*}[4em]{\rotatebox[origin=c]{90}{Fluid flow}}}     
        & \includegraphics[width=0.15\textwidth]{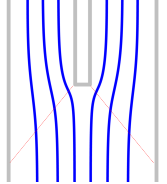} 
        & \includegraphics[width=0.25\textwidth]{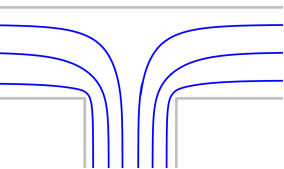}  
        & \includegraphics[width=0.25\textwidth]{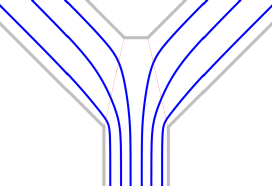} \\
        & best & worst & reasonable \\ & & & \\
        
        \parbox[t]{2mm}{\multirow{2}{*}[5em]{\rotatebox[origin=c]{90}{Electric current}}}
        & \includegraphics[width=0.13\textwidth]{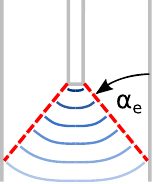}  
        & \includegraphics[width=0.25\textwidth]{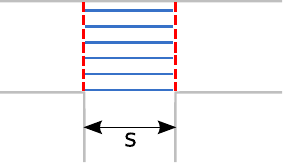}   
        & \includegraphics[width=0.25\textwidth]{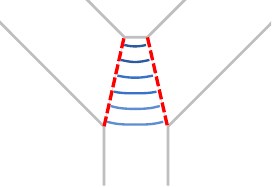} \\
        & worst & best & reasonable \\
        \bottomrule
    \end{tabular}
    \label{tab:concepts}
\end{table}

In recent studies, a distinction is made between flow-through electrolyzers, flow-by electrolyzers \cite{Hashemi2015, Hashemi2019, Pang2020, Yang2021b}, capillary-fed electrolyzers \cite{Hodges2022, Hoang2023} and decoupled electrolyzers \cite{Dotan2019, Solovey2021}. Membraneless flow-through electrolyzers (FT-ME) are using the electrolyte flow through porous electrodes to separate the products H\textsubscript{2} and O\textsubscript{2} by flushing the gas out of the electrode gap into separate channels \cite{Esposito2017}. Prominent examples divide into I- and T-shape, as shown in Tab.~\ref{tab:concepts}. Due to its efficient product separation at high flow rates \cite{Neil2016}, high current densities of \SI{4}{\ampere\per\centi\metre\squared} and product purities were reported in \cite{Gillespie2018, Rajaei2021}. Therefore, the approach of FT-ME was chosen for the present study. 

The performance of an FT-ME is mainly influenced by the following general parameters. 
(i) The surface area of the electrodes $A_\mathrm{e}$ determines the current density and the electrochemical active area (ECSA). Thus, it also affects the specific current density which is critical for the overall efficiency of the process.
(ii) The width $s$ of the gap between the electrodes affects the ohmic resistance inside the cell \cite{Gillespie2015}, with decreasing gap width resulting in a lower cell potential.
(iii) The electrode angle $\alpha_\mathrm{e}$ characterizes the change of distance between the electrodes from edge to edge. As a result, it affects the homogeneity of the current between the electrodes \cite{Neil2016} and, hence, of the ohmic resistances.
(iv) The bulk flow velocity $U$ of the electrolyte is relevant for the bubble detachment at the electrode \cite{Rox2023, Khalighi2022}, the transport of bubbles away from the electrode, as well as the distribution of reactants at the electrode surfaces. On the other hand, the electrolyte flow through the entire cell experiences flow resistance which determines the hydrodynamic losses, hence the required pumps and the related energy.
Further process parameters such as conductivity of the electrolyte, working pressure $p$ and temperature $T$ also influence the efficiency of the process. In addition, the electrode morphologies \cite{Yang2022a, Rajaei2021, Yang2021}, surface properties \cite{Wang2024, Andaveh2022, Darband2019} and catalytic sites are critical for the water splitting reaction and bubble management, particularly for porous electrodes. As the focus is on the comparison of different geometrical designs, these parameters are not further discussed here. 

The influence of the introduced parameters on the electric potential $E_\mathrm{c}$ at the maximum reported current density $j_\mathrm{max}$ for FT-ME is shown in the Tab.~\ref{tab:literature_overview}. 
So far, the works of \citeauthor{Gillespie2015} \cite{Gillespie2015, Gillespie2017, Gillespie2018} reported the highest current density $j_\mathrm{max}$ of $\approx \SI{4000}{\milli\ampere\per\square\centi\metre}$ 
at a much lower electric potential $E_\mathrm{c} \approx \SI{3.25}{\volt}$ 
compared to $E_\mathrm{c}\approx \SI{4.95}{\volt}$ achieved by the work of \citeauthor{Rajaei2021} \cite{Rajaei2021}. However, since the designs vary quite a bit in size (e.g., see $A_\mathrm{e}$ or $U$) as well as in the process parameters, a complete comparison of the different works is not really possible. 

Nevertheless, it can be shown that all reported works with higher electrode angle $\alpha_\mathrm{e}$ (see Tab.~\ref{tab:concepts}) seem to perform worse compared to the FT-MEs when the electrodes are aligned parallel to each other. This configuration facilitates a uniform electric field across the electrode surfaces, promoting consistent electrochemical reactions and reducing hotspots that could lead to uneven gas production or electrode degradation, and thus, ensures homogeneous reaction~\cite{Olesen2019}. If gas production is uniform over the electrodes, the flow through the electrodes should also be uniform to ensure an efficient bubble detachment and avoidance of blockage of active site at the electrode surface due to bubbles. However, this is not achieved in these T-shaped designs with $\alpha_\mathrm{e} = \SI{0}{\degree}$, sketched in Tab.~\ref{tab:concepts}. 

On this background, the present work used a systematic approach to introduce a new Y-shaped design. Numerical simulations were performed for all concepts of FT-ME designs (I-, T-, and Y-shape) for both electrolyte flow and electrical current distribution. By a combined optimization of these parameters a new design criterion could be developed. It is based on the equilibrium between bubble generation and bubble mobilization and removal.
In addition, experimental studies were performed using Shadowgraphy and Particle Image Velocimetry (PIV) to prove the suitability of the Y-shaped design. 

\begin{landscape}
\topskip0pt
\vspace*{\fill}
\begin{table}[htb!]
    \centering
      \caption{
    Overview of recent flow-through designs and their characteristics. All quantities defined in the text.
    Annotations: *)  not provided, **) only buoyancy-driven separation of gas products without any forced flow. ***) microfluidic device.
    }
    \begin{tabular}{l|c|c|c|c|c|c|c|c}
    \toprule
       \multirow{2}{*}{authors} & \multirow{2}{*}{year} & \multirow{2}{*}{shape} & $\alpha_\mathrm{e}$ & $s$ &$A_\mathrm{e}$ & $U$ & $j_\mathrm{max}$ & $E_\mathrm{c}(j_\mathrm{max})$ \\
        & & & in \si{\degree} & in \si{\milli\metre} & in \si{\square\centi\metre} & in \si{\metre\per\second} &in \si{\milli\ampere\per\centi\metre\squared} & in \si{\volt} \\
    \specialrule{\cmidrulewidth}{0pt}{0pt}
        \citeauthor{Gillespie2015} \cite{Gillespie2015}     & 2015 & T & 0              & 0.8 \ldots 5.5    & 0.8 \ldots 2.5	& 0.075 \ldots 0.2  & 3833.63           & 3.5\\
        \citeauthor{Hartvigsen2015} \cite{Hartvigsen2015}   & 2015 & T & 0              & 0.1               & --\,*)            & 0.00045           & $150$             & $\approx 2.025$\\
        \citeauthor{Neil2016} \cite{Neil2016}               & 2016 & I & 15\ldots 90    & 0 \ldots 13       & 0.24 \ldots 0.93 	& 0 \ldots 0.198    & $\approx 220$     & $\approx 2.5$ \\
        \citeauthor{Gillespie2017} \cite{Gillespie2017}     & 2017 & T & 0              & 2.5               & 344.32 			& 0.03 \ldots 0.04  & $\approx 2700$    & 4 \\
        \citeauthor{Gillespie2018} \cite{Gillespie2018}     & 2018 & T & 0              & 2.5 \ldots 4.5    & 30.17 			& 0.05 \ldots 0.1   & $\approx  4000$   & $\approx 3.25$ \\
        \citeauthor{Davis2019} \cite{Davis2019}             & 2019 & I & 90             & 2 \ldots 45       & 0.6 				& 0.005             & $\approx 107$     & 2.7\\
        \citeauthor{Bui2020} \cite{Bui2020}                 & 2020 & I & 30             & $1 \ldots 45$     & 7.5 				& 0**)				& 50                & $\approx 2.6$\\       
        \citeauthor{Rajaei2021} \cite{Rajaei2021}           & 2021 & T & 0              & 1.1               & --\,*)			& 0.028 \ldots 0.05 & 4000             	&  $\approx 4.95$ \\     
        \citeauthor{Hadikhani2021} \cite{Hadikhani2021}     & 2021 & I & $\approx 0.8$  & 0.55 \ldots 0.69  & 0.00347 			& $\le  0.4$        & $\approx 420$     & 3\\
        \citeauthor{SamirDe2023} \cite{SamirDe2023} & 2023 & MF-T***) & 0 & 1 & -*) & 0.34 & 747 & 2.5 \\
    \bottomrule
    \end{tabular}
    
    \label{tab:literature_overview}
\end{table}
\vspace*{\fill}
\end{landscape}

\section{Material and methods}
\label{sec:methods}

\subsection{Simulation method for electrolyte flow}
\label{sec:cfd_simulation_method}
\begin{figure}[ht]
	\centering
	\hspace*{0pt}{\rlap{\subcaptionmark\label{fig:Geometry_Y_zoom}}\adjustbox{trim=-7mm 0mm 0mm 0mm, valign=t}
		{\includegraphics[height=.29\textwidth]{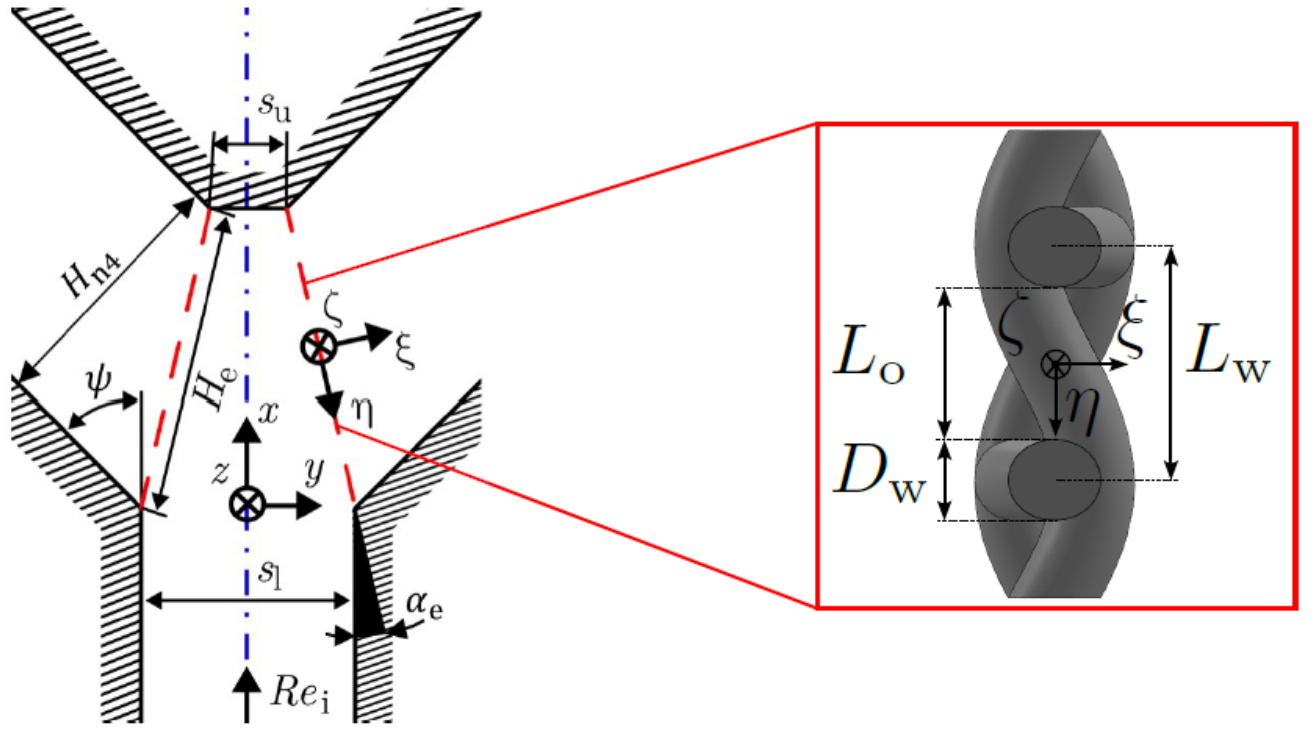}}}
	\hspace*{0pt}{\rlap{\subcaptionmark\label{fig:wire-screen_foto}}\adjustbox{trim=-7mm 0mm 0mm 0mm, clip, valign=t}
		{\includegraphics[height=.29\textwidth]{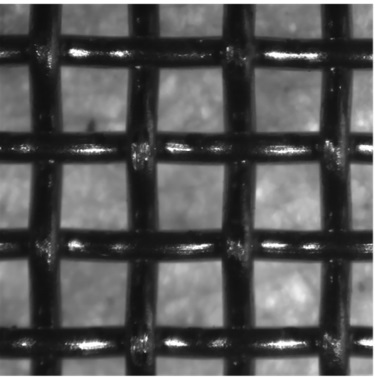}}}
	\caption{\subref{fig:Geometry_Y_zoom}~Centerplane view of the Y-junction of the Y-shape membraneless electrolyzer. Parametrization of geometry together with a global coordinate system and an electrode-fixed coordinate system. Electrodes indicated with red dashed lines and with highlighted side view of the wire 
		mesh of the electrode serving as geometry model the simulations.
		\subref{fig:wire-screen_foto}~Front view of the wire screen, photography.
	}
	\label{fig:wire-screen}
\end{figure}

The computational method used in this study to simulate the electrolyte flow in the different configurations of FT-MEs is based on the solution of the Navier-Stokes equations for single-phase, laminar, unsteady flow 
\begin{subequations}
\begin{align}
    \nabla \cdot \boldsymbol{u} &= 0      \label{eq:conti} \\
    \frac{\partial \boldsymbol{u}}{\partial t}
    + \left( \boldsymbol{u} \cdot \nabla \right) \boldsymbol{u} 
    + \frac{1}{\rho} \nabla p
    &=
    \nu \nabla^2 \boldsymbol{u}
    + \boldsymbol{S} \, ,                     \label{eq:momentum} 
\end{align}
\label{eq:navier_stokes}
\end{subequations}
with $\boldsymbol{u}$ the velocity vector, $p$ the pressure, $\rho$ the fluid density, and $\nu$ the kinematic viscosity. Density and viscosity are assumed constant.
The liquid considered is a 1M KOH solution with properties given in Tab.~\ref{tab:flow_parameters}. The electrodes consist of a wire mesh, shown in Fig.~\ref{fig:wire-screen}, with a wire diameter of $D_\mathrm{w}=\SI{0.224}{mm}$ and an aperture width of $L_\mathrm{o} = \SI{0.425}{mm}$ resulting in a porosity of $\beta = 0.429$. The electrode height $H_\mathrm{e}$ is \SI{10}{mm} for all cases. The upper distance between the electrodes $s_\mathrm{u}$ varies between \SI{1}{mm}, \SI{2.5}{mm} and \SI{7}{mm} for I-, Y- and T-shape, respectively, which results in a change of $\alpha_\mathrm{e}$ due to the constant lower electrode gap of $s_\mathrm{l}= \SI{7}{\milli\metre}$. The complete geometry of all studied FT-ME designs is shown in Fig.~\ref{fig:3D_CAD} and Tab.~\ref{tab:comparison_geometry_parameters}.

To make numerical simulations feasible without prohibitive cost, it is necessary to employ a homogenized representation of the electrode in the form of a porous medium. 
To this end, a new model was developed by the present authors specifically designed for cases characterized by sharp angles between the approaching flow and the electrode planes \cite{Schoppmann_Froehlich_model}.
This model accounts for the deflection of the flow by the porous screen and introduces as vector-valued momentum source term $\boldsymbol{S}$ in Eq.~\eqref{eq:momentum} vanishing everywhere except within a layer of thickness $T$ representing the electrode. The model is formulated in the Cartesian coordinate system $(\xi, \eta, \zeta)$ of the electrode based on its principal axes (Fig.~\ref{fig:Geometry_Y_zoom}), with $\boldsymbol{e}_i$ the unit vector pointing in direction $i=\xi, \eta, \zeta$ and $u^\mathrm{e}_i$ the corresponding velocity component. 
In this reference frame the source term reads
    \begin{equation}
        \label{eq:source_term_alternative}
        \boldsymbol{S}^\mathrm{e}
        = 
        - \left( 
        \mu\, \boldsymbol{D}^\mathrm{e} \cdot \boldsymbol{u}^\mathrm{e} 
        + \frac{\rho}{2}\, \boldsymbol{F}^\mathrm{e} \cdot 
           \left( 
            \sum_{i} \lvert u_i^\mathrm{e} \rvert \; u_i^\mathrm{e} \,\boldsymbol{e}_i 
           \right)
        \right) 
        \, 
    \end{equation}
with the tensors 
    \begin{equation}
    \label{eq:resistance_tensors_local}
        \boldsymbol{D}^\mathrm{e} =
        \begin{pmatrix}
            D & 0                       & 0            \\
            0          & f_\mathrm{D} D & 0            \\
            0          & 0            & f_\mathrm{D} D
        \end{pmatrix}
        ,\quad 
        \boldsymbol{F}^\mathrm{e} =
        \begin{pmatrix}
            F & 0                       & 0            \\
            0          & f_\mathrm{F} F & 0            \\
            0          & 0                       & f_\mathrm{F} F
        \end{pmatrix}
        \, 
    \end{equation}
reflecting the anisotropy between the coordinate direction $\xi$ and the other two directions.
The parameters $D$, $f_\mathrm{D}$, $F$, $f_\mathrm{F}$ in Eq.~\eqref{eq:resistance_tensors_local} were determined for the given electrode using fully resolved Direct Numerical Simulations of the flow through a representative volume element of the wire screen \cite{Schoppmann_Froehlich_model}. The numerical values employed are provided in Tab.~\ref{tab:calibrated_parameters}. 

Due to symmetry only one half of the full geometry ($y\geq 0$) was simulated to save computational resources.
The electrode was represented as a homogeneous zone with anisotropic properties modeled by the additional source term~\eqref{eq:source_term_alternative} in the momentum equation~\eqref{eq:momentum}.
The boundaries of the computational domain were defined by imposing a symmetry condition at $y=0$ and a no-slip condition along walls. 
A uniform velocity $U_\mathrm{p1}$ (Tab.~\ref{tab:flow_parameters}) was imposed at the inlet and a pressure-outlet condition at the outlet of the cell geometry.

The computational grid was generated with Ansys Fluent Meshing, employing a structured hexaeder grid in the bulk flow region and polyhedral prism cells along boundaries. Both regions were connected by a region of general polyhedral cells. The average cell size in the bulk flow region is $h\approx \SI{0.1}{mm}$, leading to a total number of cells of $N_\mathrm{c} \approx 3 \ldots \SI{6}{\mathrm{M}}$ cells, depending on the geometry. The porous zone representing the electrode was discretized with at least 3~cells over the thickness.

The Navier-Stokes equations \eqref{eq:navier_stokes} were solved without any turbulence modeling since the Reynolds number of the channel flow suggests a laminar solution (Tab.~\ref{tab:flow_parameters}). 
A second-order Finite-Volume Method with a pressure-based coupled algorithm implemented in the commercial package Fluent (version 2021, R2) was used. A second-order implicit scheme was chosen for time integration allowing a Courant number of $1.5$, and the solver was set to a maximum of 20 iterations per time step.

The flow field was initialized by the solution of a previous steady-state simulation.
The transient simulations were performed over 5000 time steps in total, corresponding to a total physical time of $T_\mathrm{tot} \approx 20\, T_\mathrm{ref}$ with $T_\mathrm{ref} = H_\mathrm{e} / U_\mathrm{i}$ and $U_\mathrm{i}$ as the velocity at the inlet of the junction. Averaging was undertaken for a duration of $T_\mathrm{av} \approx 12\, T_\mathrm{ref}$ after a startup time of $T_\mathrm{st} \approx 8\, T_\mathrm{ref}$.

\subsection{Simulation of current distribution}
\label{sec:simulation_current_distribution}
The current simulations were performed in two-dimensional settings, assuming uniform current distribution in $z$-direction. 
The homogenized geometry employed for the current simulation follow the CFD geometry, shown in  Fig.~\ref{fig:3D_CAD} and Tab.~\ref{tab:comparison_geometry_parameters}. 
For simplification the electrodes were represented here by straight lines confining the computational domain. The entry channel was given a length of \SI{40}{mm} suitable for the current to decay by a sufficient factor. 

For all simulations performed, the overall electrical current 
\begin{equation}
    I_\mathrm{c} = \int_{\partial\Omega_\mathrm{c}} \bm{j}\cdot \bm{n} \; \mathrm{d}S
    \label{eq:bc_cathode}
\end{equation}
was imposed,
with $\partial\Omega_\mathrm{c}$ the surface of the cathode.
Setting $I_\mathrm{c} = \SI{-1}{\ampere}$ for the cathode results in a mean current density $J_\mathrm{c} = \SI{-1}{\ampere\per\centi\metre}$. 
The potential of the anode was set to $\Phi_\mathrm{a} = \SI{0}{\volt}$. 
The walls were considered to be insulating, 
so that $\bm{j}\cdot \bm{n}=0$ was imposed there. 
The same was done at the boundary limiting the upstream duct in all geometries. A graphical overview of the applied boundary conditions for the current simulations is given in Fig.~\ref{fig:current_simulation_setup}. The conductivity of the electrolyte was set to \SI{20.13}{S\per\metre} for 1M KOH at $\SI{20}{\celsius}$ after \cite{Gilliam2007}. 

The stationary distribution of the electric current was computed using the finite element software package COMSOL Multiphysics\textsuperscript{\textregistered} V6.2 \cite{comsol} and its AC/DC Module in combination with the implemented MUMPS solver with default parameters. Linear elements were employed, so that the method is of second order in space. The numerical meshes for the \mbox{I-,} T- and Y-shaped cell contain 15344, 19744, and 16192 triangles, respectively, with refinement at the electrode surface and the electrode gap. The edge length of the triangles varies in the range of about~$\num{0.125} \ldots \SI{0.8}{\milli\metre}$.

\subsection{Experimental setup}
\label{sec:exp_setup}

The Y-shaped geometry was 3D printed with DraftGrey (Stratasys, USA) using an Objet30 Prime V5 printer (Stratasys, USA). This material is resistant to KOH which was used as an electrolyte in all experiments (1M KOH, Titripur, Merck, Germany). Two observation windows made of PMMA (polymethyl methacrylate) were installed for optical access. The electrodes, cathode and anode, consisted of woven Ni-meshes as specified above (Fig.~\ref{fig:wire-screen}). For technical reasons, the measurement section was mounted horizontally, iÌ.\ the $x$- and $y$-coordinate in Fig.~\ref{fig:Geometry_Y_zoom} were horizontal and the $z$-direction vertical, against gravity. 

To apply a constant and pulsation-free electrolyte flow to the cell, a peristaltic pump (LabV6-III, Shenchen, China) was used in combination with a pulse dampener (Lead Fluid, China). An electrochemical workstation CHI660E (CH Instruments, USA) was connected to the cell to run the electrolysis under galvanostatic condition. The electrolyte was recirculated through the system and the gas produced could only escape the process in the reservoir. As a result, a small number of micro-bubbles were recirculated under certain conditions, which can be seen in some of the figures, without being detrimental to the present findings. All experiments were carried out under ambient conditions, i.e.\ $T \approx \SI{293}{\kelvin}$, $p \approx \SI{1}{\bar}$.

To determine the velocity field of the electrolyte  in the absence of bubbles, a micro Particle Image Velocimetry ($\mu$-PIV) system was used with front illumination of the measurement plane employing a Nd-YLF Laser (\SI{527}{\nano\metre}, Photonics Industries, USA). A high-speed camera (Phantom VEO 410L \SI{1280}{px}~$\times$~\SI{800}{px}, AMETEK, USA) was employed, connected to a Stereo Discovery.V8 microscope with PlanApo S $1.0$x objective (Zeiss, Germany) at a spatial resolution of $\SI{98.26}{px\per\milli\metre}$. The frame rate was varied with the fluid velocity in the range $\num{2800}\ldots\SI{5200}{\hertz}$. 
Polystyrene particles ($\rho_\mathrm{t} = \SI{1050}{\kilo\gram\per\metre\cubed}$, microParticles GmbH, Germany) were used as tracer particles with a mean diameter of $d_\mathrm{t} = \SI{33.03}{\micro\metre}$ and an incorporated fluorescent dye (rhodamine, red fluoroscent: excitation and emission at $\SI{530}{\nano\metre}$ resp.\ $\SI{607}{\nano\metre}$ wavelength). Their size was chosen to yield a Stokes number well below 1 and a sufficient particle image size of at least $\SI{4}{px}$. The PIV-measurements were carried out in three different measuring planes, one in the centre of the cell and $\pm \SI {2.5}{\milli\metre}$ shifted in $z$-direction. Here, only centerplane data are reported.
Details on the image processing of the PIV data can be found in Sec.~\ref{sec:appendix_piv_processing}.

High-speed shadowgraphy imaging was used to obtain detailed information about the evolving H\textsubscript{2} and O\textsubscript{2} bubbles and their transport out of the electrode gap. 
A camera of the type  OS-7 S3 (\SI{1920}{px}~$\times$~\SI{1280}{px}, IDT, USA) and a FUSION objective (Optem, USA) with a magnification of 2 were used. The spatial resolution was \SI{123}{px\per\milli\metre} with a TH2 LED panel (CCS, Japan) as background illumination. 

As a measure of the void fraction, gray value distributions (GVD) were calculated in slices parallel to the electrode. To this end, each image was first divided by an averaged background image to enhance the contrast between bubbles and electrolyte. Then, the image was binarized using the iso-data threshold implemented in scikit image 0.21.0 and the arithmetic mean over \SI{10}{px} in $\xi$-direction to reduce noise. The resulting GVD, denoted $G$ here, then represents a measure for the presence of bubbles in form of a smoothed integral in spanwise direction. By definition, $0 \le G \le 1$. Note that with the shadowgraphy device employed a single bubble did not generate black, but an intermediate gray value, so that several bubbles in spanwise direction, one behind the other, were needed to yield $G=1$. Due to the saturation entailed, once a certain number of bubbles reached $G$ is no longer proportional to the void fraction while smaller void fractions are well represented by G.

\section{Results and discussion}
\label{sec:results}
In the following, the naming scheme of the FT-ME concepts follows the combination of the shape, i.e. I-, T- or Y-shape, (Tab.~\ref{tab:concepts}) and the upper electrode distance $s_\mathrm{u} = \SI{1}{mm}, \SI{7}{mm}, \SI{2.5}{mm}$: I1, T7, Y2.5, etc.

\subsection{Electrolyte flow field}
\label{sec:results_cfd}
\begin{figure}[ht!]
\centering
    \hspace*{0pt}{\rlap{\subcaptionmark\label{fig:I1_streamlines}} \adjustbox{trim=-5mm 0mm 0mm 0mm, valign=t}{\includegraphics[width=.62\textwidth]{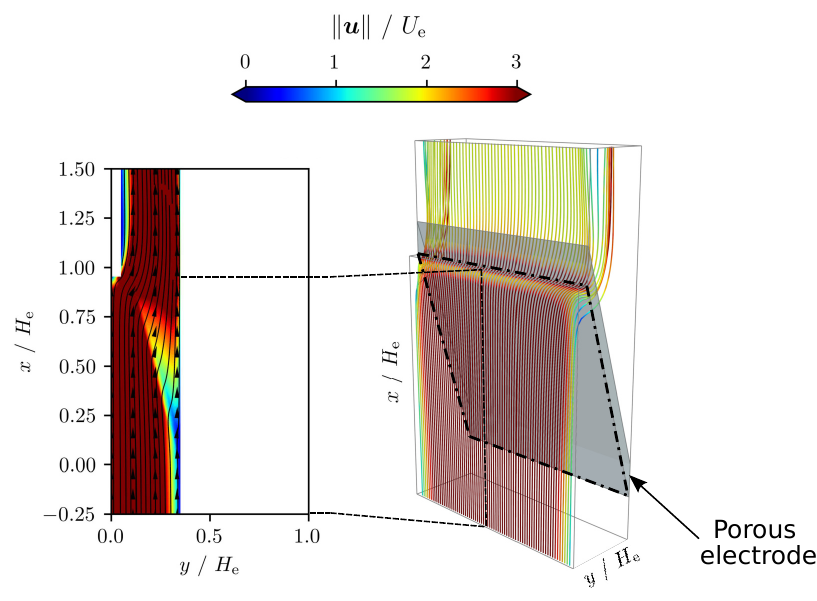}}}\\
    \hspace*{0pt}{\rlap{\subcaptionmark\label{fig:T7_streamlines}} \adjustbox{trim=-5mm 0mm 0mm 0mm, valign=t}{\includegraphics[width=.62\textwidth]{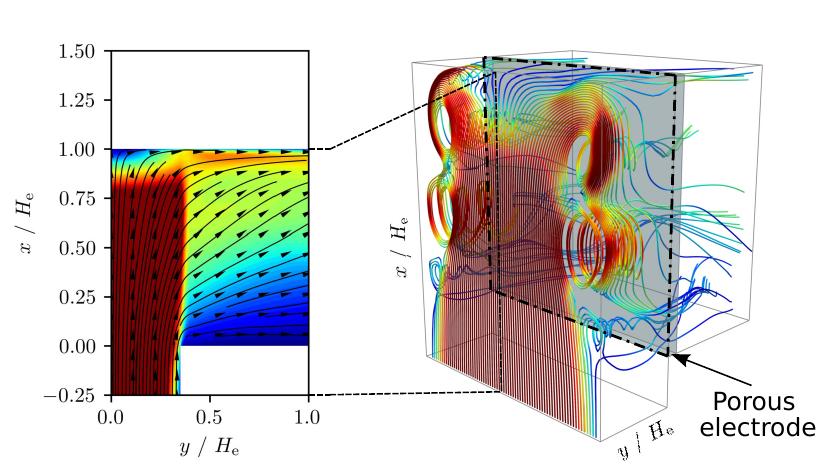}}}\\
    \hspace*{0pt}{\rlap{\subcaptionmark\label{fig:Y2p5_streamlines}} \adjustbox{trim=-5mm 0mm 0mm 0mm, valign=t}{\includegraphics[width=.62\textwidth]{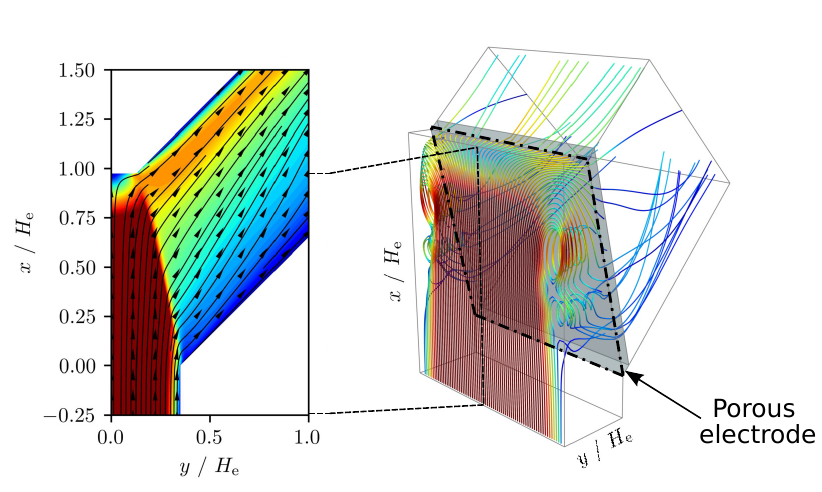}}}
\caption{
Streamlines and mean velocity obtained for the different variants of the flow-through electrolyzers.
\subref{fig:I1_streamlines}~Electrolyzer I1,
\subref{fig:T7_streamlines}~electrolyzer T7,
\subref{fig:Y2p5_streamlines}~electrolyzer Y2.5.
Left column:
Magnitude of time-averaged velocity $\lVert \langle \boldsymbol{u} \rangle \rVert$ together with streamlines in $xy$-plane at $z = 0$ (centerplane of flow cell).
Right column:
Three-dimensional view with selected streamlines of instantaneous flow
emitted along a line of points in $z$-direction at $x=-0.2/H_\mathrm{e}$ and $y=0.01/H_\mathrm{e}$, 
and colored with the velocity magnitude according to the same scale as in the left column.
Homogenized electrode is rendered gray with highlighted upstream plane (dash-dotted line).
Only a part of computational domain is shown. 
} \label{fig:streamlines}
\end{figure}

The left column in Fig.~\ref{fig:streamlines} provides the streamlines in the centerplane and the magnitude of the velocity. The latter is normalized by the mean velocity, $U_\mathrm{e}$, through the electrode 
\begin{equation}
    U_\mathrm{e} = \frac{U_\mathrm{i}}{2} \, \frac{s_\mathrm{l}}{H_\mathrm{e}} \;,
\end{equation}
with $U_\mathrm{i}$ the velocity at the inlet of the junction. Both remain unchanged across all configurations investigated 
(Tab.~\ref{tab:flow_parameters}). 

\subsubsection*{I1-Electrolyzer}
In Fig.~\ref{fig:I1_streamlines} it is obvious that the overall deflection of the flow is small for the I1-electrolyzer. The angle of the flow with the electrode plane, on the other hand, is sharp, and the sidewall inhibits stronger turning of the flow. The streamlines feature a turning point, first being deflected by the electrode towards the normal of the electrode when passing the wire screen, then deflected in opposite direction by the presence of the sidewall. As a result, the flow through the electrode is stronger in the top compared to the bottom, demonstrated by Fig.~\ref{fig:elec_inlet_I1_xi_av} displaying the time-averaged normal velocity component through the electrode with values around zero in a broad lower region. The slower oncoming near-wall flow upstream of the electrode also contributes to this feature.  

Furthermore, the streamlines in Fig.~\ref{fig:I1_streamlines} do not show any vortices or perturbations. Only small 3D effects can be discerned behind the electrode near the sidewalls. This is also the case for the velocity magnitude at the upstream side of the electrode in Fig.~\ref{fig:elec_inlet_I1_mag}, where only little disturbance by the sidewalls at $\zeta/H_\mathrm{e} = \pm 0.5$ is visible. Fig.~\ref{fig:elec_inlet_I1_xi_rms} shows the RMS value of the velocity normal through the electrode and reveals the absence of any unsteadiness in this plane. 

\subsubsection*{T7-Electrolyzer}
Fig.~\ref{fig:T7_streamlines} shows a substantial deceleration of the incoming flow in the stagnation region between the electrodes at the upper end for the T7-electrolyzer. The total deflection of the flow is large and not fully accomplished within the area of view displayed. It occurs mainly at the upper end, while in the lower part it is small.
An overall deceleration of the flow in the junction is expected from the increase of cross-sectional area and is well visible in the part downstream of the electrode.

\begin{figure}[!ht]
    \centering
    
    \begin{tabular}{lccc}
        & \textbf{I1} &\textbf{T7} & \textbf{Y2.5} \\
        \parbox[t]{2mm}{\multirow{1}{*}[-1em]{\rotatebox[origin=c]{90}{$\lVert \langle \boldsymbol{u} \rangle \rVert\ /\ U_\mathrm{e}$}}}   &  {\rlap{\subcaptionmark\label{fig:elec_inlet_I1_mag}}\adjustbox{trim=-5mm 0mm 0mm 0mm, clip, valign=t}{\includegraphics[width=0.24\textwidth]{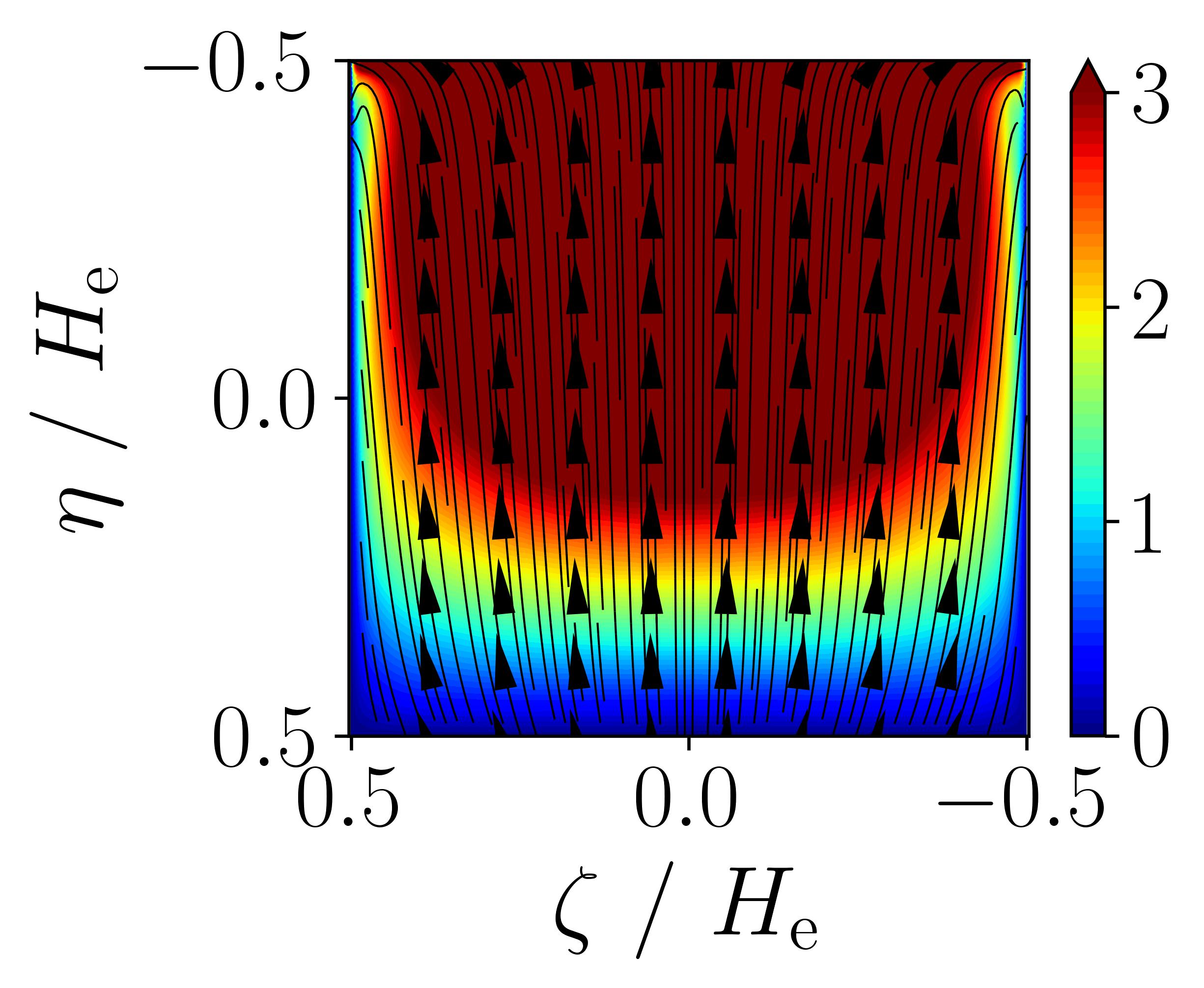}}}         
        & {\rlap{\subcaptionmark\label{fig:elec_inlet_T7_mag}}\adjustbox{trim=-5mm 0mm 0mm 0mm, clip, valign=t}{\includegraphics[width=0.24\textwidth]{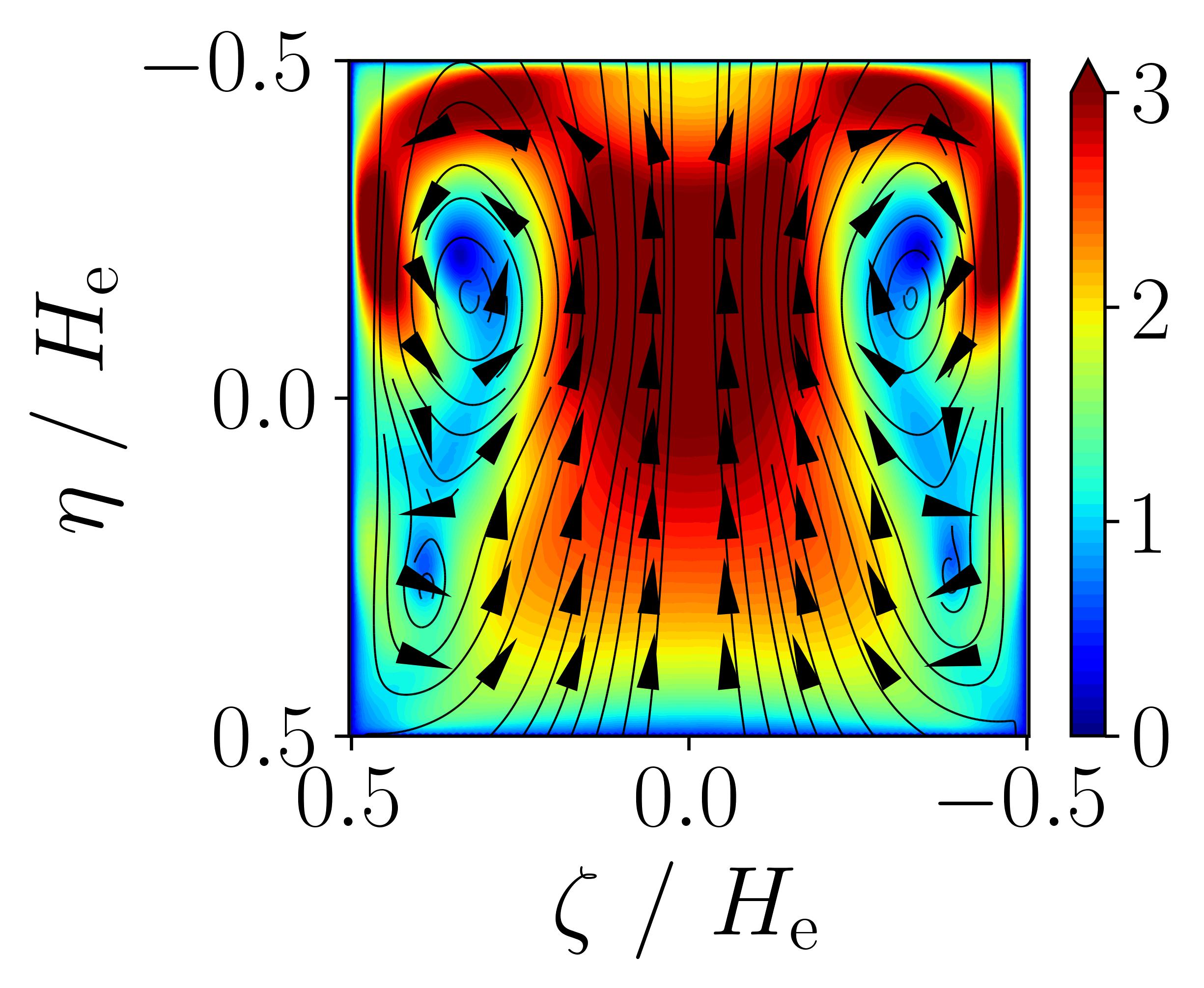}}}      
        & {\rlap{\subcaptionmark\label{fig:Y_elec_inlet_Y2p5_mag}}\adjustbox{trim=-5mm 0mm 0mm 0mm, clip, valign=t}{\includegraphics[width=0.24\textwidth]{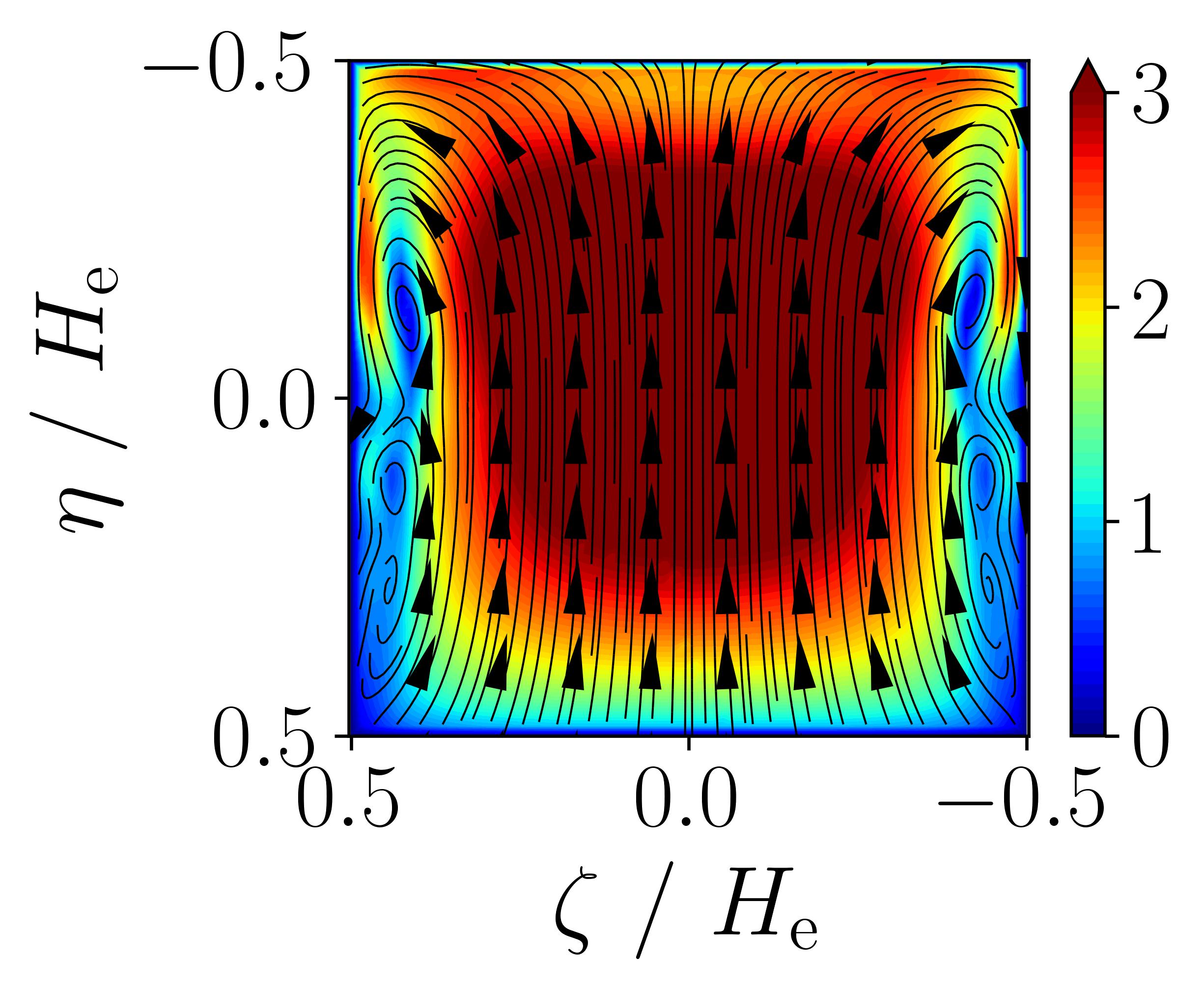}}}  \\     
        \parbox[t]{2mm}{\multirow{1}{*}[-1em]{\rotatebox[origin=c]{90}{$\langle u_\mathrm{\xi} \rangle\ /\ U_\mathrm{e}$}}}   &   {\rlap{\subcaptionmark\label{fig:elec_inlet_I1_xi_av}}\adjustbox{trim=-6mm 0mm 0mm 0mm, clip, valign=t}{\includegraphics[width=0.24\textwidth]{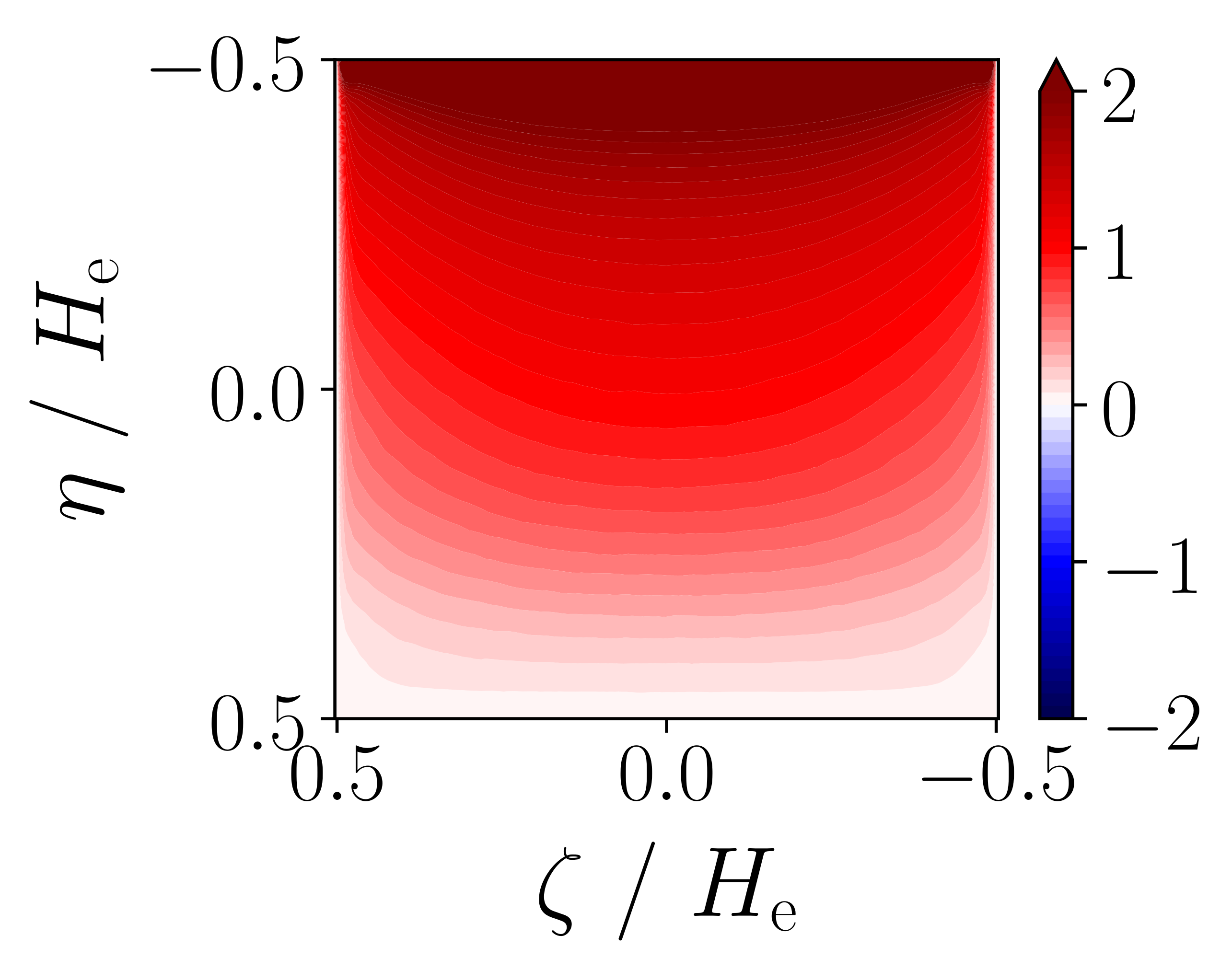}}}
        & {\rlap{\subcaptionmark\label{fig:elec_inlet_T7_xi_av}}\adjustbox{trim=-6mm 0mm 0mm 0mm, clip, valign=t}{\includegraphics[width=0.24\textwidth]{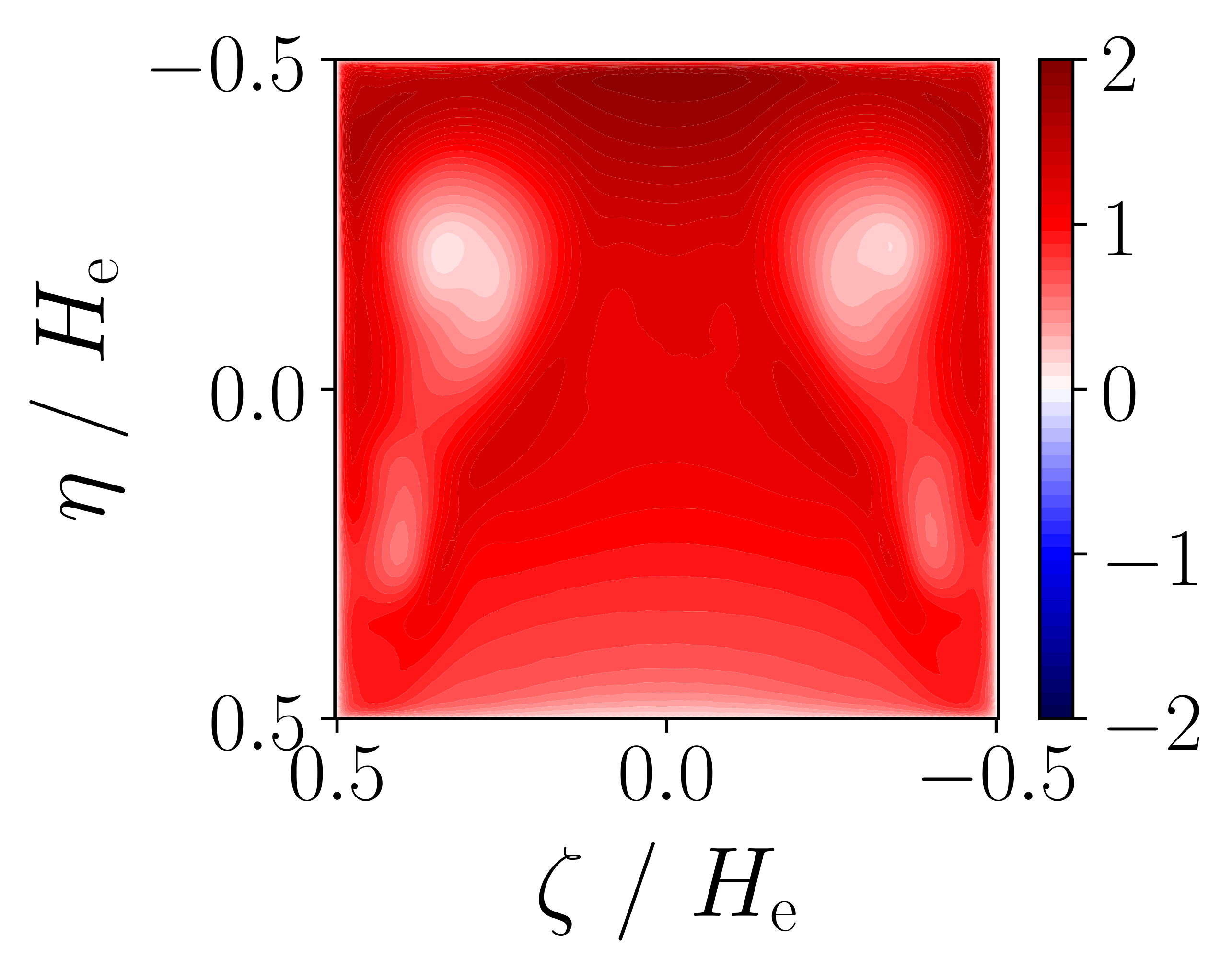}}}
        & {\rlap{\subcaptionmark\label{fig:Y_elec_inlet_Y2p5_xi_av}}\adjustbox{trim=-6mm 0mm 0mm 0mm, clip, valign=t}{\includegraphics[width=0.24\textwidth]{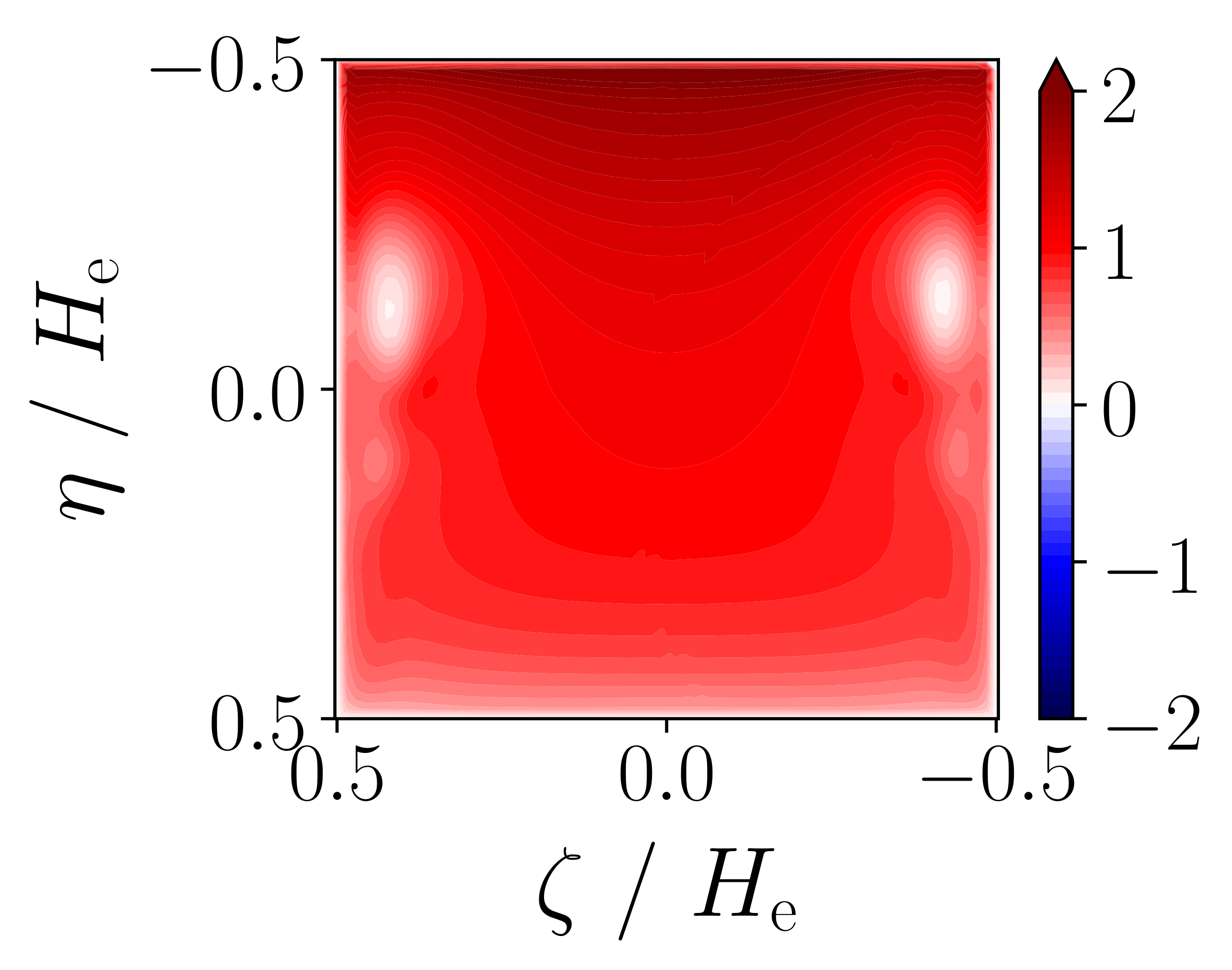}}} \\
        \parbox[t]{2mm}{\multirow{1}{*}[-2em]{\rotatebox[origin=c]{90}{$ u_\mathrm{\xi}\  /\ U_\mathrm{e}$}}}   &  {\rlap{\subcaptionmark\label{fig:elec_inlet_I1_xi}}\adjustbox{trim=-6mm 0mm 0mm 0mm, clip, valign=t}{\includegraphics[width=0.24\textwidth]{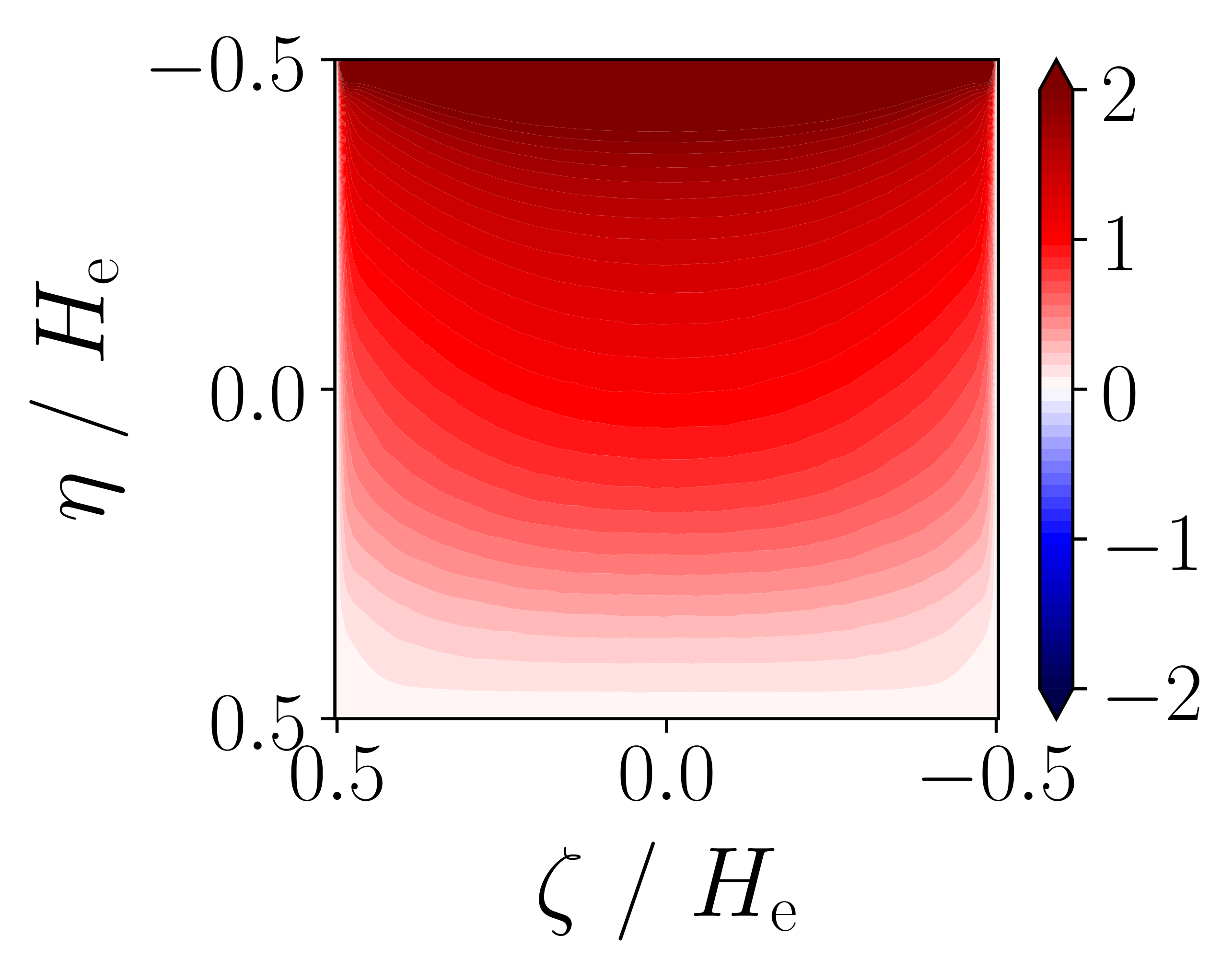}}}
        & {\rlap{\subcaptionmark\label{fig:elec_inlet_T7_xi}}\adjustbox{trim=-6mm 0mm 0mm 0mm, clip, valign=t}{\includegraphics[width=0.24\textwidth]{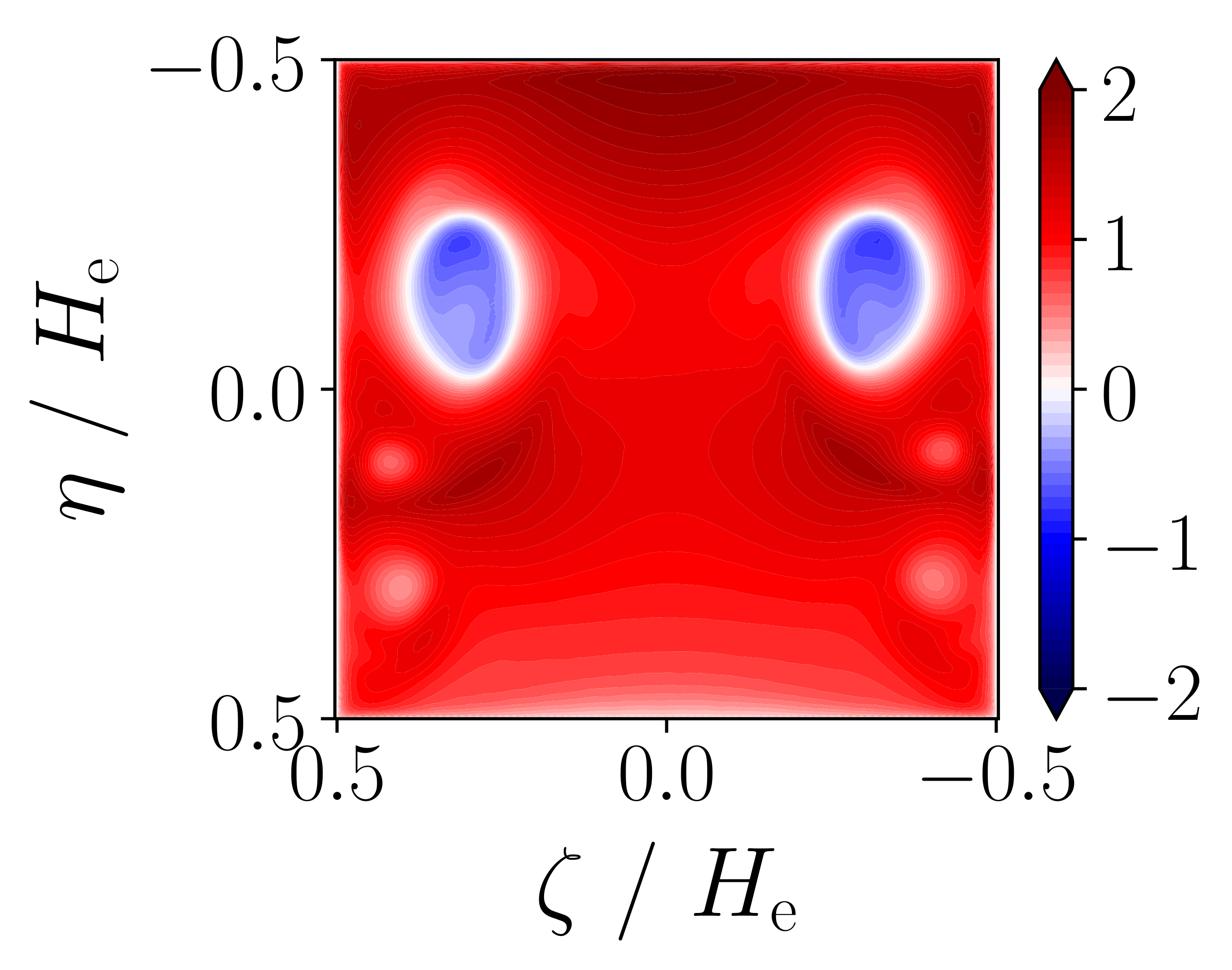}}}
        & {\rlap{\subcaptionmark\label{fig:Y_elec_inlet_Y2p5_xi}}\adjustbox{trim=-6mm 0mm 0mm 0mm, clip, valign=t}{\includegraphics[width=0.24\textwidth]{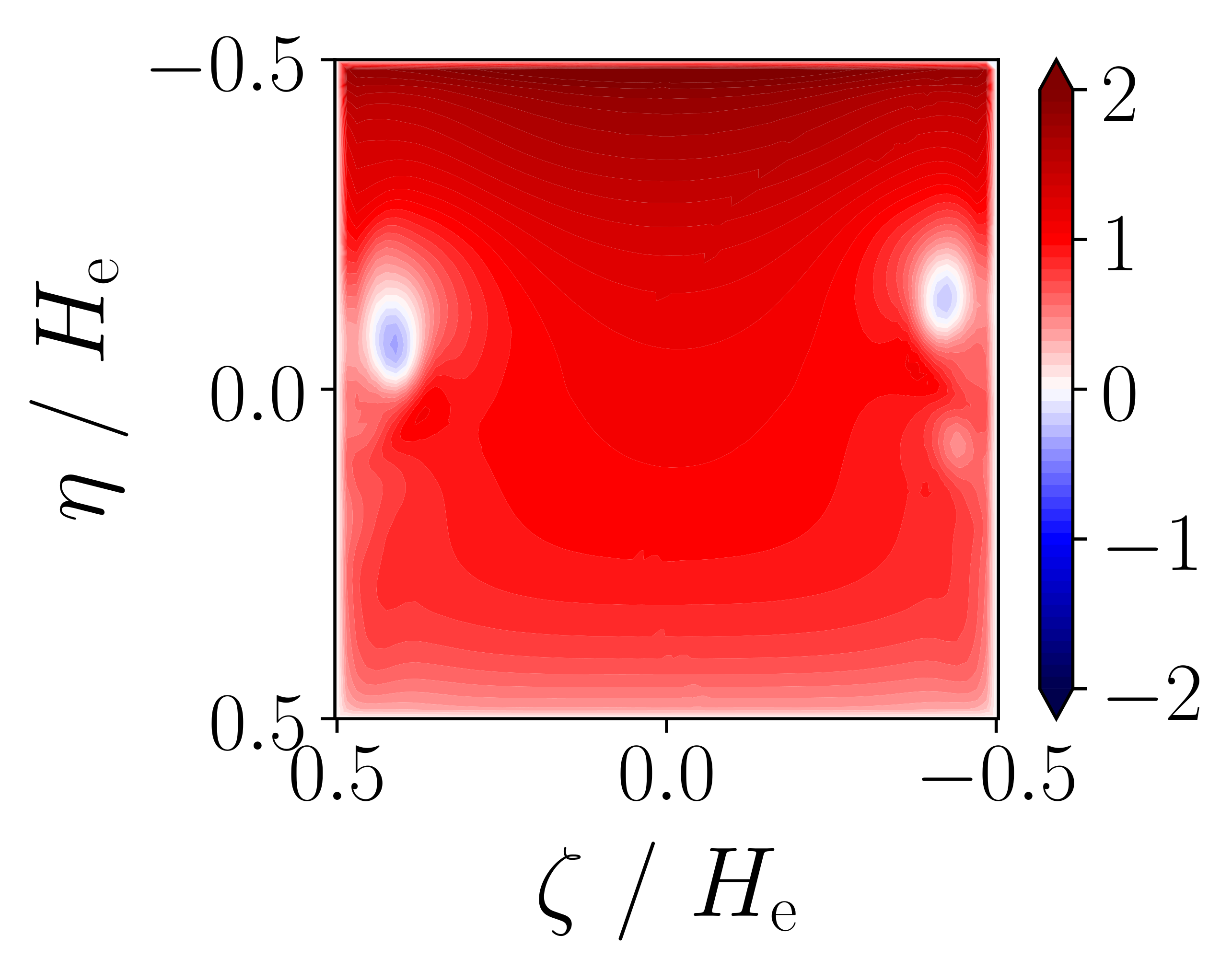}}} \\
        \parbox[t]{2mm}{\multirow{1}{*}[-1em]{\rotatebox[origin=c]{90}{$ \sqrt{\langle u_\mathrm{\xi}' u_\mathrm{\xi}' \rangle }\ /\ U_\mathrm{e}$}}}   &   {\rlap{\subcaptionmark\label{fig:elec_inlet_I1_xi_rms}}\adjustbox{trim=-6mm 0mm 0mm 0mm, clip, valign=t}{\includegraphics[width=0.24\textwidth]{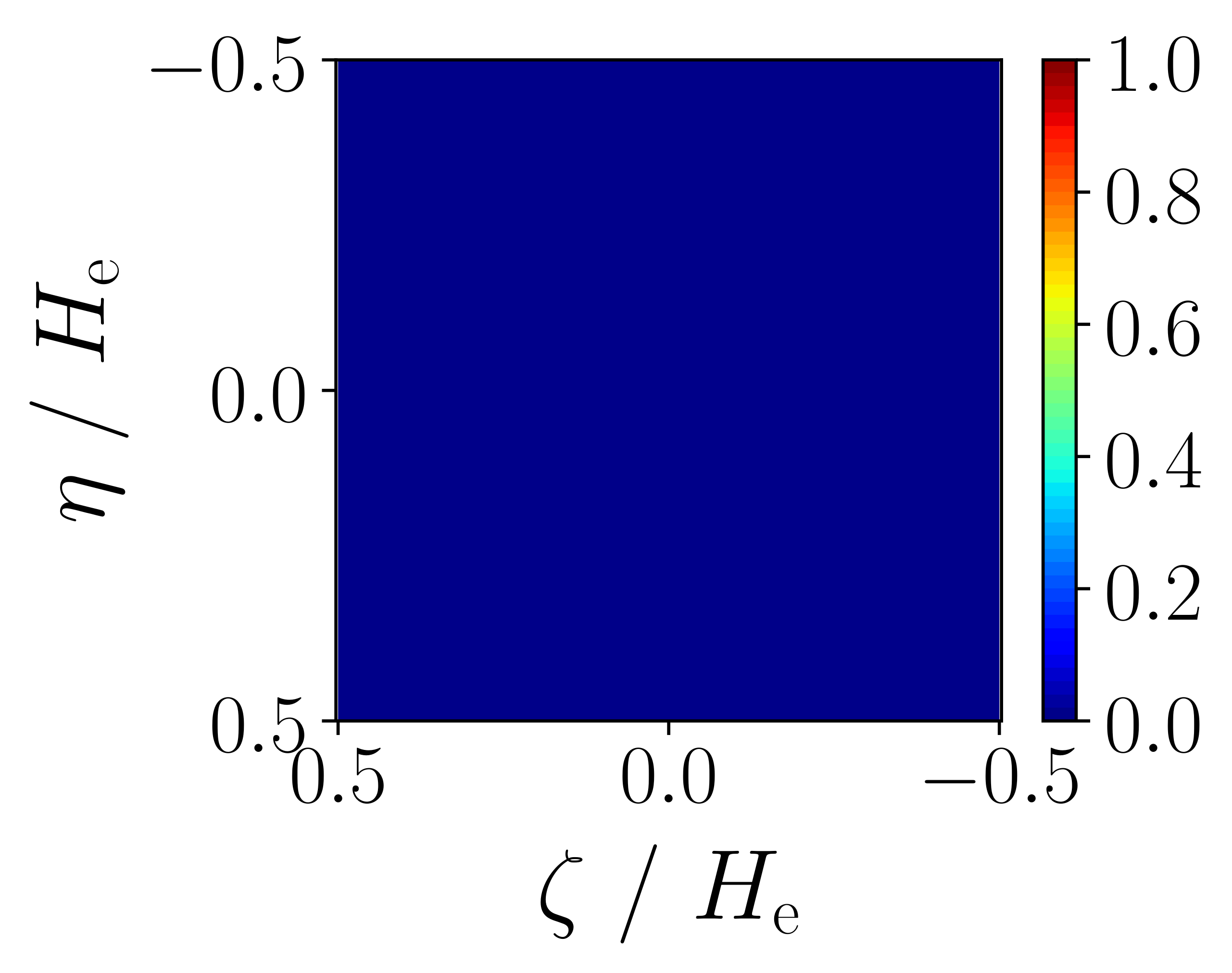}}}  
        & {\rlap{\subcaptionmark\label{fig:elec_inlet_T7_xi_rms}}\adjustbox{trim=-6mm 0mm 0mm 0mm, clip, valign=t}{\includegraphics[width=0.24\textwidth]{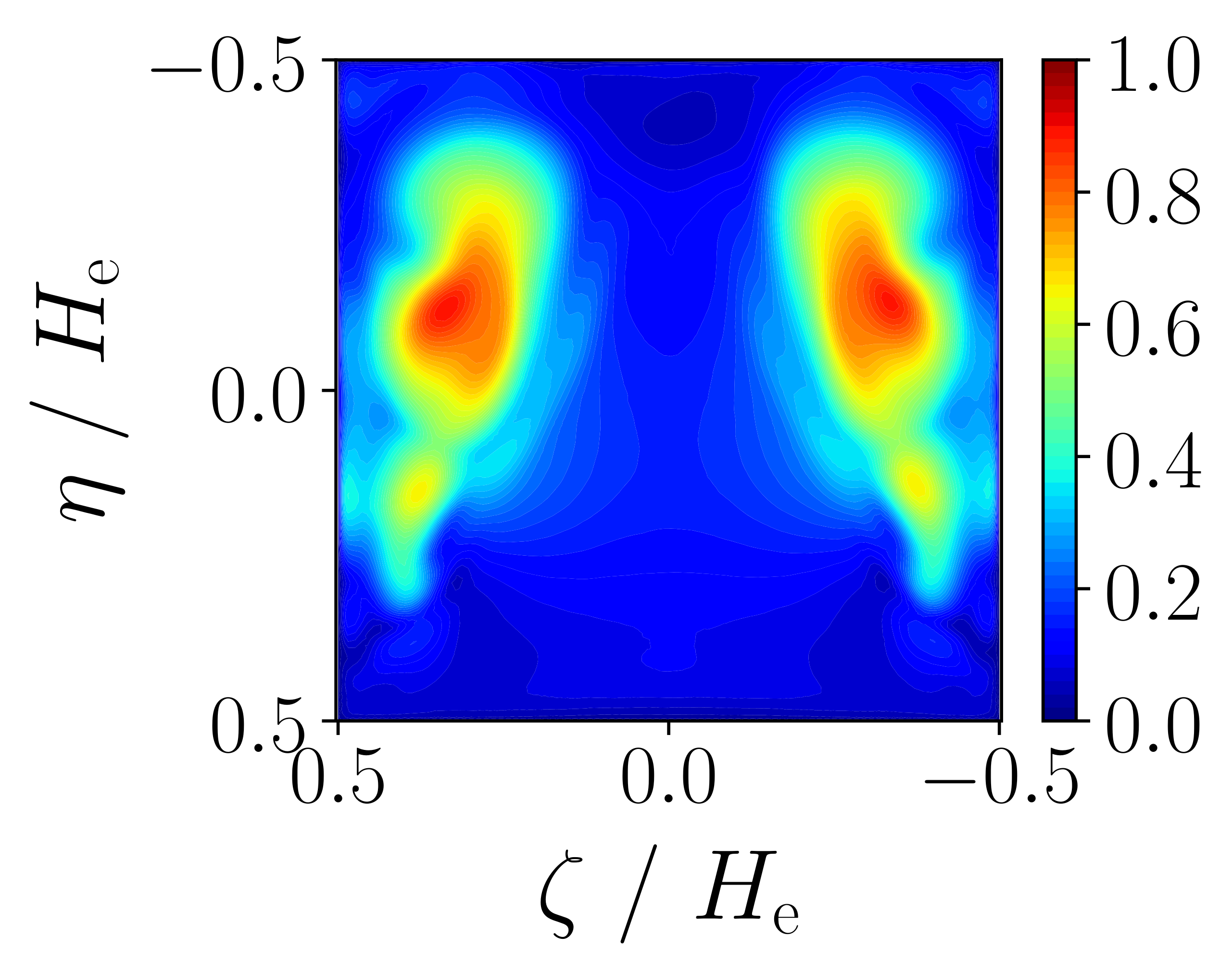}}} 
        & {\rlap{\subcaptionmark\label{fig:Y_elec_inlet_Y2p5_xi_rms}}\adjustbox{trim=-6mm 0mm 0mm 0mm, clip, valign=t}{\includegraphics[width=0.24\textwidth]{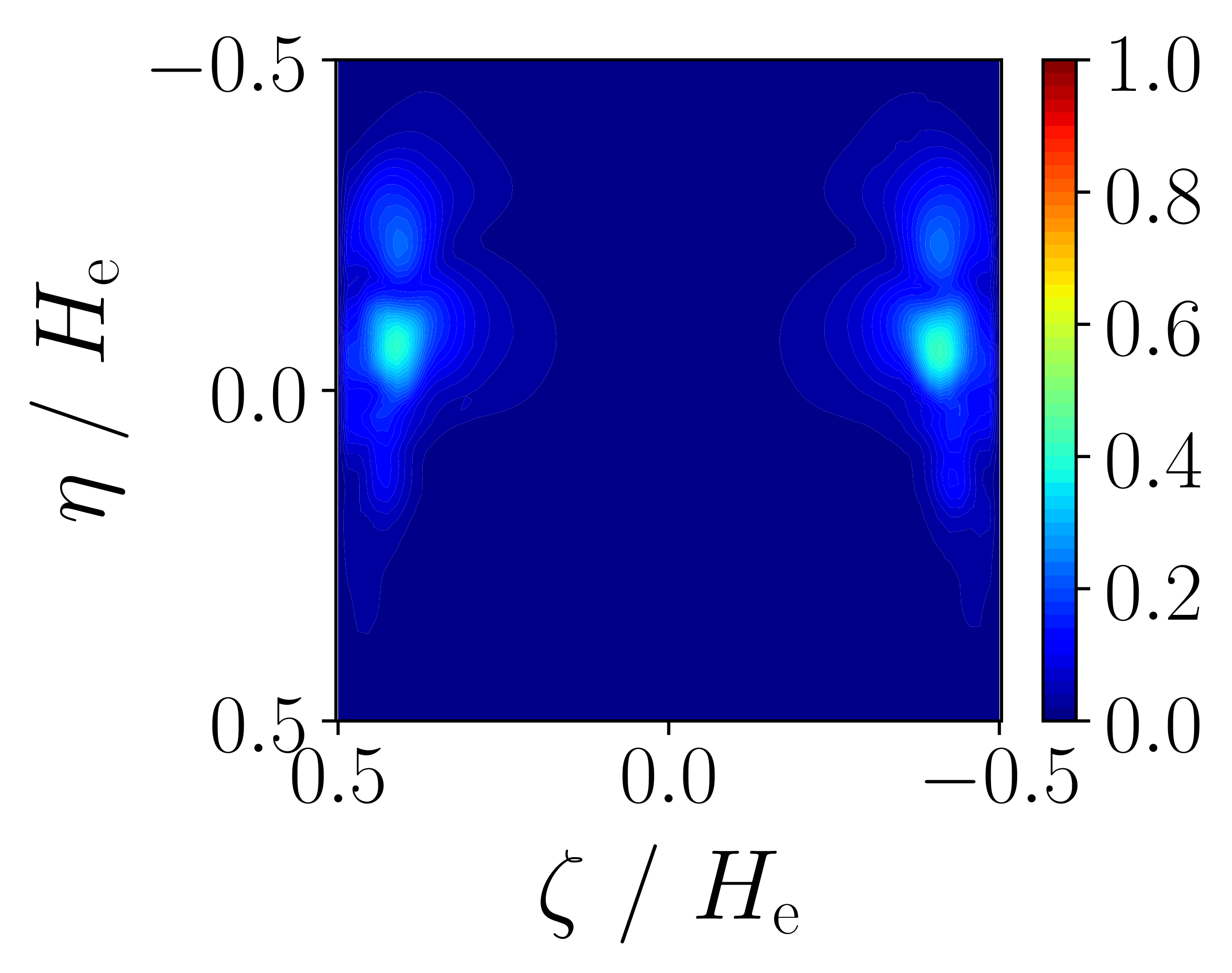}}}
    \end{tabular}
    \caption{Flow in $\zeta\eta$-plane at $\xi = -T/2$ (upstream side of electrode, see highlighted area in Fig.~\ref{fig:streamlines}) for all electrolyzers. From upper to lower row: Magnitude of time-averaged velocity $\lVert \langle \boldsymbol{u} \rangle \rVert$ and streamlines, time-averaged normal velocity component $\langle u_\xi \rangle$, instantaneous normal velocity component $u_\xi$ at an instant with the strongest reverse flow observed over the simulated time and temporal standard deviation of $u_\xi$.} \label{fig:Y_elec_inlet}
\end{figure}

Fig.~\ref{fig:elec_inlet_T7_mag} reveals a strong secondary flow in the upstream electrode plane. In the center, a strong tangential component in upward direction is seen, impinging on the upper wall and then turning downwards, leading to two main symmetrical vortices with almost vanishing values of flow velocity in their center regions. As a consequence, the non-uniformity of the flow is larger compared to the I1-electrolyzer.

The instantaneous 3D streamlines in Fig.~\ref{fig:T7_streamlines} demonstrate that the flow in the T7-electrolyzer is fairly irregular throughout the entire cell. 
Twisting and spiralling streamlines witness strong secondary flow which changes in magnitude over the domain and is highly unsteady. The RMS value of the electrode-normal velocity component in Fig.~\ref{fig:elec_inlet_T7_xi_rms} reaches substantial values around $0.8\,U_\mathrm{e}$. The contour plot in Fig.~\ref{fig:elec_inlet_T7_xi} displays the instantaneous electrode-normal velocity at an instant when strong back flow through the electrode occurs in the vortex cores, with a local magnitude up to around $U_\mathrm{e}$. Such a situation is highly detrimental for the correct mobilization of bubbles generated at the electrodes. The local back flow through the electrode may push the bubbles into the electrode gap, which is highly undesired since prone to generate gas crossover. Animations reveal that the spots with upstream directed flow arise and decay periodically in time with only slight lateral displacement. 

\subsubsection*{Y2.5-Electrolyzer}
As an intermediate between the I1- and the T7-concept, the outlet ducts of the Y-electrolyzer cell form an angle of $\SI{45} {\degree}$ with the inlet, as shown in Tab.~\ref{tab:concepts} above.
An additional parameter is the inclination of the electrodes, also meriting consideration as it bears opportunities for optimization. In the present configuration this angle was altered without changing the electrode size ($H_\mathrm{e} \times H_\mathrm{e}$) by changing the upper distance between the electrodes, $s_\mathrm{u}$ (cf.\ Fig.~\ref{fig:Geometry_Y_zoom}). More specifically, the distances $s_\mathrm{u} = \SI{1}{mm}, \SI{2.5}{mm}, \SI{5}{mm}$ were chosen to provide a suitable screening of the parameter space. The corresponding simulation cases are labeled Y1, Y2.5, and Y5. With increasing $s_\mathrm{u}$ the electrode is more and more parallel to the incoming flow, so that for small values, i.e. in the case Y1, the situation is expected to resemble the one of I1, whereas the situation of case Y5 is expected to be closer to a case T5 (not investigated). In the following, only Y2.5 is discussed as it shows the best results for design criterion
introduced later on. 
The remaining cases can be found in 
Fig.~\ref{fig:appendix_Y_streamlines} and Fig.~\ref{fig:appendix_Y_elec_inlet} of the supplementary material. 

The simulation result in Fig.~\ref{fig:Y2p5_streamlines} 
and Fig.~\ref{fig:Y_elec_inlet} demonstrate that 
the non-uniformity of the flow through the electrode is somewhat larger than for I1
but much smaller than for T7. 
The snapshot in Fig.~\ref{fig:Y_elec_inlet_Y2p5_xi} 
shows slight backflow, but this is the instant with the worst situation. 
The unsteadiness of the flow is small, which was additionally checked using video animations, so that the flow might be addressed as transitional. The variance in Fig.~\ref{fig:Y_elec_inlet_Y2p5_xi_rms} is small in magnitude and concentrated in two small spots on either side backing this statement.

\subsection{Current distribution}
\label{sec:results_current_distribution}
\begin{figure}[htb]
	\centering
    \hspace*{0pt}{\rlap{\subcaptionmark\label{fig:current_streamline_a}}\adjustbox{trim=-5mm 1mm 0mm 0mm, valign=t}
    {\includegraphics[height=60mm]{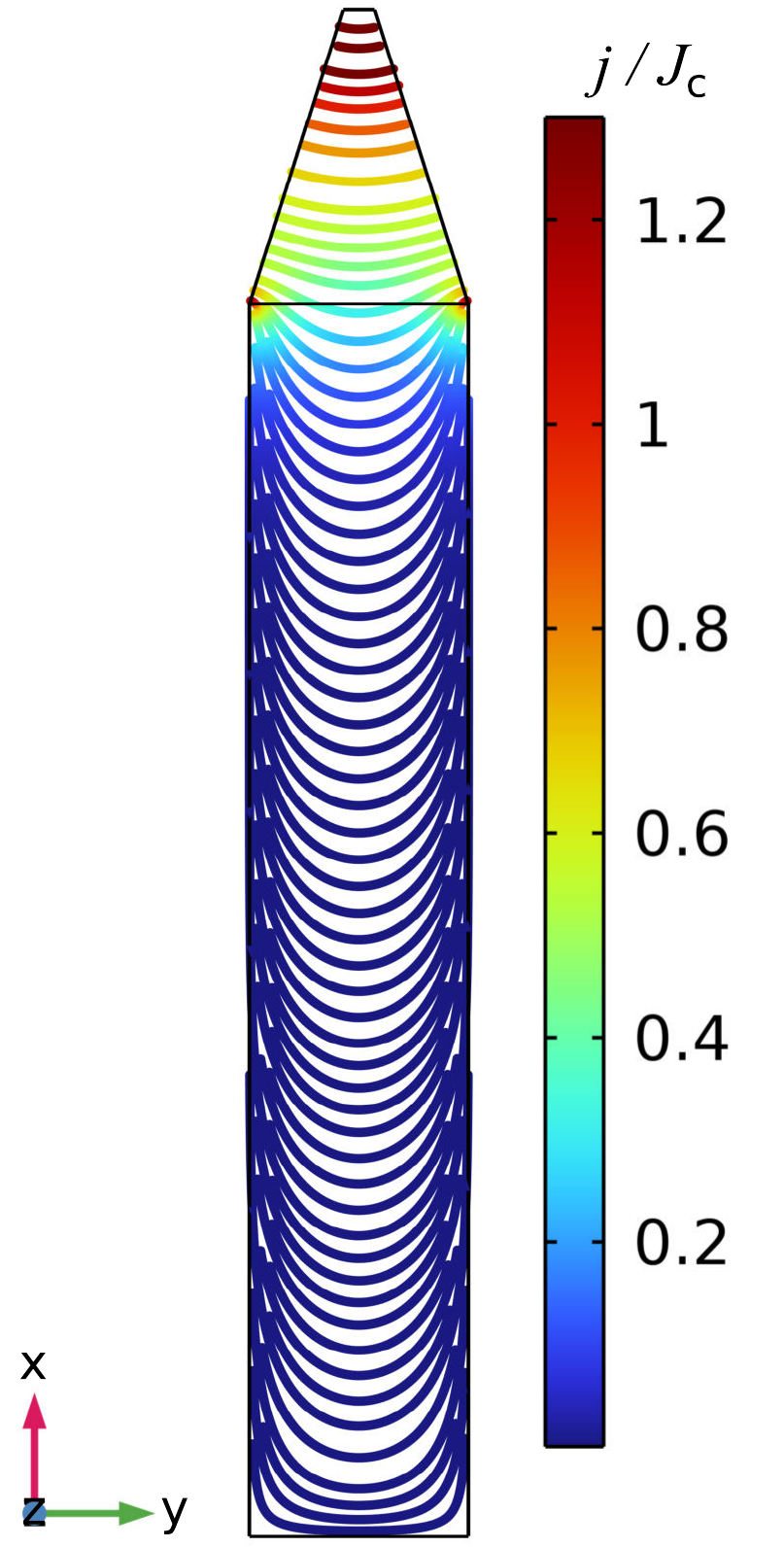}}} 
    \hspace*{0pt}{\rlap{\subcaptionmark\label{fig:current_streamline_b}}\adjustbox{trim=-5mm 1mm 0mm 0mm, valign=t}
    {\includegraphics[height=60mm]{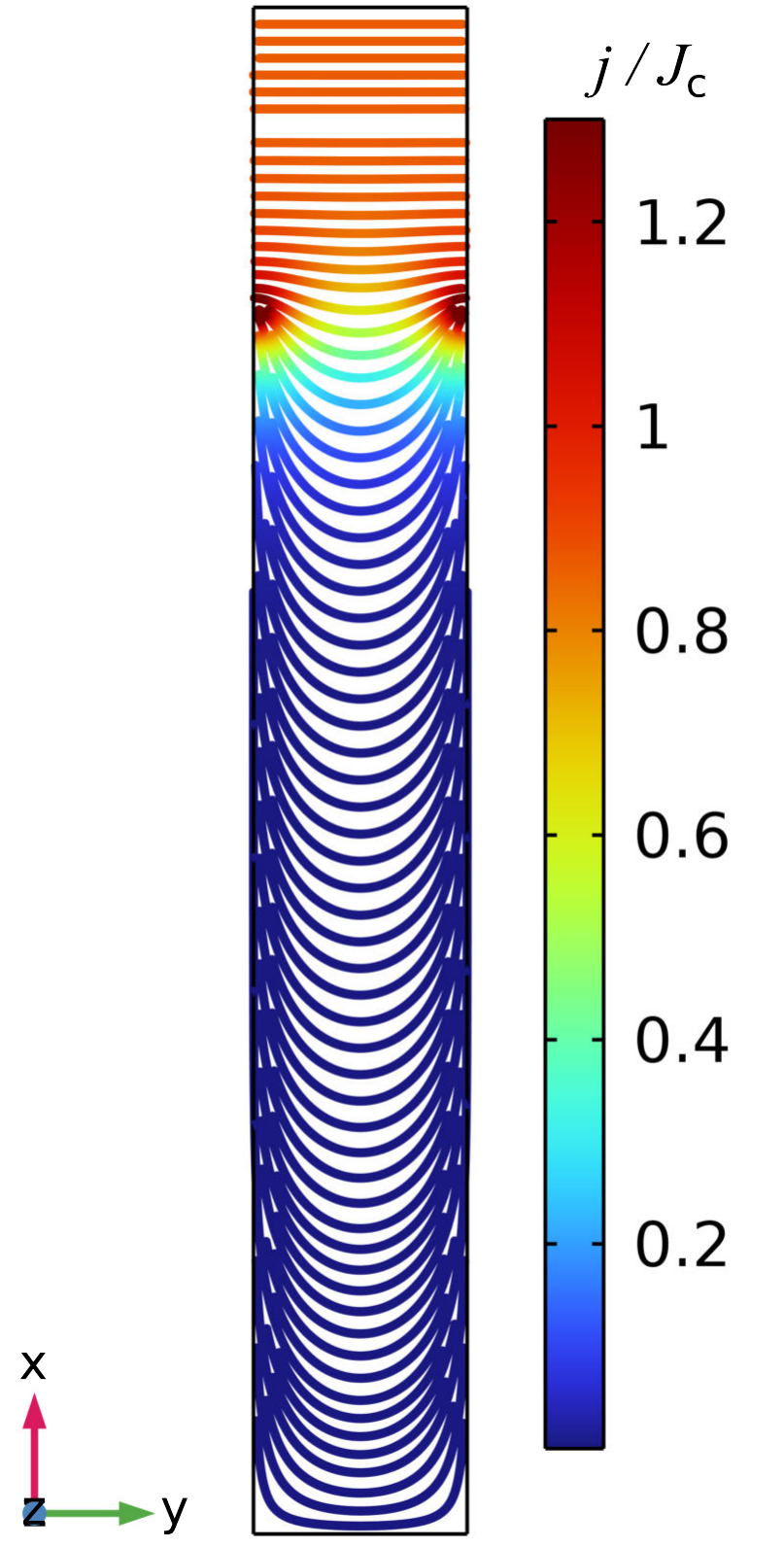}}} 
    \hspace*{0pt}{\rlap{\subcaptionmark\label{fig:current_streamline_c}}\adjustbox{trim=-5mm 1mm 0mm 0mm, valign=t}
    {\includegraphics[height=60mm]{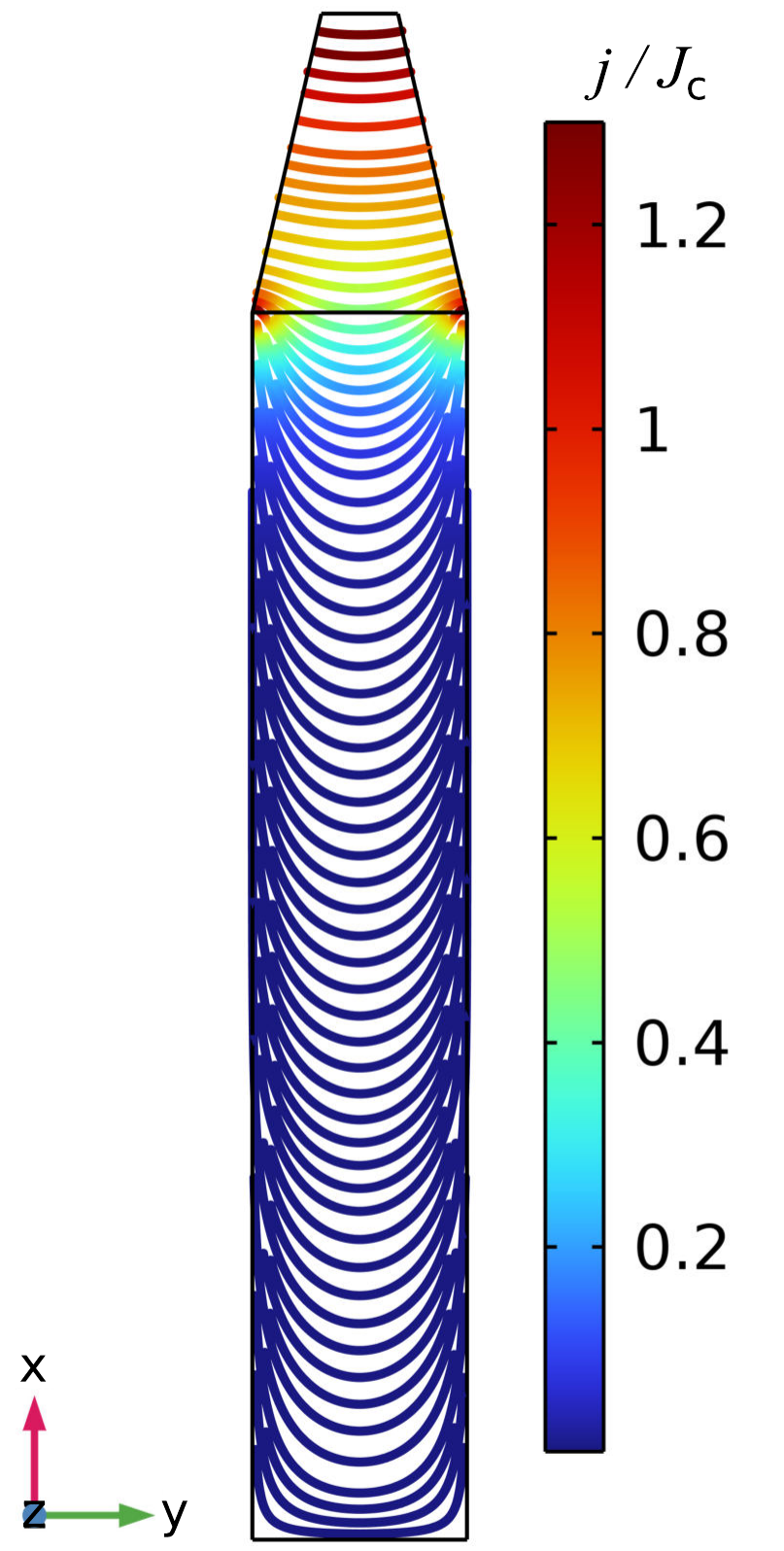}}} 
    \caption{Stationary current density distribution inside the \subref{fig:current_streamline_a}~I1, \subref{fig:current_streamline_b}~T7, and \subref{fig:current_streamline_c}~Y2.5 electrolyzer cell at an applied current of $ I_\mathrm{c} = \SI{-1}{\ampere}$.}
	\label{fig:current_streamlines}
\end{figure}

The calculated two-dimensional current density distribution normalized by the mean value $J_c$
is shown in Fig.~\ref{fig:current_streamlines}. 
The results 
support the ranking of the different types in Tab.~\ref{tab:concepts} in terms of homogeneity of the current distribution. The T-shape electrolyzer displays the most homogeneous $j$ distribution,
the I-shaped electrolyzer the least homogeneous current distribution.
%

The presence of the entry duct introduces singularities at the lower end of the electrodes. Here, a concentration of the field lines appears leading to 
a  peak in the current density at the lower end of the electrode.
The resulting profiles of the centerplane current density $j$ 
along the coordinate $\eta$ 
are shown in Fig.~\ref{fig:comp_TIY_flow_j}a. 
For the I-electrolyzer, the profile peaks at both ends of the electrode, 
showing the smallest peak of all electrolyzers at the lower edge and the highest peak amongst all at the upper end.
Over a wide central region, the current density distribution has the strongest gradient since the electrodes are most strongly angled.
In contrast, the T7-electrolyzer exhibits an almost uniform current density distribution across the electrode with the exception of the lower edge.
Here, it has its only peak which is the highest of all, however.
This is a result of the alignment of the electrodes and the boundary conditions employed on the electrodes, with Eq.~\eqref{eq:bc_cathode} imposing an integral condition for the current density at the cathode, but still allowing a non-uniform distribution,
which is realistic.
As a measure of uniformity, the homogeneity index  
\begin{equation}
    H = \frac{j_\mathrm{max} - j_\mathrm{min}}{J_\mathrm{c}}
\end{equation}
was calculated with the absolute maximum minimum 
of the current density distribution, $j_\mathrm{max}$
and $j_\mathrm{min}$, respectively.
and the applied current density at electrode surface, $J_\mathrm{c}$ 
from $I_\mathrm{c}$ in \ref{eq:bc_cathode}). 
For all cases two different locations were studied, along a line in the middle of the electrode gap ($y = 0$) and over the entire surface of the electrode ($\xi = 0$). 
The results are compiled in Tab.~\ref{tab:hom_index_current_simulation}.
%
\begin{table}[h]
\caption{
Homogeneity index $H$ in the center of the electrode gap ($y = 0$) and at the surface of the electrode ($\xi = 0$).
}
\label{tab:hom_index_current_simulation}
\centering
\begin{tabular}{ l c c }
    \toprule
        Electrolyzer    & $H(y = 0)$    & $H(\xi = 0)$  \\
    \midrule
        I1		        &  2.57        & 2.91   \\
        T7		        &  0.86        & 2.97   \\
        Y2.5            &  1.45        & 1.55   \\
    \bottomrule
\end{tabular}
\end{table}

In agreement with Fig.~\ref{fig:current_streamlines},
the T-shaped electrolyzer has the best, hence smallest value $H(y = 0)$.
However, the local singularities observed in Fig.~\ref{fig:current_streamlines} and Fig.~\ref{fig:comp_TIY_flow_j}a  at the lower end of the electrodes  lead to a large and, thus, undesired homogeneity index $H(\xi=0)$ for T7. The value obtained for I1 is nearly identical, though for a different reason. Its poor performance can be attributed to the considerable increase of current density towards the upper end of the electrode since I1 features the smallest upper gap width among all geometries investigated. In comparison to T7 and I1, the homogeneity index is smaller for Y2.5 by a factor of two, hence compensating the issues at the ends best. According to Faraday's law $j \propto \frac{\mathrm{d}n}{\mathrm{d}t}$, a more homogeneous electrochemical reaction is to be expected for the Y-shaped cell.

\subsection{New design criterion}
\label{sec:new_design_criterion}
\begin{figure}
%
\hspace*{0pt}{\rlap{\subcaptionmark\label{fig:comp_TIY_j}}\adjustbox{trim=-5mm 0mm 0mm 0mm, clip, valign=t}
{\includegraphics[height=.5\textwidth]{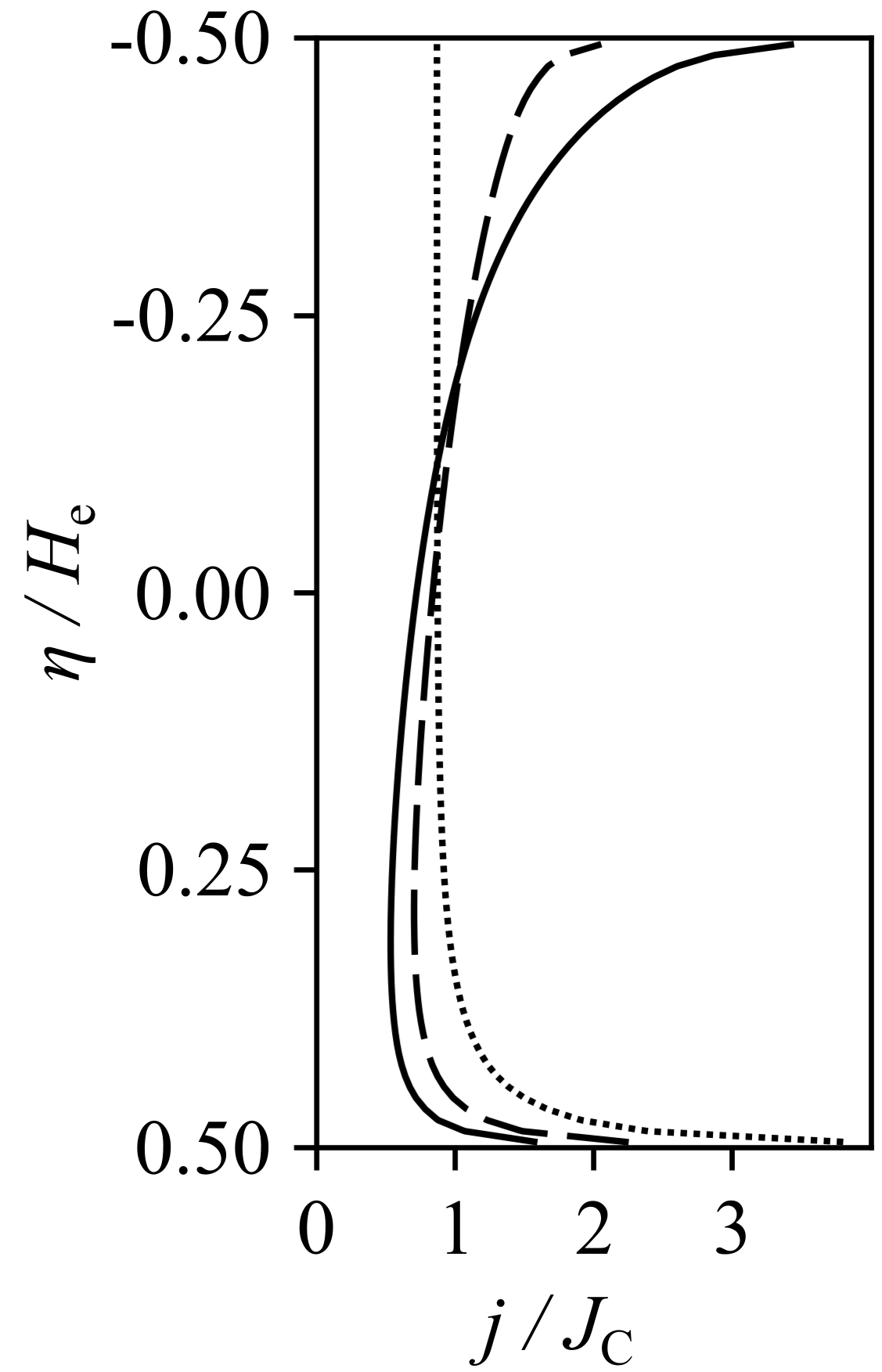}}} 
\hspace*{0pt}
{\rlap{\subcaptionmark\label{fig:comp_TIY_u_xi}}\adjustbox{trim=-5mm 0mm 0mm 0mm, clip, valign=t}
{\includegraphics[height=.5\textwidth]{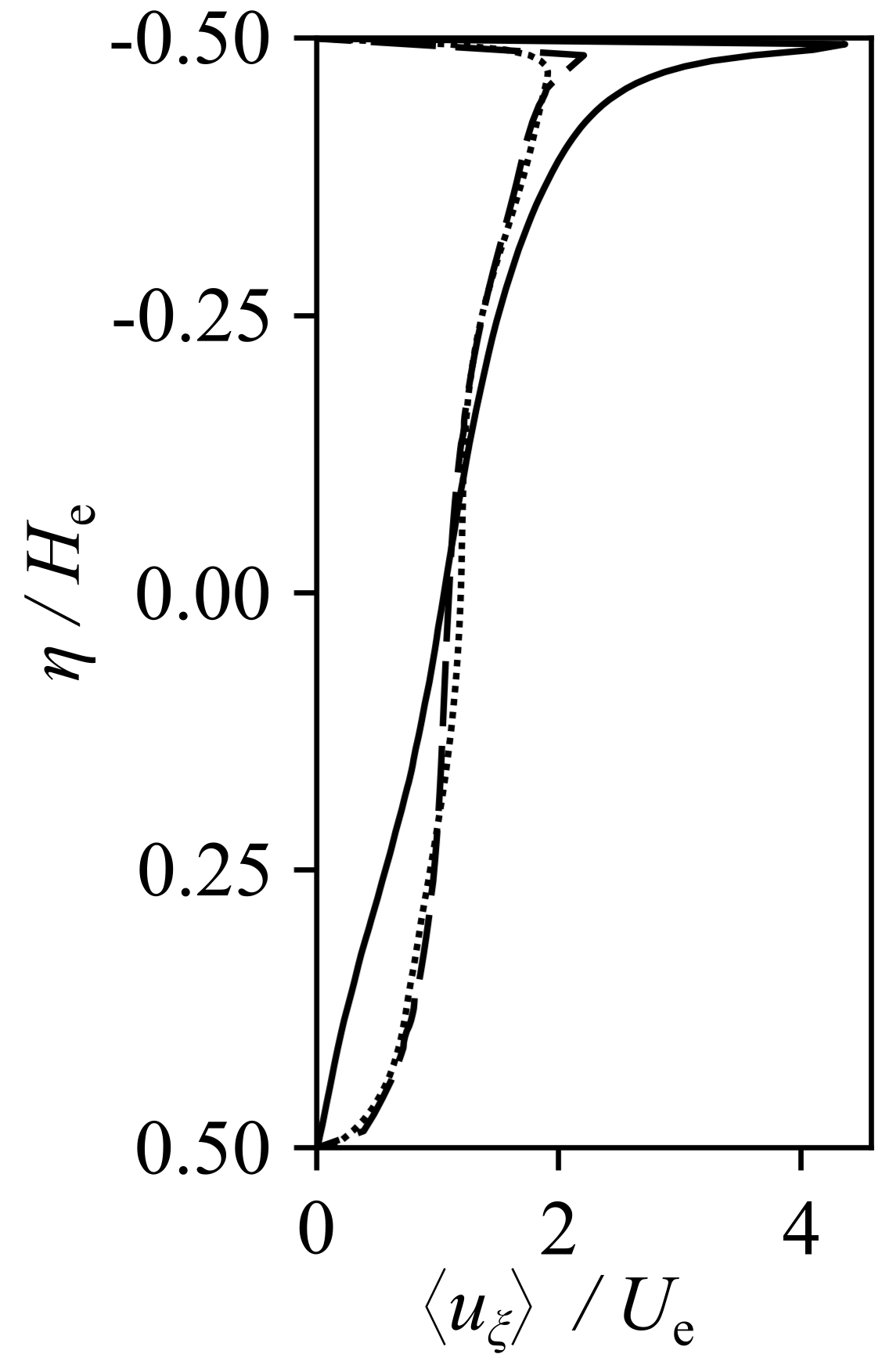}}} \\
\hspace*{0pt}{\rlap{\subcaptionmark\label{fig:comp_TIY_u_xi_j}}\adjustbox{trim=-5mm 0mm 0mm 0mm, clip, valign=t}
{\includegraphics[height=.5\textwidth]{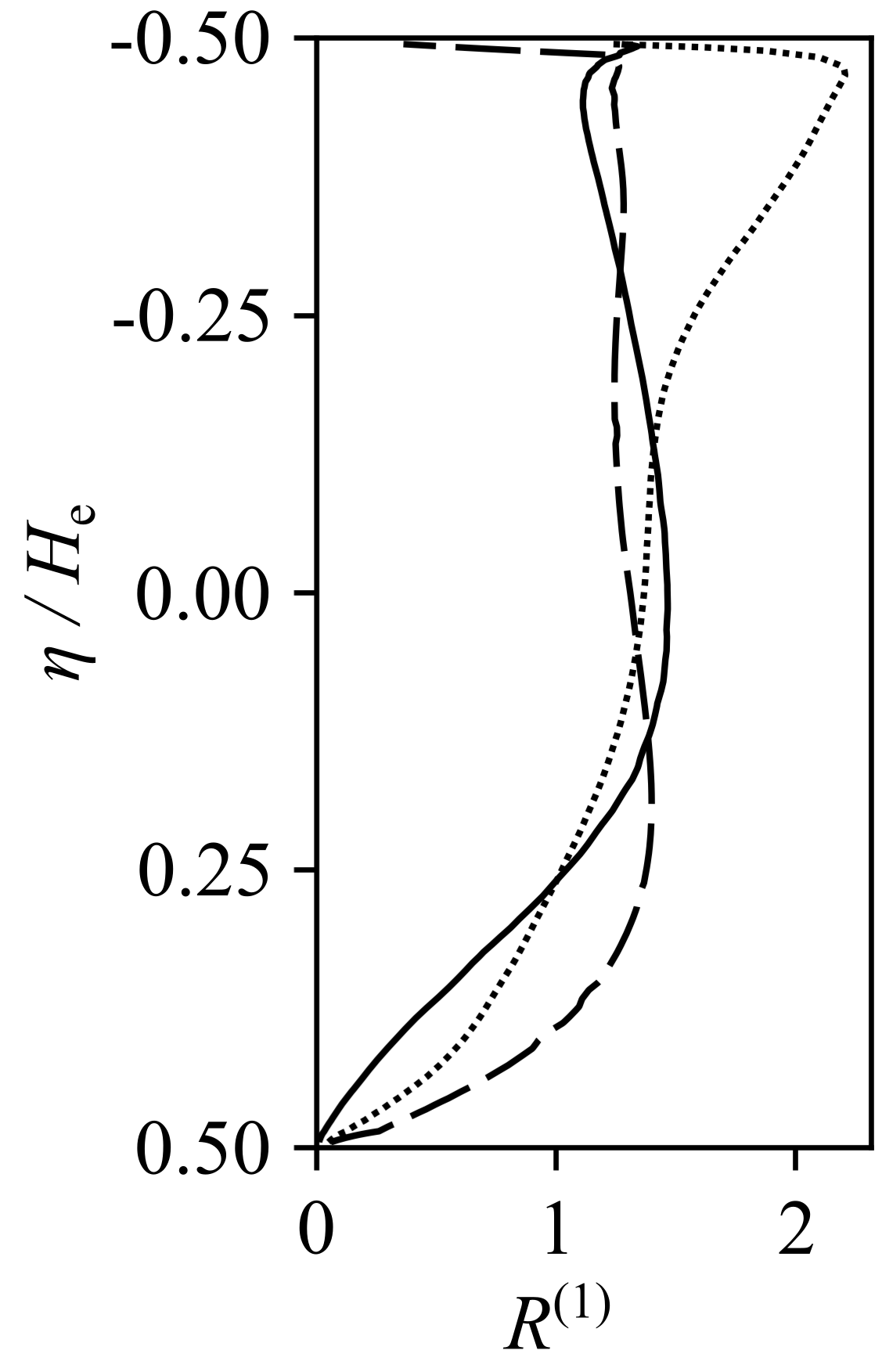}}} 
\hspace*{0pt}{\rlap{\subcaptionmark\label{fig:comp_TIY_u2_xi_j}}\adjustbox{trim=-5mm 0mm 0mm 0mm, clip, valign=t}
{\includegraphics[height=.5\textwidth]{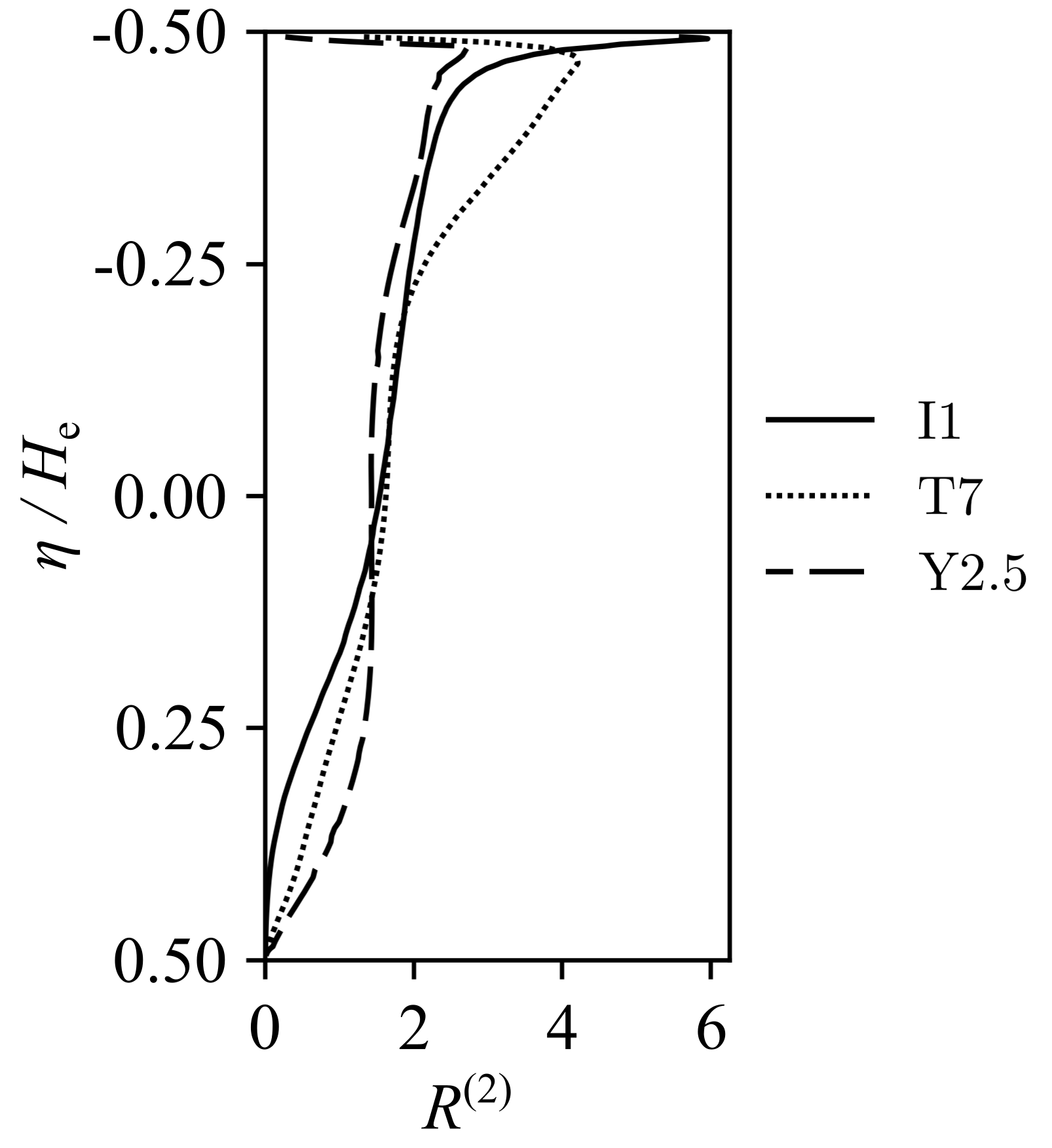}}} 
\caption{ 
Evaluation of design criteria along the centerline with $\zeta=0$ on the upstream side of the electrode at $\xi=-T/2$.
Results for I1, T7, and Y2.5 in each plot, as indicated by the line styles.
In all plots, the direction of the coordinate $\eta$ was inverted 
for better readability, with the lower edge of the electrode at the bottom and the upper edge at the top.
\subref{fig:comp_TIY_j}~Current density $j$,
\subref{fig:comp_TIY_u_xi}~Time-averaged normal velocity component $\langle u_\xi \rangle$,
\subref{fig:comp_TIY_u_xi_j}~ratio of time-averaged normal velocity to current density, $R^{(1)}$
\subref{fig:comp_TIY_u2_xi_j}~ratio of time-averaged normal velocity squared to current density, $R^{(2)}$. 
} \label{fig:comp_TIY_flow_j}
\end{figure}

As was shown in Figs.~\ref{fig:current_streamlines} and \ref{fig:comp_TIY_flow_j}a, when the electrodes are angled, more current is transmitted through the upper electrode gap region due to smaller electrode distance, whereas lower current density occurs in the lower region with a larger electrode distance.
For the case of parallel electrodes as for the T-electrolyzer, however, end effects dominate.
Such regions are critical, which becomes obvious when inspecting the
velocity profiles of the time-averaged normal velocity $\langle u_{\xi} \rangle$, 
plotted in Fig.~\ref{fig:comp_TIY_u_xi} for all geometries. 
The velocity vanishes at the channel walls, $\eta = \pm H_\mathrm{e}/2$, resulting from the no-slip condition imposed there. This is disadvantageous for electrolysis since it creates critical regions where gas bubbles can grow to a rather large extent without being removed by the flow, thus increasing the risk of gas crossover. This behavior of the flow at the channel walls can generally not be avoided, but it can be optimized so that there is a rapid increase of the flow velocity with increasing distance from the walls. 
%

From the analysis of the current density conducted above it can be concluded that the design goals of  uniform flow and uniform current density, cannot be achieved at the same time for any of the considered geometries. It even seems that, by construction, flow cannot be uniform in opposite directions through parallel electrodes, and if flow is uniform, current cannot be uniform. 
Therefore, a new design criterion is introduced here, loosely formulated as
\begin{equation} 
\label{eq:design_ratio_by_words}
  R
  =
  \frac{\textrm{force on bubbles}}{\textrm{bubble generation rate}}
  \overset{!}{=}
  \mathrm{const.}
\end{equation} 
over the electrode.
The bubble generation rate in the denominator is proportional to the moles of gas produced per time unit, $\mathrm{d}n / \mathrm{d}t$, which is proportional to the current density $j$.
The electrode-normal fluid velocity determines the bubble removal rate by fluid forces on bubbles, appearing in the numerator of Eq.~\eqref{eq:design_ratio_by_words}. 
In case of low bubble Reynolds number $Re_\mathrm{b} = D_\mathrm{b} u_\mathrm{b} / \nu$, 
with $D_b$ the bubble diameter and $u_b$ the magnitude of the relative velocity between fluid an bubble,
the fluid forces are proportional to $u_\mathrm{b}$.
The velocity magnitude at the position of the bubble scales with $U_e$, so that this is used as a reference value here.
For viscosity-dominated forces it is appropriate to define the dimensionless design ratio 
\begin{equation} 
\label{eq:ratio_u_j}
  R^{(1)}
  =
  \frac{ \langle u_\xi \rangle / U_\mathrm{e}} {j /J_\mathrm{c}}
  \overset{!}{=}
  \mathrm{const.}
\end{equation} 
with $\langle u_\xi \rangle$ the time-averaged velocity normal to the electrode, $j$ the local current density, $U_\mathrm{e}$ the mean velocity through the electrode, and $J_\mathrm{c}$ the mean current density imposed on the cathode.
It is deliberately disregarded here whether individual bubbles of a certain size located at a given position on a complex electrode geometry are mobilized or not. Instead, the scale of the electrode extension is addressed and thresholds of mobilization covered by the normalization factors. 
The value $R^{(1)}=1$ would be achieved for perfectly uniform flow and current density distribution since both, time-averaged velocity $\langle u_\xi \rangle(\zeta, \eta)$ and current density $j(\zeta, \eta)$, are normalized with their cross-sectional averages, $U_\mathrm{e}$ and $J_\mathrm{c}$, respectively. 

From another perspective, the central objective of FT-ME optimization is to enable the largest possible gas production rate without risking gas crossover for a given hydrodynamic flow rate through the electrolyzer.
The dimensional expression $U_\mathrm{e} / J_\mathrm{c}$ then describes a mean removal-to-production ratio of gas bubbles. For the present study a characteristic design with $U_\mathrm{e} / J_\mathrm{c} = \SI{6.8}{cm^3/(\ampere \second)}$
was realized.

While Eq.~\eqref{eq:ratio_u_j} is suitable for viscous-type forces on the bubbles, pressure forces are proportional to the square of the local velocity. 
The latter dominate for larger bubbles and/or higher velocities, as both increase $Re_b$.
In such a case it is appropriate to define the design criterion 
\begin{equation}
\label{eq:ratio_u_j_square}
    R^{(2)}
    =
    \frac{ \langle u_\xi \rangle^2 / U_\mathrm{e}^2} {j / {J_\mathrm{c}}  }
    \overset{!}{=}
    \mathrm{const.}
\end{equation}
for optimization,
with an analogous interpretation of the pre-factor as for $R^{(1)}$.

An obvious generalization can be obtained by a blending between both dependencies, employing the general drag coefficient
in (\ref{eq:c_d_sphere}) which valid in both regimes.
In what follows, the limiting cases \eqref{eq:ratio_u_j} and \eqref{eq:ratio_u_j_square} are considered.

The new design criterion \eqref{eq:design_ratio_by_words} with its $Re$-dependent formulations \eqref{eq:ratio_u_j} and \eqref{eq:ratio_u_j_square} was evaluated along a sampling line in the center of the electrode plane at $\zeta=0$ on the upstream side, $\xi=-T/2$, yielding $R^{(1)}(\zeta=0,\eta)$ in Fig.~\ref{fig:comp_TIY_u_xi_j} and $R^{(2)}(\zeta=0,\eta)$ in Fig.~\ref{fig:comp_TIY_u2_xi_j}. 
The Reynolds number of the wire, accounting for blockage effects, is 
$Re_w = D_w \, U_e /(\nu \beta)$ and equals 33 in the present case.
Assuming bubbles of size $D_b = 0.1 \ldots 1.0 D_w$ results in
$Re_b = 3.3 \ldots 33$. 
For the smaller ones this would motivate \eqref{eq:ratio_u_j}
and for the larger ones \eqref{eq:ratio_u_j_square}.
In the experiments reported below $D_b=\SI{0.15}{mm}$, 
giving $Re_b=22$.

Focusing on $R^{(1)}$ first, the different electrolyzers can be compared as follows.
For the I1-electrolyzer, $R^{(1)}$ is only gently increasing from the lower electrode edge, then showing a large plateau of global maximum in the middle of the electrode and slightly decreasing again further upwards despite a small peak close to the upper electrode edge. 
The T7-Electrolyzer performs even worse, with $R^{(1)}$ increasing almost continuously from the lower to the upper edge of the electrode and peaking shortly before it. 
For the Y-design, in contrast, $R^{(1)}$ increases most rapidly from the lower edge and saturates in a large plateau which extends to the upper edge. Hence, Y2.5 shows the most uniform distribution of $R^{(1)}$ along the sampling line among all geometries investigated, without the presence of any sharp peaks.

Analogous conclusions can be drawn for $R^{(2)}$ along the sampling line, shown in Fig.~\ref{fig:comp_TIY_u2_xi_j}. 
The profiles of the T7- and Y2.5-electrolyzer have a similar shape as for $R^{(1)}$, but for the I1-electrolyzer the situation is even worse since a sharp peak of $R^{(2)}$ occurs close to the upper edge of the electrode.

While the current density is uniform in $\zeta$, the electrolyte velocity is not, as seen in Sec.~\ref{sec:results_cfd}, 
so that the results in Fig.~\ref{fig:comp_TIY_flow_j} remain qualitative, used here for illustrating the principle.
For a final verdict, 
the design ratios $R^{(1)}$ and $R^{(2)}$ need to be evaluated all over the electrode area. 
 
This was done for $\xi=-T/2$ and all values of $\eta$ and $\zeta$, with results reported in Tab.~\ref{tab:std_skew_zeta_eta_plane}. Here, the standard deviation 
$\sigma_{\zeta\eta}$ 
provides information about the uniformity achieved with a criterion.
The values 
demonstrate that in terms of  $R^{(1)}$ the Y-electrolyzer 
has the lowest standard deviation, which is $\SI{33}{\%}$ lower compared to T7 and $\SI{25}{\%}$ lower compared to I1.
Hence, this electrolyzer exhibits the most uniform design ratio,
i.e. the force of the electrolyte flow and the rate of gas production are balanced best across the entire electrode.
The same conclusion can be drawn with $R^{(2)}$ 
yielding $\sigma_{\zeta\eta}$ being approximately $\SI{36}{\%}$ lower for Y2.5 compared to both I1 and T7.

Ideally the standard deviation $\sigma_{\zeta\eta}$ should vanish, 
indicating a constant design ratio.
This is not the case, even for Y2.5, e.g. due to the presence of walls as mentioned above.
The small remaining risk of hotspots where $R$ would be too small,
resulting in the possibility of gas crossover, is discussed further in the SI.
At the present stage we can conclude that the analysis of both newly introduced design ratios undertaken here 
demonstrates the superiority of the Y-shaped design compared to the I- and T-design.

\subsection{Validation of single phase flow}
\label{sec:validation_singlephase}
\begin{figure}
\centering
%
\hspace*{0pt}{\rlap{\subcaptionmark\label{fig:comparison_vel_exp}}\adjustbox{trim=-5mm 1mm 0mm 0mm, clip, valign=t}
{\includegraphics[width=0.4\textwidth]{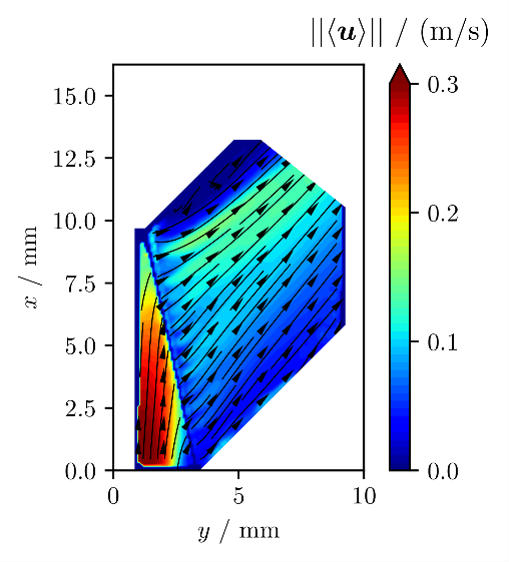}}} \hspace*{2mm}
\hspace*{0pt}{\rlap{\subcaptionmark\label{fig:comparison_vel_sim_porous}}\adjustbox{trim=-5mm 1mm 0mm 0mm, clip, valign=t}
{\includegraphics[width=0.4\textwidth]{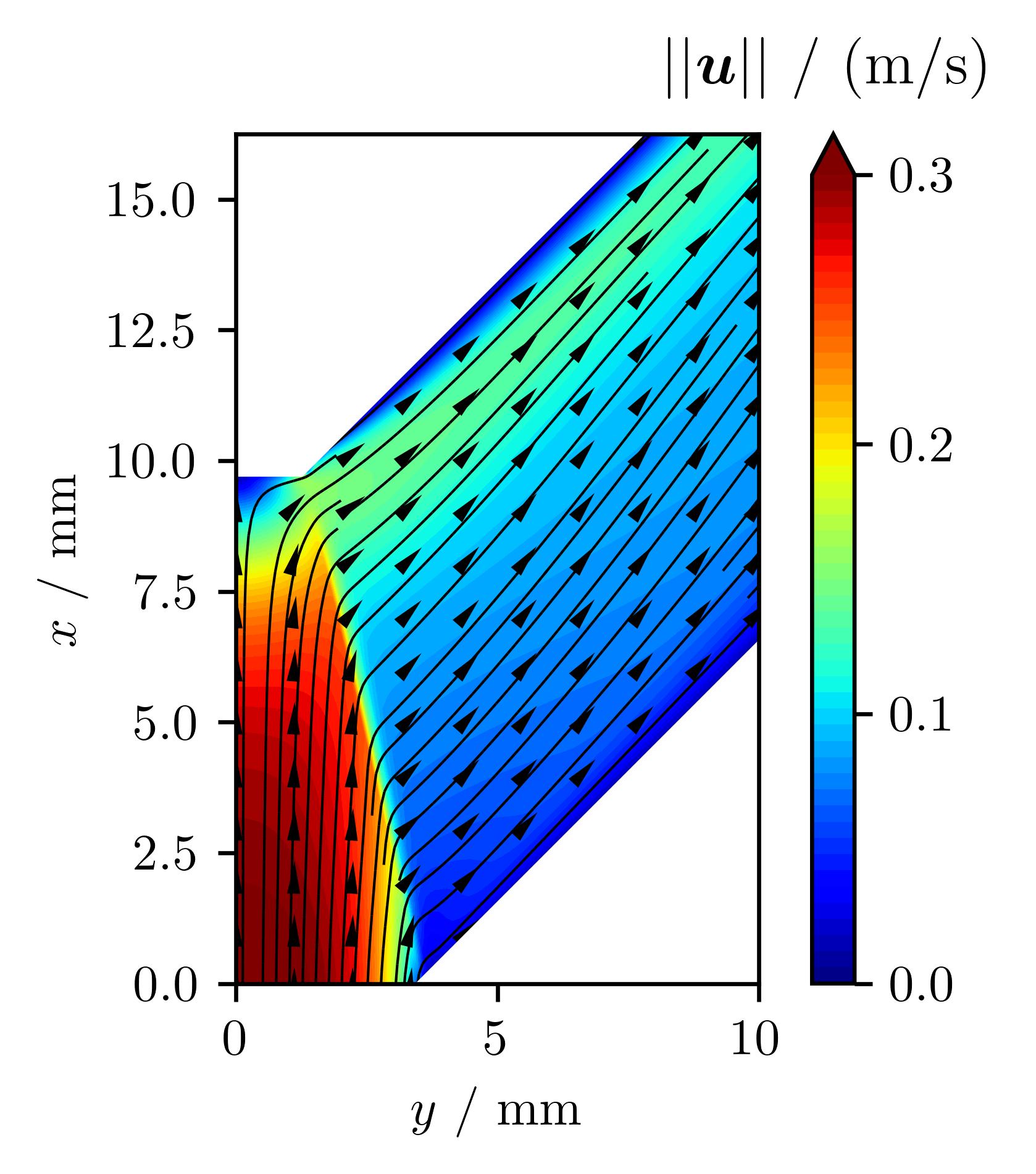}}} \hspace*{5mm}
\caption{ 
Magnitude of time-averaged velocity $\lVert \langle \boldsymbol{u}\rangle \rVert$ and streamlines in the $xy$-centerplane ($z=0$) for the region downstream of the electrode.
\subref{fig:comparison_vel_exp}~Experimental result,
\subref{fig:comparison_vel_sim_porous}~simulation result. The windows where data are available differ for technical reasons.
} 
\label{fig:comparison_vel}
\end{figure}

As a proof-of-concept for the Y2.5-electrolyzer, 
experiments were performed to validate the CFD simulations and  to gain insights into the multiphase flow during electrolysis,
in particular the bubble removal rate. 
Fig.~\ref{fig:comparison_vel_exp} shows experimental results for the magnitude of the mean velocity vector in dimensional units, as well as streamlines of the mean flow. Massive flow reorientation over the electrode can be observed, together with a reduction of velocity magnitude in the centerplane. The downstream flow is fairly regular and well directed towards the outlet in the lower part of the duct and the center. At the upper end, a small recirculation zone is visible.

Fig.~\ref{fig:comparison_vel_sim_porous} provides the simulation data under the same conditions. Technical limitations of the measurement window result in apparently slightly different  upstream flows. However, 
the simulation captures the experimental flow field quite well. The deceleration created by the stagnation point between the electrodes is similar, although a bit weaker in the simulation. Upstream of the lower end of the electrode the flow seems to be somewhat slower, but this results from the difficulty of the measurement close to the wall and the electrode. Behind the electrode the flow is more evenly distributed over the cross section in the simulation, and the recirculation zone is absent.

The insert of Fig.~\ref{fig:exp_setup} shows that the electrodes were mounted in grooves at the lower end, while at the upper end they were held in slits traversing the upper wall for being connected to the electric sources. 
This mounting caused a small shortcut of the fluid flow at the upper end of the electrode and, hence, is responsible for the visible differences in Fig.~\ref{fig:comparison_vel}.
It disturbs the flow through the electrode at the upper end by a small crosswise jet moving the main flow away from the wall, thus generating the small recirculation zone observed. At the lower downstream wall somewhat larger velocities can be discerned in the experiment which very likely result from the grooves holding the electrode and reducing the resistance of the electrode in these locations. 
The deviations are moderate and suggest that the conclusions drawn from the simulation campaign still hold.

\subsection{
Multiphase flow in the Y-shaped Electrolyzer}
\label{sec:impression_multiphase}

In galvanostatic measurements 
an overview of the formation and transport of bubbles from the electrode gap 
was obtained.
The flow rate was adjusted so as to provide data at 
$Re=Re_i=1000$, $1500$ and $3000$ 
which refers to the junction inlet. 
In all cases a current density of $J_\mathrm{c} = \SI{-100} {\milli\ampere\per\centi\metre\squared}$ was employed. 
Minimal gas crossover, effective bubble detachment,
and decent transport out of the electrode gap were observed. 
In the following, only the results for $Re = 1500$ at its initial phase of the electrolysis are presented.

\begin{figure}[!ht]
	\begin{subfigure}[t]{\textwidth}
        \centering
		\includegraphics[width=0.9\textwidth]{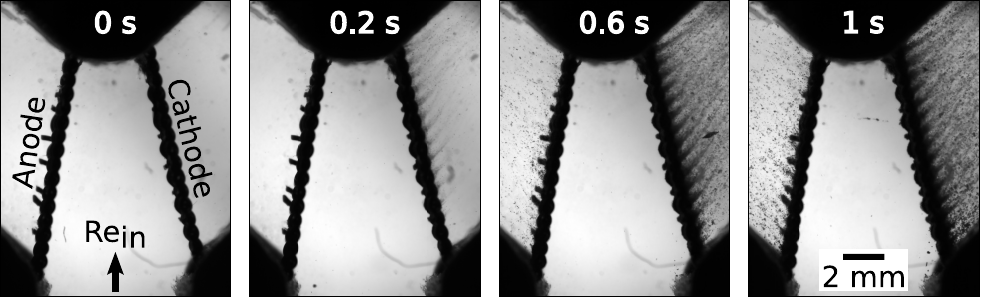}
	\end{subfigure}
	\caption{Image sequence representing the start-up phase of the electrolyzer showing the situation around both electrodes (region~I in Fig.~\ref{fig:exp_setup}) for $Re = 1500$, $j = \SI{-100}{\milli\ampere\per\centi\metre\squared}$ and $t= 0, \ldots, \SI{1.0}{\second}$.}
	\label{fig:shadow_timeseries_gap}
\end{figure}

Due to the inhomogeneous electrode gap varying from \num{2.5} to \SI{7}{\milli\metre}, the H\textsubscript{2} evolution starts at the upper position of the electrode and then successively extends to increasing distances $s$, as shown in Fig.~\ref{fig:shadow_timeseries_gap}.
From $t=\SI{0.2}{\second}$ on, H\textsubscript{2} bubbles are recognized at the cathode, more numerous at the upper end than at the lower end. The flow through the openings of the wire screen transports them in downstream direction creating streaks of higher bubble concentration. Bubble growth and transport successively occur at positions with wider gap and eventually cover the entire electrode. At about $t=\SI{1.0}{\second}$ the situation is stationary.
The earlier start of bubble generation at the upper end with the smaller gap is indicative of the higher current establishing there as already shown in Sec.~\ref{sec:results_current_distribution}. In the developed state the gas flow rate is fairly uniform behind the electrodes, as witnessed by the almost uniform presence of bubbles, with superimposed streaks for reasons of the mesh geometry. This hints to an equilibrium between bubble generation and removal.
Similar observations can be made at the anode, where the O\textsubscript{2} bubbles are larger and less numerous resulting in a longer growth time of the bubbles and, hence, delay in their presence downstream of the anode.
No crossover is seen in these pictures. Conducting measurements over substantially longer time intervals was hampered by the recirculation of the electrolyte. This will be amended in future experiments.

\begin{figure}[!ht]
    \centering
    \includegraphics[width=\textwidth]{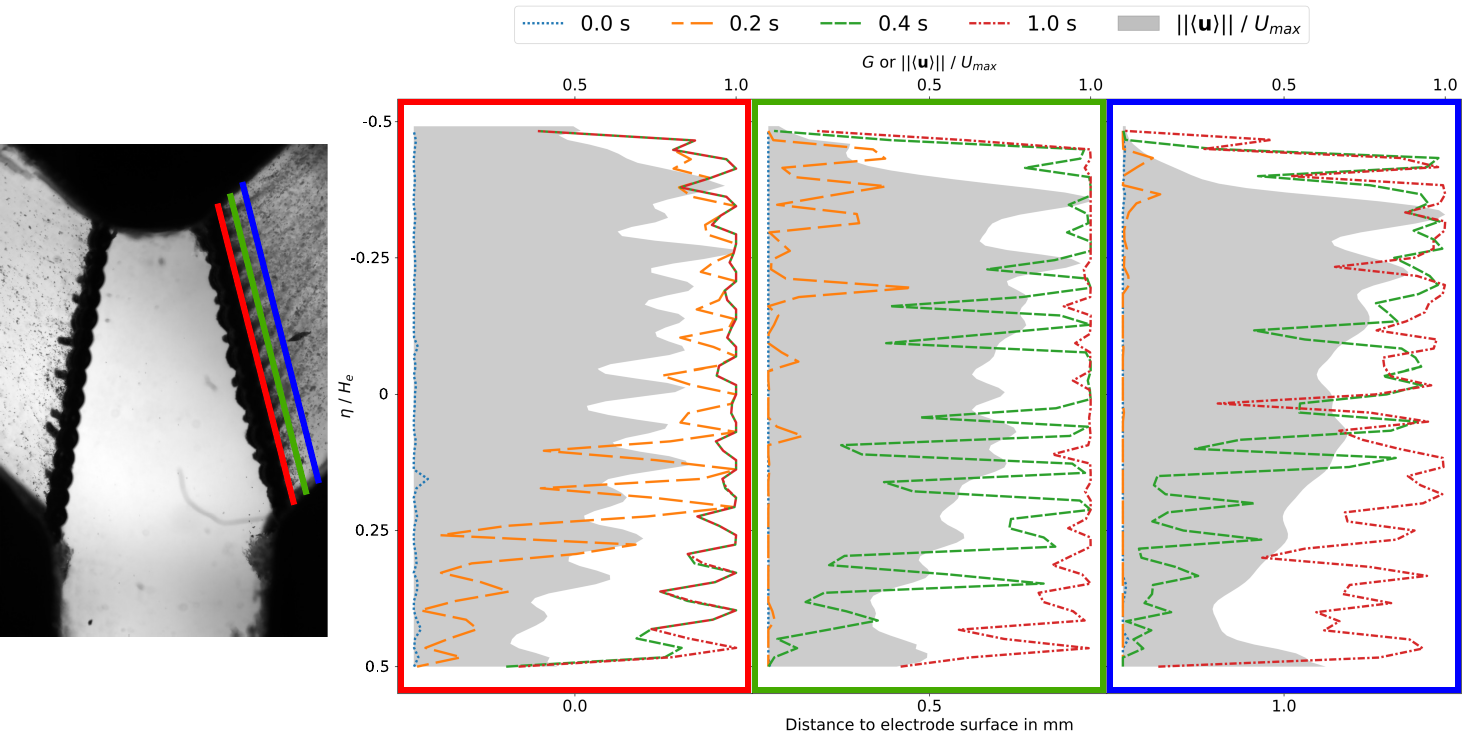}

    \caption{
    Presence of bubbles in terms of the gray value distributions (GVD) for $Re = 1500$, different times and different positions behind the cathode. Each subfigures assembles plots at different positions behind the cathode, $\SI{0}{mm}$, $\SI{0.5}{mm}$, $\SI{1}{mm}$ from left to right, as indicated below the respective graphs. The lines show $G$ along the electrode obtained from the measurements at different times after starting the current, $\SI{0}{s}$, $\SI{0.2}{s}$, $\SI{0.4}{s}$, $\SI{1.0}{s}$. The shaded area shows the fluid velocity of the single-phase flow.
    }
	\label{fig:void_fraction}
\end{figure}

Fig.~\ref{fig:void_fraction} provides a detailed, quantitative account of the transport of H\textsubscript{2} bubbles behind the cathode during startup of the electrolysis reaction at three different distances, $\SI{0}{mm}$, i.e.\ right behind the electrode, $\SI{0.5}{mm}$ and $\SI{1}{mm}$. 
The shaded gray area displays the single phase fluid velocity at the respective positions. It is seen that with increasing distance the local jets generated by the openings in the mesh are leveled out, generating a much smoother profile. 
The bubbles are small in this experiment ($d_\mathrm{} \le \SI{150}{\micro\metre}$), so that their impact on the flow remains limited. Hence, the single phase fluid flow can be used as a good approximation of the electrolyte flow in the presence of bubbles. The fluid forces of the electrolyte mobilize the bubbles and transport them from the streamwise side of the electrode through the wire mesh into the duct behind the electrode. The different curves in each plot report the GVD indicative of the void fraction at the given positions for different instants in time. The blue curve for $t=\SI{0}{\second}$ is barely seen as the value is $0$ almost throughout.
It is obvious that close to the electrode, at $t=\SI{0.2}{\second}$ the upper part of the electrode is covered with bubbles, the lower part not yet entirely, with maxima occurring in line with transporting electrolyte velocity and minima in between. Soon afterwards, at $t=\SI{0.4}{\second}$, the electrode back is almost uniformly covered with bubbles.
At a distance of $\SI{0.5}{mm}$ behind the electrode, the same behavior is seen, just with a delay in time due to the lower velocity (gray shadow) reaching $G \approx 1$ until $t=\SI{1.0}{\second}$.
Also, the coverage at a distance of $\SI{1}{mm}$ is not to $\SI{100}{\%}$, but fluctuates in space since here the bubbles are transported away from the electrode in more irregular manner. At the upper end, the curves for $t=\SI{0.4}{\second}$ and $t=\SI{1.0}{\second}$ are similar so that here saturation is already achieved around $t=\SI{0.4}{\second}$ while this takes place somewhat later in the lower part. The observations made for $Re = 1500$ carry over to the other cases, as shown in Fig.~\ref{fig:appendix_void_fraction}.

Overall, Fig.~\ref{fig:void_fraction} reflects the streakiness of the bubble presence seen in Fig.~\ref{fig:shadow_timeseries_gap} in a more quantitative way. It also highlights the transient start-up phase in the bottom area of the cathode, but for $t=\SI{1.0}{\second}$ shows uniform bubble presence with superimposed irregular fluctuations. Hence, bubble generation appears to be fairly homogeneous along the electrode with all three Reynolds numbers investigated.

\section{Conclusions}
\label{sec:conclusion}
The influence of the shape of membraneless flow-through electrolyzers (FT-ME) on the hydrodynamics and the water electrolysis process was investigated by combining numerical simulations with a first experimental proof-of-concept study. The aim was to propose a cheap electrolyzer concept and to demonstrate that it achieves a homogeneous electrochemical reaction over the entire electrode area. 

Using a newly developed homogenization model for the flow-through electrode, CFD simulations of the single-phase flow and numerical simulations of the current distribution were performed for the new Y-shaped geometry. This geometry can be seen as a sweet spot between the I- and T-shaped FT-MEs reported in the literature. It was found that both cell designs established in the literature, I- and T-shape, have either a poor current or a poor electrolyte velocity distribution. The new Y-shaped cell, on the other hand, is convincing on both quantities. 

Moreover, a novel design criterion was proposed. It was demonstrated that naive criteria of even current distribution and even flow through cannot be fulfilled simultaneously. The new criterion can be understood as a removal-to-production ratio of gas bubbles, where the electrolyte velocity determines the bubble removal rate and the current density the amount of generated gas. 
To differentiate between the dominant viscous-type forces on smaller bubbles or at low electrolyte velocities, and pressure forces on larger bubbles or at higher electrolyte velocities, $Re$-dependent formulations of the design criterion were introduced. By keeping these as constant as possible, a homogeneous reaction could be achieved for the Y-shaped cell as predicted by the simulations. 
This was proven by corresponding experiments showing homogeneous gas distribution across the entire electrode. The developed criterion appears to offer a suitable methodology for evaluating a FT-ME and potentially other electrolyzers in a comparable manner. This allows for the targeted optimization of process parameters. 

Future work will analyze in more detail the multiphase flow and especially the gas phase and its purity.

\section*{Conflicts of interest}
There are no conflicts to declare.

\section*{Acknowledgments}
This project is supported by the Federal State of Saxony in terms of the "European Regional Development Fund" (H2-EPF-HZDR), the Helmholtz Association Innovation pool project "Solar Hydrogen", the Faculty of Mechanical Science and Engineering of TU Dresden, and the Hydrogen Lab of the School of Engineering
of TU Dresden.
The authors gratefully acknowledge the computing time provided by them on the high-performance computers ``Barnard'' at the NHR Center NHR@TUD. This is funded by the Federal Ministry of Education and Research and the state governments participating on the basis of the resolutions of the GWK for the national high-performance computing at universities (www.nhr-verein.de/unsere-partner).

\settocbibname{References}
\bibliography{references}

\clearpage

\newcommand\tabrowsep{\vspace*{2mm}}
\newcommand\FigHeightElec{34mm}
\newcommand\FigHeightXY{55mm}
\newcommand\FigHeightDDD{55mm}
\renewcommand{\thefigure}{S\arabic{figure}}
\renewcommand{\thetable}{S\arabic{table}}
\renewcommand{\theequation}{S\arabic{equation}}
\renewcommand{\thesection}{S\arabic{section}}
\setcounter{section}{0}
\setcounter{equation}{0}
\setcounter{figure}{0}
\setcounter{table}{0}
\section*{Supplemental material}
\section{Full geometric details of the investigated electrolyzer geometries}
\label{sec:appendix_geometry}

\begin{figure}[hb]
\centering
\hspace*{0pt}{\rlap{\subcaptionmark\label{fig:3D_CAD_I}}\adjustbox{trim=-5mm 0mm 0mm 0mm, clip, valign=t}
{\includegraphics[height=123mm]{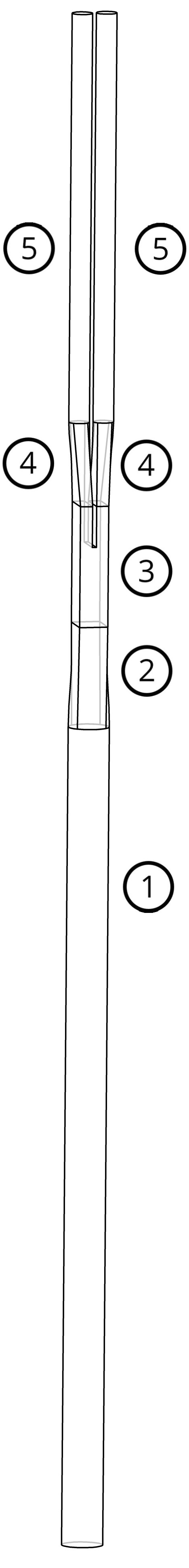}}} \hspace*{5mm}
\hspace*{0pt}{\rlap{\subcaptionmark\label{fig:3D_CAD_T}}\adjustbox{trim=-5mm 0mm 0mm 0mm, clip, valign=t}
{\includegraphics[height=123mm]{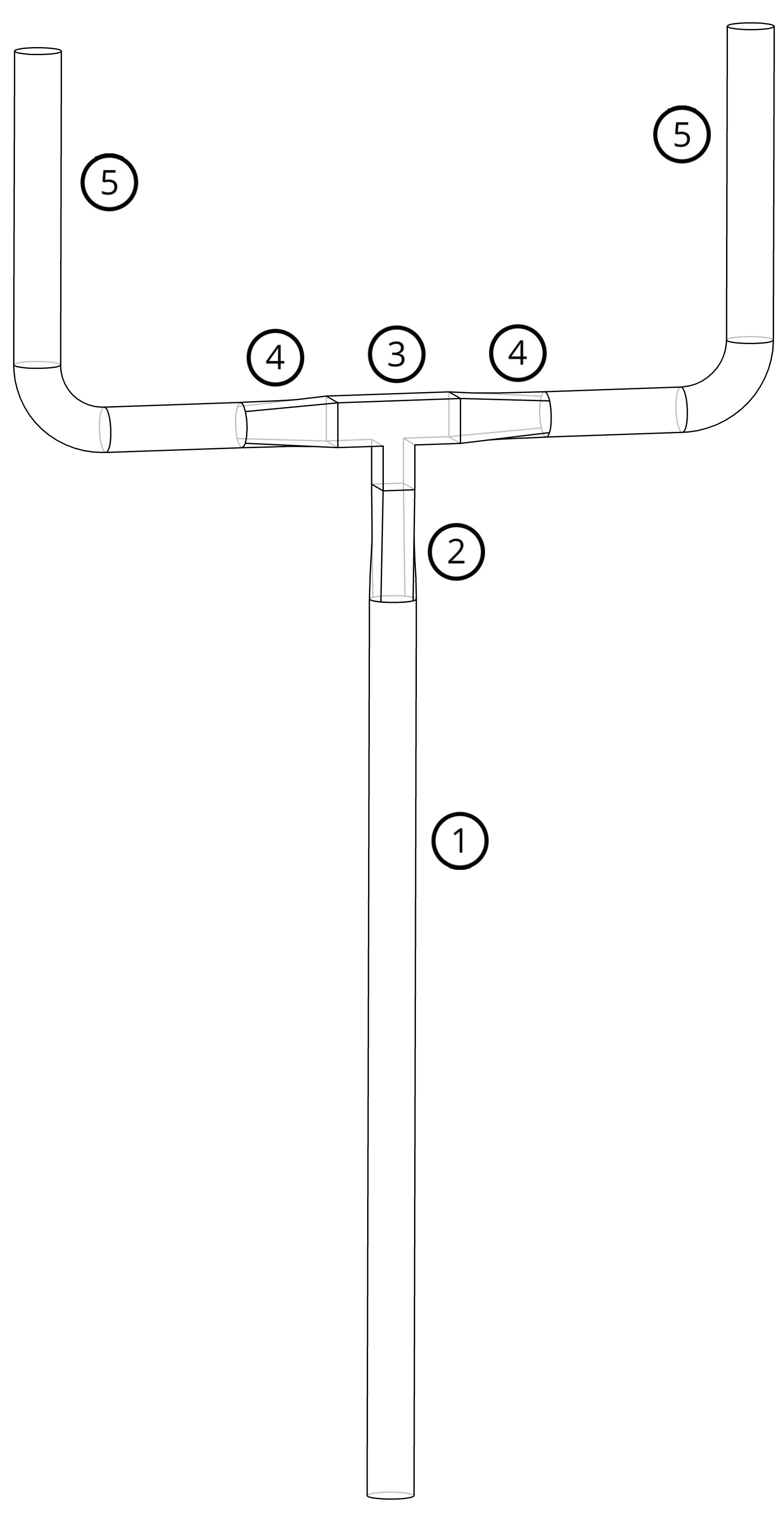}}}
\hspace*{0pt}{\rlap{\subcaptionmark\label{fig:3D_CAD_Y}}\adjustbox{trim=-5mm 0mm 0mm 0mm, clip, valign=t}
{\includegraphics[height=123mm]{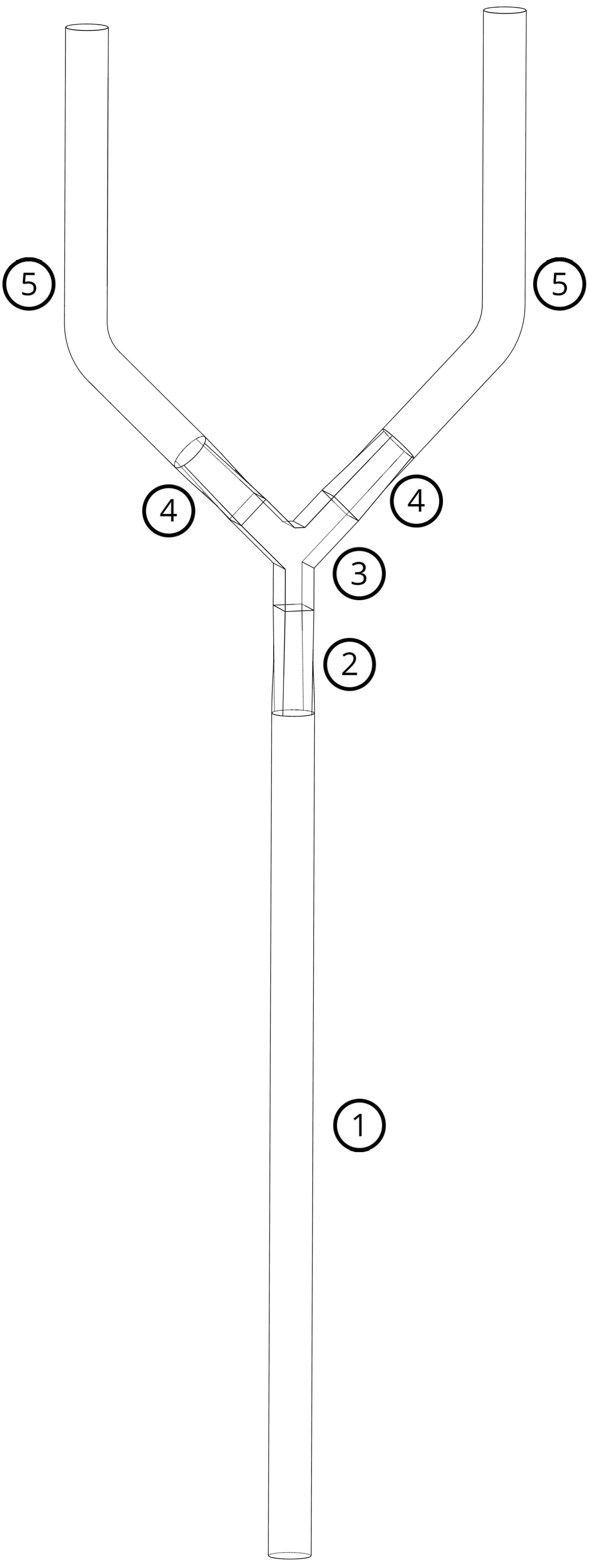}}} \hspace*{5mm}
\caption{
Three-dimensional CAD models of entire electrolyzers.
\subref{fig:3D_CAD_I}~I1 electrolyzer,
\subref{fig:3D_CAD_T}~T7 electrolyzer and
\subref{fig:3D_CAD_Y}~Y2.5 electrolyzer.
}
\label{fig:3D_CAD}
\end{figure}

The entire device consists of a feeding system, the 
cell with the electrodes, and an exhaust system.
Figure~\ref{fig:3D_CAD} illustrates the overall geometries of the 
configurations investigated, with a generic I-shaped, T-shaped and Y-shaped design.
These were chosen to maximize comparability between the cases. 
Geometrical data for each of the numbered components are reported in Tab.~\ref{tab:comparison_geometry_parameters}.

The inlet pipe~\circled{1} with circular cross section connects to the inlet peripheral devices and provides flow conditioning. 
The flow in this pipe is laminar and develops to a good extent over the length of the pipe.
Over the subsequent nozzle~\circled{2} the cross sectional area is reduced resulting in a slight acceleration of the flow.
The main purpose of the nozzle is to accomplish the change in cross sectional shape from circular to rectangular.
The entire upstream part with pipe~\circled{1} and nozzle~\circled{2} is identical for all cases simulated and the inlet of the electrolyzer cell.

The cell~\circled{3} contains two identical flow-through electrodes and accomplishes the separation of the electrolyte stream into two streams, as illustrated in Fig.~1a.
The exit angle~$\psi$ of the outlet of this section varies for the I-, the T- and the Y-design.
The junction containing the electrodes has a rectangular cross section to allow for plane windows in the experiment
minimizing deflection of light for optical access, needed for PIV and shadowgraphy.
The electrodes are plane and made of woven wire screens (Fig.~1b).
They have identical cross-sectional size 
$H_\mathrm{e}\times H_\mathrm{e}$
in all designs, for comparability.
The design study was conducted such that the conditions upstream of the electrodes, 
i.e.\ the feeding system and junction inlet, 
as well as the cross-sectional extension of the electrodes 
are unchanged to allow sound comparison.
This amounts to conserving the distance of the electrodes at the lower edge, $s_\mathrm{l}$, 
 and only the distance of the upper edges, $s_\mathrm{u}$, depends on the shape of the electrolyzer.

Downstream of the cell a connecting part~\circled{4} is positioned at each outlet of the junction to return to the circular cross section of the respective outlet pipe.
Differences in the connected exhausts result from the different angles of the junction while keeping the same electrode extension.
The outlet pipe~\circled{5} has a constant circular cross section and accomplishes the connection to the outlet peripheral devices.
The flow is diverted to the same direction as the inlet, which requires a bow for the T- and the Y-design in addition to the straight pipe segments.

\begin{table}
\caption{Geometrical parameters of the electrolyzer geometries I1, T7 and Y2.5 used for the flow simulations.
}
\label{tab:comparison_geometry_parameters}
\begin{tabular}{ c c r@{\,}l r@{\,}l r@{\,}l @{\qquad}l}
	\toprule
    ID   & quantity           & I1 & & T7 & & Y2.5 & & description \\
    \midrule
    \circled{1} & $D_\mathrm{p1}$    & \tikzmark{start1}\quad\qquad  &  &  \tikzmark{end1}\;    $\num{10}$    & $\si{mm}$ \;\tikzmark{start2} &  & \quad\tikzmark{end2}                        & diameter of pipe                        \\
    \circled{1} & $L_\mathrm{p1}$    & \tikzmark{start3}\quad\qquad  &  &  \tikzmark{end3}\;    $\num{200}$   & $\si{mm}$ \;\tikzmark{start4}   &  & \quad\tikzmark{end4}                        & length of pipe                          \\

    \vspace{-0mm} \\

    \circled{2} & $A_\mathrm{n2,i}$  & \tikzmark{start17}\quad\qquad &  & \tikzmark{end17}\;   $\num{78.54}$ & $\si{mm^2}$ \;\tikzmark{start18} &  &  \quad\tikzmark{end18}                     & nozzle inlet cross sectional area              \\
    \circled{2} & $L_\mathrm{n2}$    & \tikzmark{start5}\quad\qquad  &  &  \tikzmark{end5}\;    $\num{25}$    & $\si{mm}$ \;\tikzmark{start6}    &  &  \quad\tikzmark{end6}                      & length of nozzle                        \\
    \circled{2} & $A_\mathrm{n2,o}$  & \tikzmark{start19}\quad\qquad &  & \tikzmark{end19}\;   $\num{70}$    & $\si{mm^2}$ \;\tikzmark{start20} &  &  \quad\tikzmark{end20}                     & nozzle outlet cross sectional area             \\

    \vspace{-0mm} \\
    
    \circled{3} & $s_\mathrm{u}$     & $\num{1}$  &$\si{mm}$       & $\num{7}$    &$\si{mm}$      & $\num{2.5}$     & $\si{mm}$         & upper gap size of junction              \\
    \circled{3} & $s_\mathrm{l}$     & \tikzmark{start7}\quad\qquad  &  &   \tikzmark{end7}\;    $\num{7}$     & $\si{mm}$ \;\tikzmark{start8}  &  &   \quad\tikzmark{end8}                        & lower gap size of junction              \\
    \circled{3} & $H_\mathrm{e}$     & \tikzmark{start15}\quad\qquad &  &   \tikzmark{end15}\;   $\num{10}$    & $\si{mm}$ \;\tikzmark{start16} &  &  \quad\tikzmark{end16}                       & cross-sectional extension of electrode \\
    \circled{3} & $W_\mathrm{j}$     & \tikzmark{start9}\quad\qquad  &  &  \tikzmark{end9}\;    $\num{10}$    & $\si{mm}$ \;\tikzmark{start10} &  &   \quad\tikzmark{end10}                       & width of junction in $z$-direction   \\
    \circled{3} & $D_\mathrm{h}$     & \tikzmark{start11}\quad\qquad &  &   \tikzmark{end11}\;   $\num{8.24}$  & $\si{mm}$ \;\tikzmark{start12} &  &  \quad\tikzmark{end12}                       & hydraulic diameter of junction inlet    \\
    \circled{3} & $\psi$             & $\num{0}$  &$\si{\degree}$  & $\num{90}$   & $\si{\degree}$   & $\num{45}$      & $\si{\degree}$  & junction angle of global flow deflection         \\

    \vspace{-0mm} \\
    
    \circled{4} & $H_\mathrm{c4}$    & $\num{3}$      &$\si{mm}$     & $\num{10}$    &$\si{mm}$          & $\num{8.48}$    & $\si{mm}$           & height of rectangular cross-section      \\
    \circled{4} & $A_\mathrm{c4,i}$  & $\num{30}$     &$\si{mm^2}$   & $\num{100}$   &$\si{mm^2}$    & $\num{84.8}$    & $\si{mm^2}$       & inlet area of connecting part       \\
    \circled{4} & $L_\mathrm{c4}$    & \tikzmark{start13}\quad\qquad &  &  \tikzmark{end13}\;   $\num{20}$    & $\si{mm}$ \;\tikzmark{start14} &  &  \quad\tikzmark{end14} & length of connecting part \\
    \circled{4} & $A_\mathrm{c4,o}$  & $\num{19.63}$  &$\si{mm^2}$   & $\num{78.54}$ &$\si{mm^2}$    & $\num{78.54}$  & $\si{mm^2}$       & outlet area of connecting part   \\
    
    \vspace{-0mm} \\
    
    \circled{5} & $D_\mathrm{p5}$    & $\num{5}$  &$\si{mm}$       & $\num{10}$    &$\si{mm}$      & $\num{10}$    & $\si{mm}$             & diameter of pipe                        \\
    \circled{5} & $L_\mathrm{p5}$    & $\num{100}$&$\si{mm}$       & $\num{123.6}$&$\si{mm}$      & $\num{111.8}$& $\si{mm}$       & length of pipe                          \\
    \circled{5} & $R_\mathrm{p5}$    &            &  n.a.          & $\num{15}$    &$\si{mm}$      & $\num{15}$    & $\si{mm}$       & curvature radius of curved segment      \\
    \bottomrule
\end{tabular}
\tikz[remember picture] \draw[overlay] ([yshift=.35em]pic cs:start1) -- ([yshift=.35em]pic cs:end1);
\tikz[remember picture] \draw[overlay] ([yshift=.35em]pic cs:start2) -- ([yshift=.35em]pic cs:end2);
\tikz[remember picture] \draw[overlay] ([yshift=.35em]pic cs:start3) -- ([yshift=.35em]pic cs:end3);
\tikz[remember picture] \draw[overlay] ([yshift=.35em]pic cs:start4) -- ([yshift=.35em]pic cs:end4);
\tikz[remember picture] \draw[overlay] ([yshift=.35em]pic cs:start5) -- ([yshift=.35em]pic cs:end5);
\tikz[remember picture] \draw[overlay] ([yshift=.35em]pic cs:start6) -- ([yshift=.35em]pic cs:end6);
\tikz[remember picture] \draw[overlay] ([yshift=.35em]pic cs:start7) -- ([yshift=.35em]pic cs:end7);
\tikz[remember picture] \draw[overlay] ([yshift=.35em]pic cs:start8) -- ([yshift=.35em]pic cs:end8);
\tikz[remember picture] \draw[overlay] ([yshift=.35em]pic cs:start9) -- ([yshift=.35em]pic cs:end9);
\tikz[remember picture] \draw[overlay] ([yshift=.35em]pic cs:start10) -- ([yshift=.35em]pic cs:end10);
\tikz[remember picture] \draw[overlay] ([yshift=.35em]pic cs:start11) -- ([yshift=.35em]pic cs:end11);
\tikz[remember picture] \draw[overlay] ([yshift=.35em]pic cs:start12) -- ([yshift=.35em]pic cs:end12);
\tikz[remember picture] \draw[overlay] ([yshift=.35em]pic cs:start13) -- ([yshift=.35em]pic cs:end13);
\tikz[remember picture] \draw[overlay] ([yshift=.35em]pic cs:start14) -- ([yshift=.35em]pic cs:end14);
\tikz[remember picture] \draw[overlay] ([yshift=.35em]pic cs:start15) -- ([yshift=.35em]pic cs:end15);
\tikz[remember picture] \draw[overlay] ([yshift=.35em]pic cs:start16) -- ([yshift=.35em]pic cs:end16);
\tikz[remember picture] \draw[overlay] ([yshift=.35em]pic cs:start17) -- ([yshift=.35em]pic cs:end17);
\tikz[remember picture] \draw[overlay] ([yshift=.35em]pic cs:start18) -- ([yshift=.35em]pic cs:end18);
\tikz[remember picture] \draw[overlay] ([yshift=.35em]pic cs:start19) -- ([yshift=.35em]pic cs:end19);
\tikz[remember picture] \draw[overlay] ([yshift=.35em]pic cs:start20) -- ([yshift=.35em]pic cs:end20);
\end{table}
\section{Anisotropic porosity model in tensor formulation}
\label{sec:appendix_model}

\begin{figure}[h]
\centering
\hspace*{1cm}
{\rlap{\subcaptionmark\label{fig:wire-screen_side-view_for_homogenization_appendix}}\adjustbox{trim=-5mm 1mm 0mm 0mm, valign=t}
{\includegraphics[height=.29\textwidth]{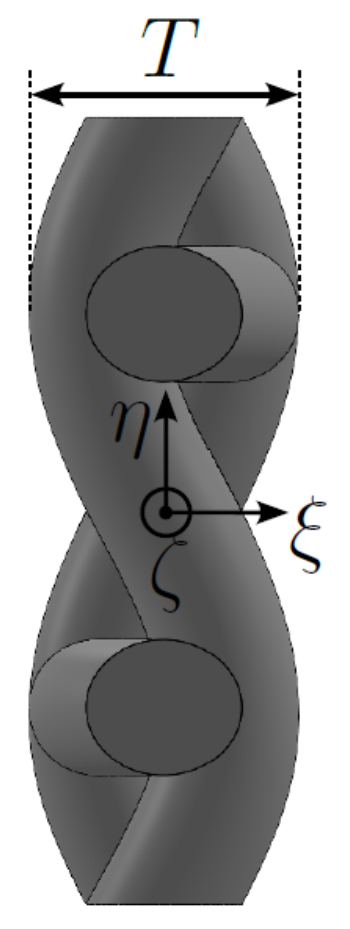}}} 
\hspace*{1cm}
{\rlap{\subcaptionmark\label{fig:wire-screen_homogeneous_appendix}}\adjustbox{trim=-5mm 1mm 0mm 0mm, valign=t}
{\includegraphics[height=.29\textwidth]{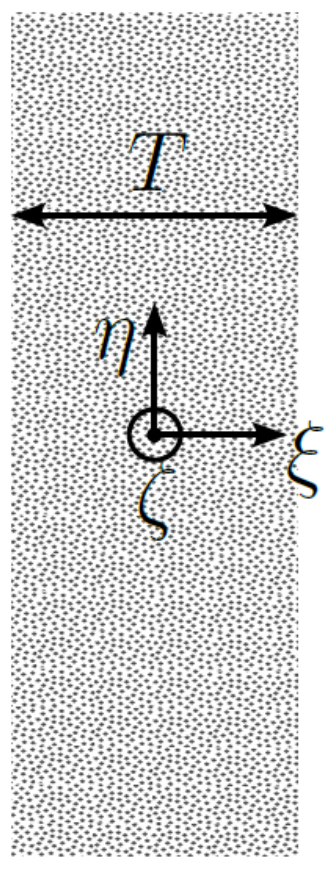}}} \\
\caption{Illustration of the homogenized model.
\subref{fig:wire-screen_side-view_for_homogenization_appendix}~Side view of the original wire screen.
\subref{fig:wire-screen_homogeneous_appendix}~Side view of the homogenized model of the wire screen.
}
\label{fig:wire-frame_for_homogenization_appendix}
\end{figure}

\begin{figure}[!ht]
	\centering
	\includegraphics[width=0.3\textwidth]{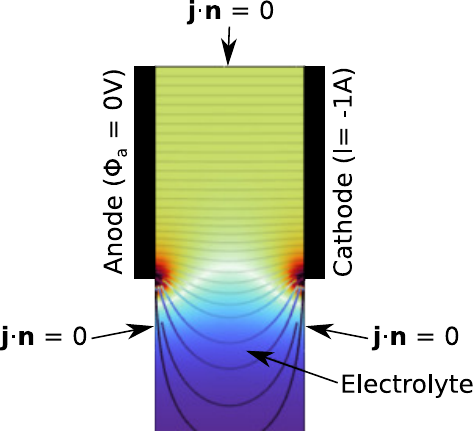}
	\caption{Schematic drawing of the applied boundary conditions for the current simulation in COMSOL. Exemplarily shown is the T7-electrolyzer.
	}
	\label{fig:current_simulation_setup}
\end{figure}

Equation~(2) is the defining equation for the 
homogeneous model, with
the tensors $\boldsymbol{D}^\mathrm{e}$ and $\boldsymbol{F}^\mathrm{e}$ 
assembling viscous and inertial resistance coefficients, respectively.
To determine the parameters by Direct Numerical Simulations, the thickness $T$ of the porous zone was first fixed to twice the wire diameter, 
as motivated by the envelope of the wire screen in Fig.~\ref{fig:wire-screen_side-view_for_homogenization_appendix}.
Then, the geometry of Fig.~\ref{fig:wire-screen_side-view_for_homogenization_appendix} was employed 
for fully resolved flow simulations from which the parameter values were obtained.
This is a non-trivial task and will be reported in~\cite{Schoppmann_Froehlich_model}.

%
%
%
\begin{table}[!h]
	\caption{Calibrated parameter values of the porous layer which served as a homogenized model for the electrode.}
	\label{tab:calibrated_parameters}
	\centering
	\begin{tabular}{ l@{$\,=\,$}r@{\,}ll  }
		\toprule
		$T$           &  \; $\num{0.448e-3}$ & $\si{m}$        & thickness of porous layer, set to $T=2 D_\mathrm{w}$           \\
		$D$           &  $\num{2.69e+8}$  & $\si{m^{-2}}$   & viscous resistance in Darcy tensor  \\  
		$F$               &  $\num{6002}$     & $\si{m^{-1}}$     & inertial resistance in Forchheimer tensor    \\ 
		$f_\mathrm{D}$    &  $\num{0.55}$     &                   & anisotropy factor for Darcy tensor \\
		$f_\mathrm{F}$    &  $\num{0.07}$     &                   & anisotropy factor for Forchheimer tensor \\ 
		\bottomrule
	\end{tabular}
\end{table}
%
\section{Configuration data for the flow simulations}
\label{sec:appendix_fluid_flow}

This appendix provides definitions and values for properties of the fluid, characteristic velocities and Reynolds numbers.
The nominal Reynolds number is the Reynolds number at the junction inlet.
The values provided in Tab.~\ref{tab:flow_parameters} were used for all simulations reported in this study.

\begin{table}[!h]
	\caption{
		Dimensional and non-dimensional flow parameters applied in numerical simulations independent of electrolyzer shape.
	}
	\label{tab:flow_parameters}
	\centering
	\begin{tabular}{ ll@{$\,=\,$}cll  }
		\toprule
		& $\rho$             &                                                                                    & $\phantom{=} \SI{1050}{kg\per\meter\cubed}$ & density of electrolyte \tabrowsep\\
		& $\mu$              &                                                                                    & $\phantom{=} \SI{1.12e-3}{\pascal\second}$  & dynamic viscosity of electrolyte \tabrowsep\\
		& $\nu$              &  $\dfrac{\mu}{\rho}$                                                               & $= \SI{1.07e-6}{m^2/s}$                     & kinematic viscosity of electrolyte \tabrowsep\\
		\midrule
		& $U_\mathrm{i}$     &  $\dfrac{Re_\mathrm{i}\,\nu}{D_\mathrm{h}}$                                        & $= \SI{0.194}{m/s}$                         & mean velocity at inlet of junction \tabrowsep\\
		& $U_\mathrm{p1}$    &  $\dfrac{s_\mathrm{l} \, W_\mathrm{j}}{\frac\pi4 D_\mathrm{p1}^2} \,U_\mathrm{i}$  & $= \SI{0.174}{m/s}$                         & mean velocity of flow in inlet pipe \tabrowsep\\ 
		& $U_\mathrm{e}$     &  $U_\mathrm{i} \; \dfrac{s_\mathrm{l}}{2\,H_\mathrm{e}}$                           & $= \SI{0.068}{m/s}$                         & mean velocity through electrode \tabrowsep\\
		\midrule
		
		& $Re_\mathrm{i}$     &  $\dfrac{U_\mathrm{i} \, D_\mathrm{h}}{\nu}$                                      & $= \num{1500}$         & Reynolds number at junction inlet            \tabrowsep\\
		& $Re_\mathrm{p1}$    &  $\dfrac{U_\mathrm{p1} \, D_\mathrm{p1}}{\nu}$                                    & $= \num{1623}$         & Reynolds number of flow in inlet pipe \tabrowsep\\
		\bottomrule
	\end{tabular}
\end{table}
\section{Boundary conditions of current simulation in COMSOL}
\label{sec:appendix_current_simulation}

\section{Composition of total hydrodynamic drag force on bubbles}
\label{sec:drag_coefficient}

\begin{figure}[hb]
	\centering
	\includegraphics[width=95mm]{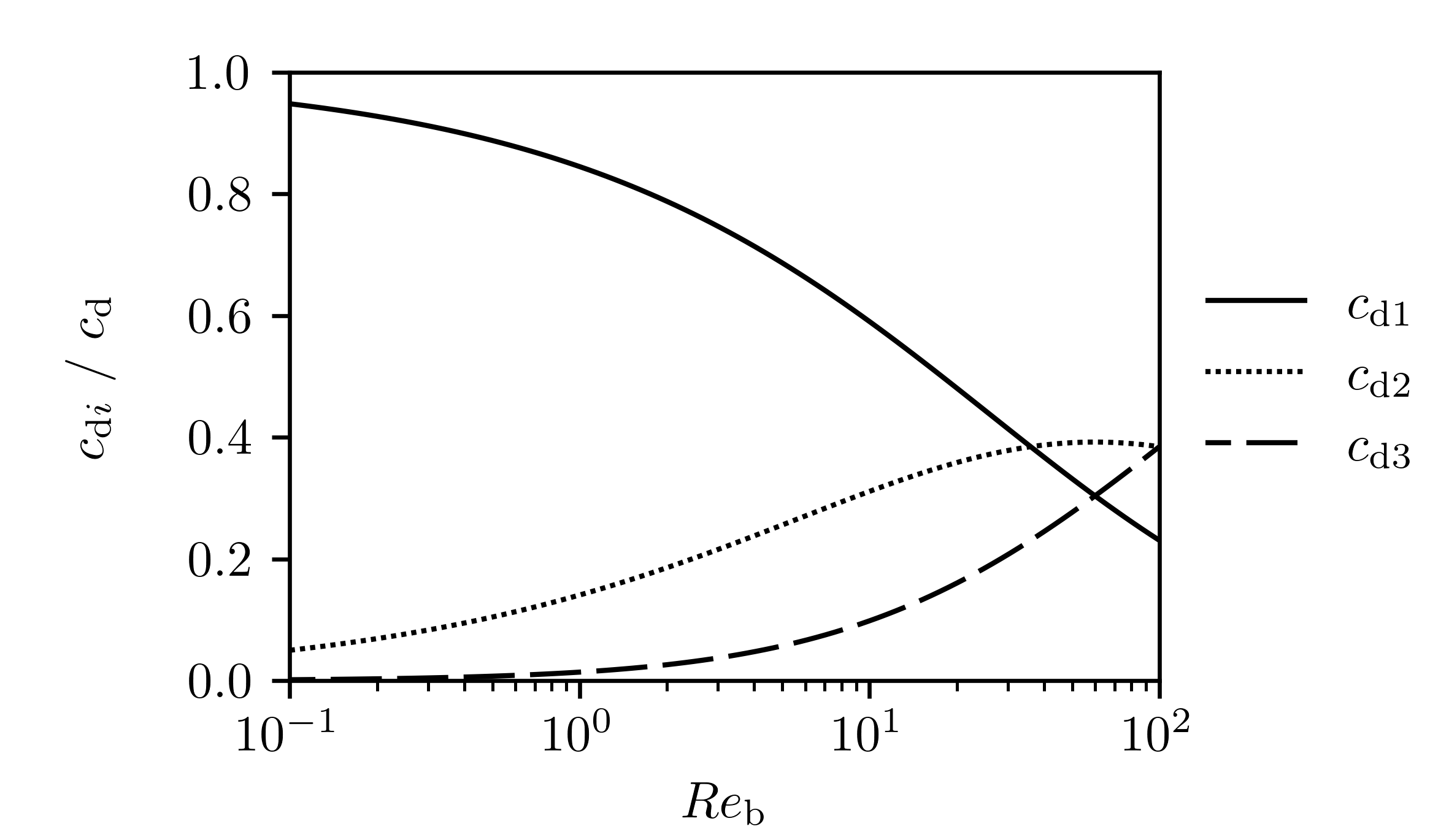}
	\caption{Contribution of different terms to total drag coefficient for laminar flow around sphere in unbounded domain.}
	\label{fig:c_d_sphere}
\end{figure}

To assess the drag regime of bubbles in an FT-ME, 
a configuration of laminar flow around a sphere in an unbounded domain is taken as a reference.
It is understood that this approach may oversimplify the much more complex problem of gas bubble formation on a flow-through wire screen electrode. 
However, it provides a rough impression of the situation, which is appropriate at this stage of investigation.
Based on these considerations, the drag coefficient can be described by the relation
\begin{equation}
	\label{eq:c_d_sphere}
	c_\mathrm{d}(Re_\mathrm{b})
	=
	\underbrace{\frac{24}{Re_\mathrm{b}} }_{c_\mathrm{d1}} 
	+ 
	\underbrace{ \frac{4}{\sqrt{Re_\mathrm{b}}}  }_{c_\mathrm{d2}} 
	+ 
	\underbrace{0.4}_{c_\mathrm{d3}} 
\end{equation}
holding for $Re_\mathrm{b} < \num{2e+5}$ according to \citeauthor{Kaskas1964}~\cite{Kaskas1964}, taken over 
by \citeauthor{Brauer1971}~\cite{Brauer1971}.
Here, the bubble Reynolds number 
$Re_\mathrm{b} = D_\mathrm{b} u_\mathrm{b} / \nu$
is calculated with a characteristic flow velocity 
$u_\mathrm{b}=U_\mathrm{e}/\beta$ 
to which a bubble is subjected. 
As can be seen from Eq.~\ref{eq:c_d_sphere},
the total drag coefficient consists of three different terms.
Their contribution is quantified in Tab.~\ref{tab:c_d_sphere} and illustrated in Fig.~\ref{fig:c_d_sphere} for a realistic range of $Re_\mathrm{b}$.
%
\begin{table}[hb]
	\caption{
		Contribution of different terms to total drag coefficient for laminar flow around sphere in unbounded domain.
	}
	\label{tab:c_d_sphere}
	\centering
	\begin{tabular}{ c | c c c c c c}
		\toprule
		$Re_\mathrm{b}$	                  &  $0.1$         &  $1$           &  $10$          &  $33$          \\
		\hline
		$c_{\mathrm{d}1} / c_\mathrm{d}$  &  $\SI{95}{\%}$ &  $\SI{85}{\%}$ &  $\SI{59}{\%}$ &  $\SI{40}{\%}$ \\
		$c_{\mathrm{d}2} / c_\mathrm{d}$  &  $\SI{5}{\%}$  &  $\SI{14}{\%}$ &  $\SI{31}{\%}$ &  $\SI{38}{\%}$ \\
		$c_{\mathrm{d}3} / c_\mathrm{d}$  &  $\SI{0}{\%}$  &  $\SI{1}{\%}$  &  $\SI{10}{\%}$ &  $\SI{22}{\%}$ \\
		\bottomrule
	\end{tabular}
\end{table}
%
\newpage
\section{Influence of upper electrode gap on electrolyte flow in Y-shape}
\label{sec:appendix_y_simulations}

\begin{figure}[ht]
	\centering
	%
	\hspace*{0pt}{\rlap{\subcaptionmark\label{fig:Y_streamlines_Y1_xy}}\adjustbox{trim=-5mm 0mm 0mm 0mm, clip, valign=t}{
			\includegraphics[height=\FigHeightXY]{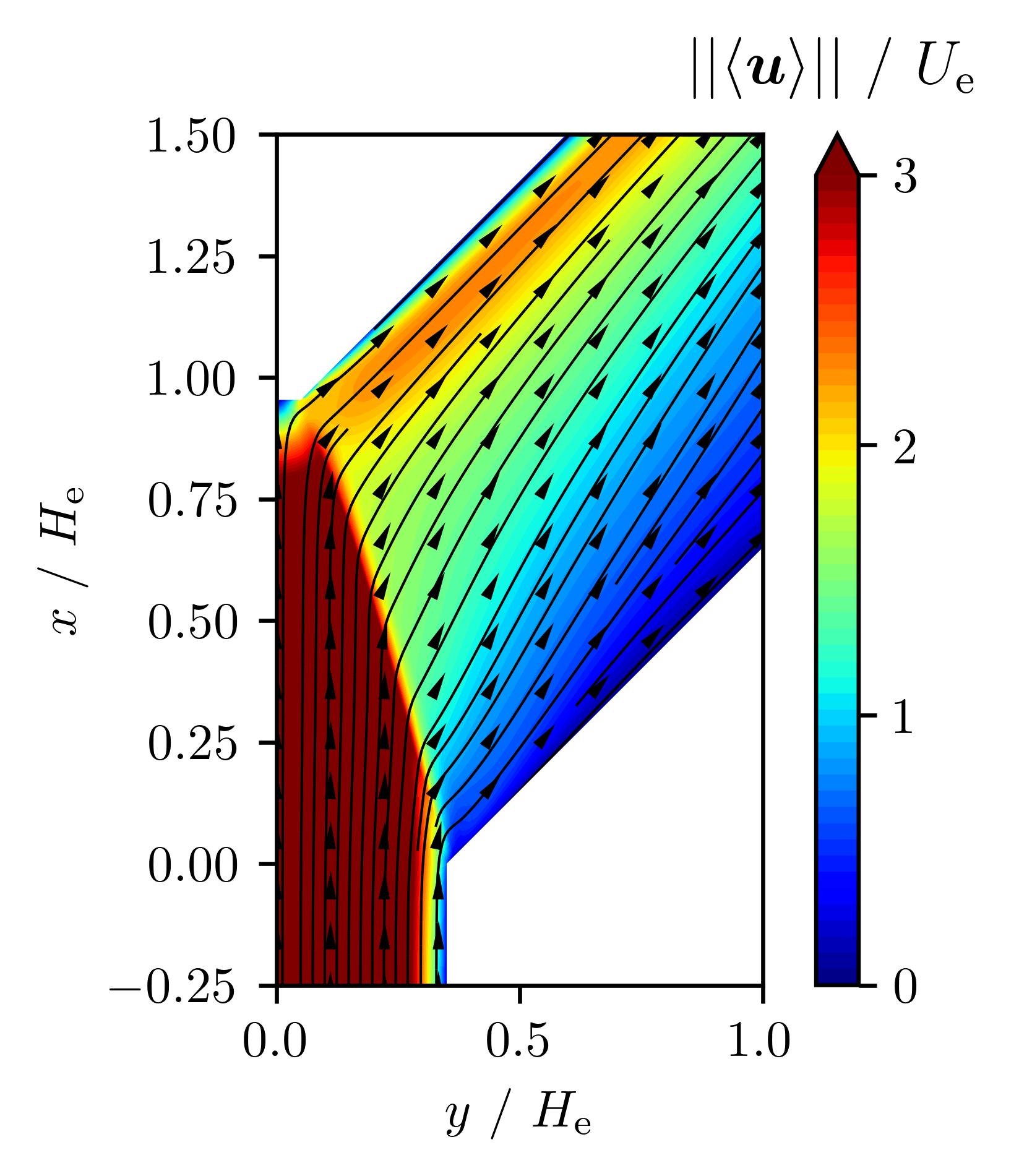}}} \hspace{5mm}
	\hspace*{0pt}{\rlap{\subcaptionmark\label{fig:Y_streamlines_Y1_3D}}
		\adjustbox{trim=-5mm 0mm 0mm 0mm, clip, valign=t}{
			\adjustbox{Clip= 5mm 0mm 5mm 0mm, valign=t}{
				\includegraphics[height=\FigHeightDDD,trim=2mm 0mm 0mm 20mm,clip]{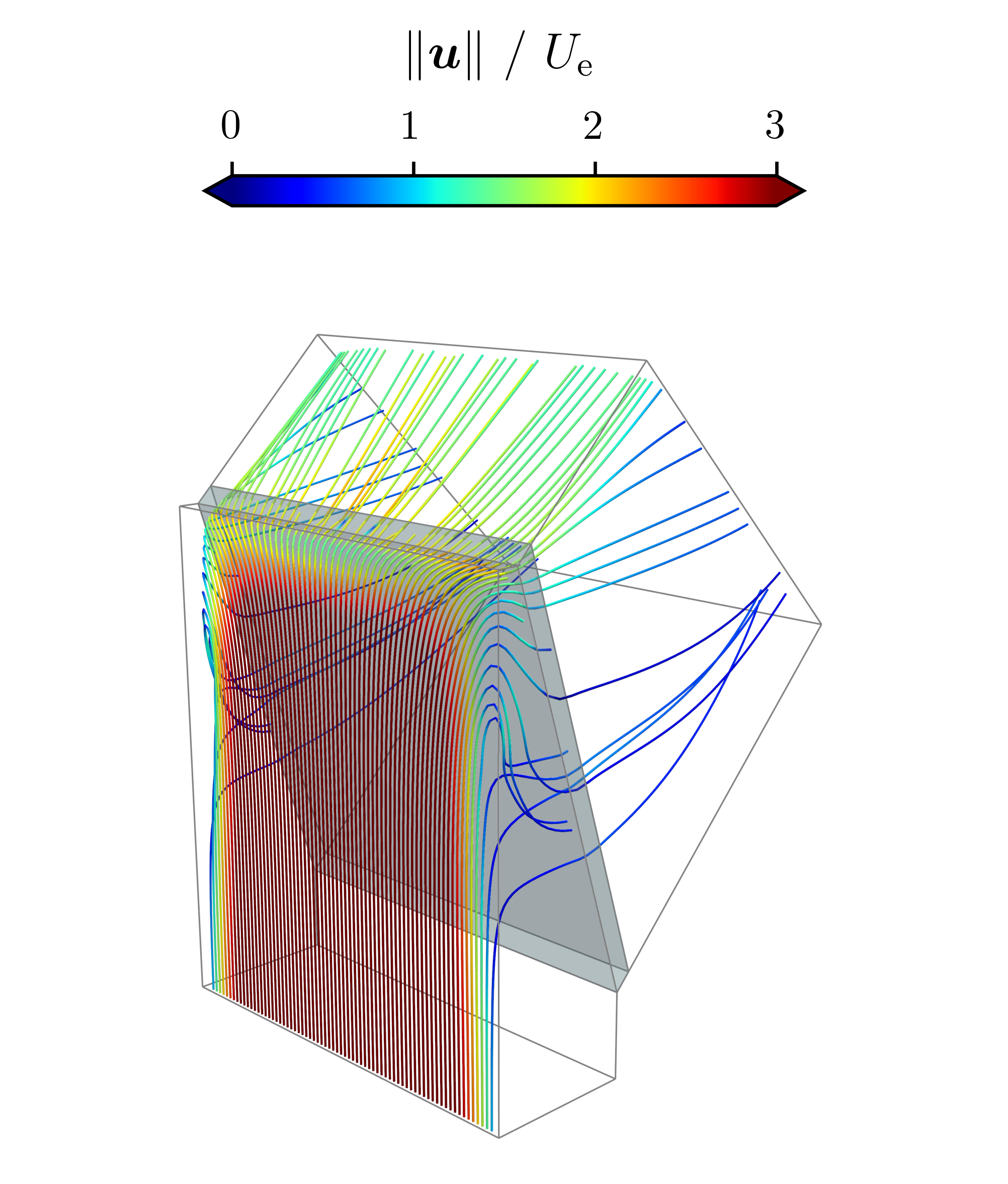}}}} \\
	%
	
	\hspace*{0pt}{\rlap{\subcaptionmark\label{fig:Y_streamlines_Y5_xy}}\adjustbox{trim=-5mm 0mm 0mm 0mm, clip, valign=t}{
			\includegraphics[height=\FigHeightXY]{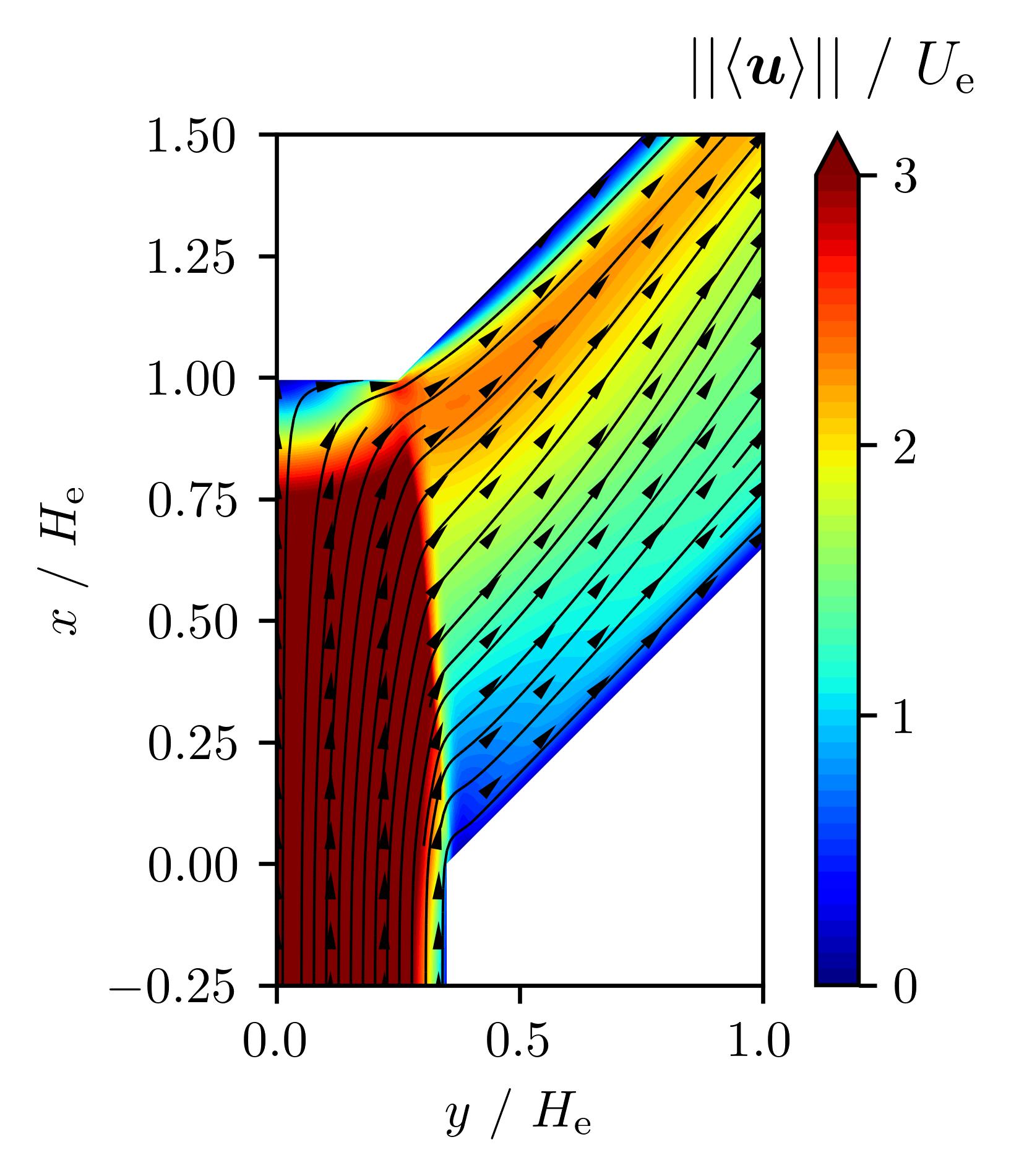}}} \hspace{5mm}
	\hspace*{0pt}{\rlap{\subcaptionmark\label{fig:Y_streamlines_Y5_3D}}
		\adjustbox{trim=-5mm 0mm 0mm 0mm, clip, valign=t}{
			\adjustbox{Clip= 5mm 0mm 5mm 0mm, valign=t}{
				\includegraphics[height=\FigHeightDDD,trim=2mm 0mm 0mm 20mm,clip]{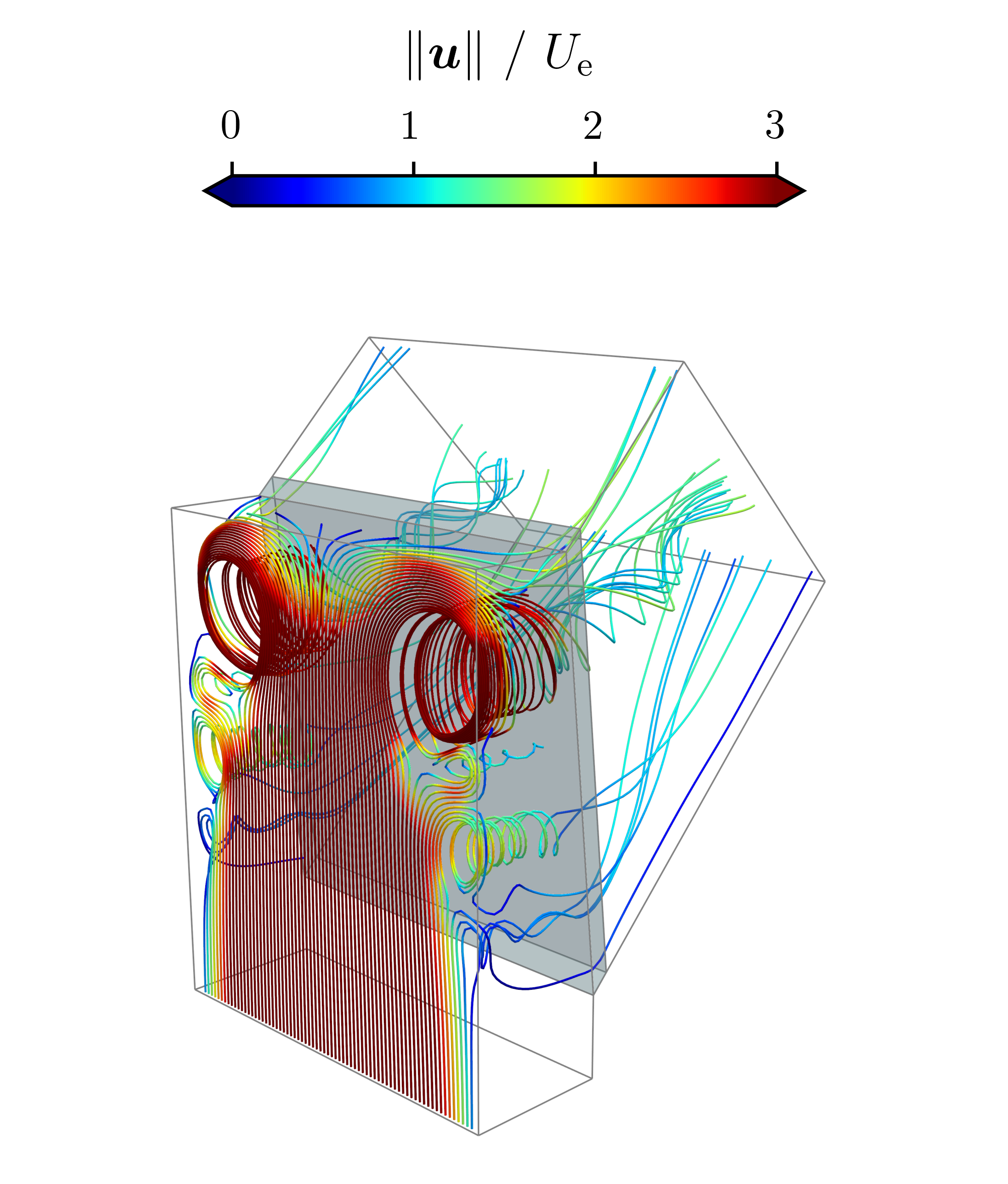}}}} \\
	\caption{
		Streamlines and mean velocity obtained for two variants of the Y-electrolyzer:
		\subref{fig:Y_streamlines_Y1_xy} - \subref{fig:Y_streamlines_Y1_3D}~Electrolyzer Y1,
		\subref{fig:Y_streamlines_Y5_xy} - \subref{fig:Y_streamlines_Y5_3D}~electrolyzer Y5.
		Left column:
		Magnitude of time-averaged velocity $\lVert \langle \boldsymbol{u} \rangle \rVert$ together with streamlines in $xy$-plane at $z = 0$ (centerplane of flow cell).
		Right column:
		Three-dimensional view with selected streamlines of instantaneous flow
		emitted along a line of points in $z$-direction at $x=-0.2/H_\mathrm{e}$ and $y=0.01/H_\mathrm{e}$, 
		and colored with the velocity magnitude according to the same scale as in the left column.
		Homogenized electrode rendered gray.
		Only part of computational domain shown.
	} \label{fig:appendix_Y_streamlines}
\end{figure}

\begin{figure}[ht]
	\centering
	\begin{tabular}{lcc}
		& \textbf{Y1} & \textbf{Y5} \\
		%
		\parbox[t]{2mm}{\multirow{1}{*}[-1em]{\rotatebox[origin=c]{90}{ $\lVert \langle \boldsymbol{u} \rangle \rVert\ /\ U_\mathrm{e}$ }}}   &
		{\rlap{\subcaptionmark\label{fig:Y_elec_inlet_Y1_mag}}\adjustbox{trim=-4mm 0mm 0mm 0mm, clip, valign=t}{
				\includegraphics[height=\FigHeightElec,trim=0mm 0mm 0mm 0mm,clip]{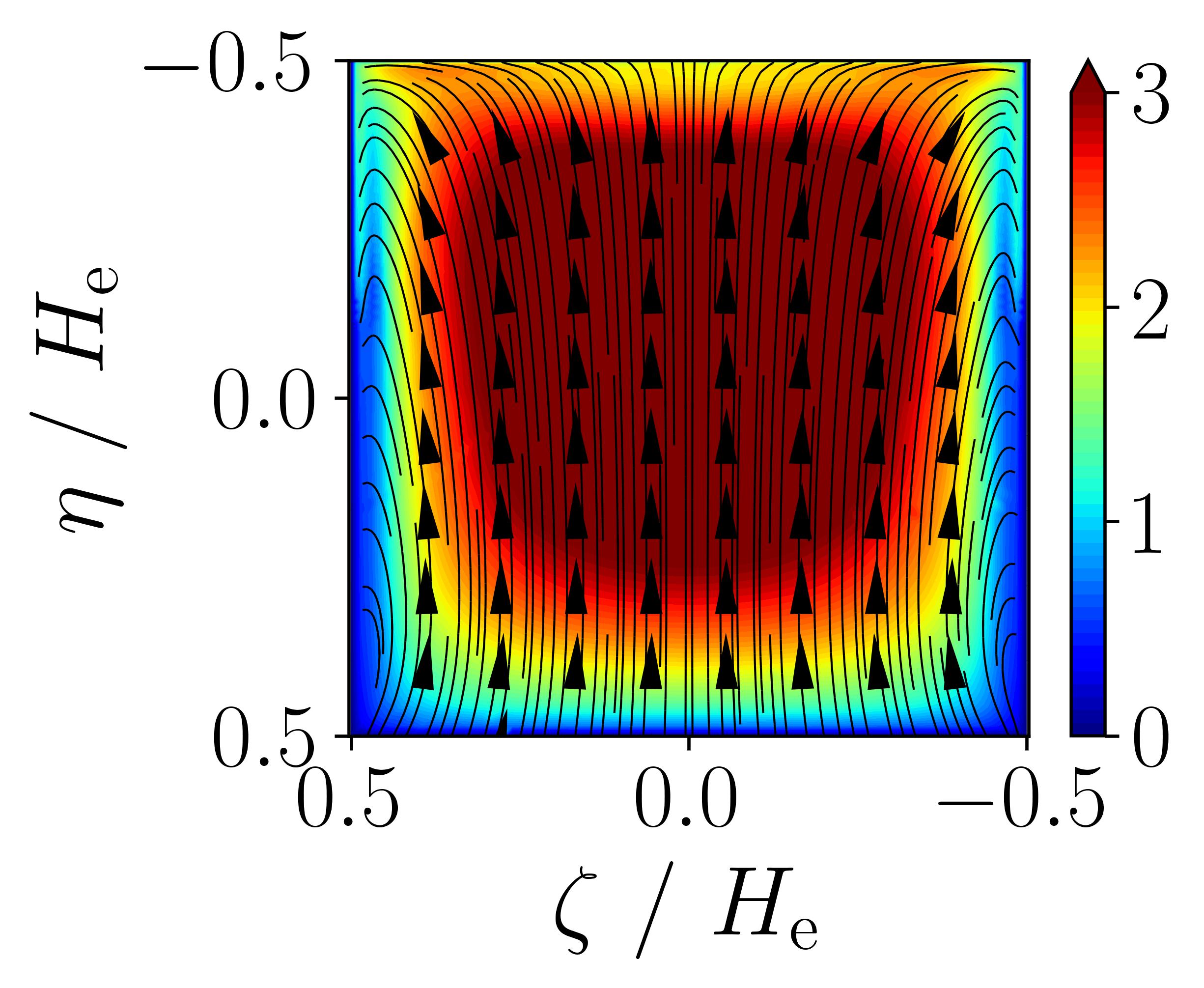}}} &
		{\rlap{\subcaptionmark\label{fig:Y_elec_inlet_Y5_mag}}\adjustbox{trim=-4mm 0mm 0mm 0mm, clip, valign=t}{
				\includegraphics[height=\FigHeightElec,trim=0mm 0mm 0mm 0mm,clip]{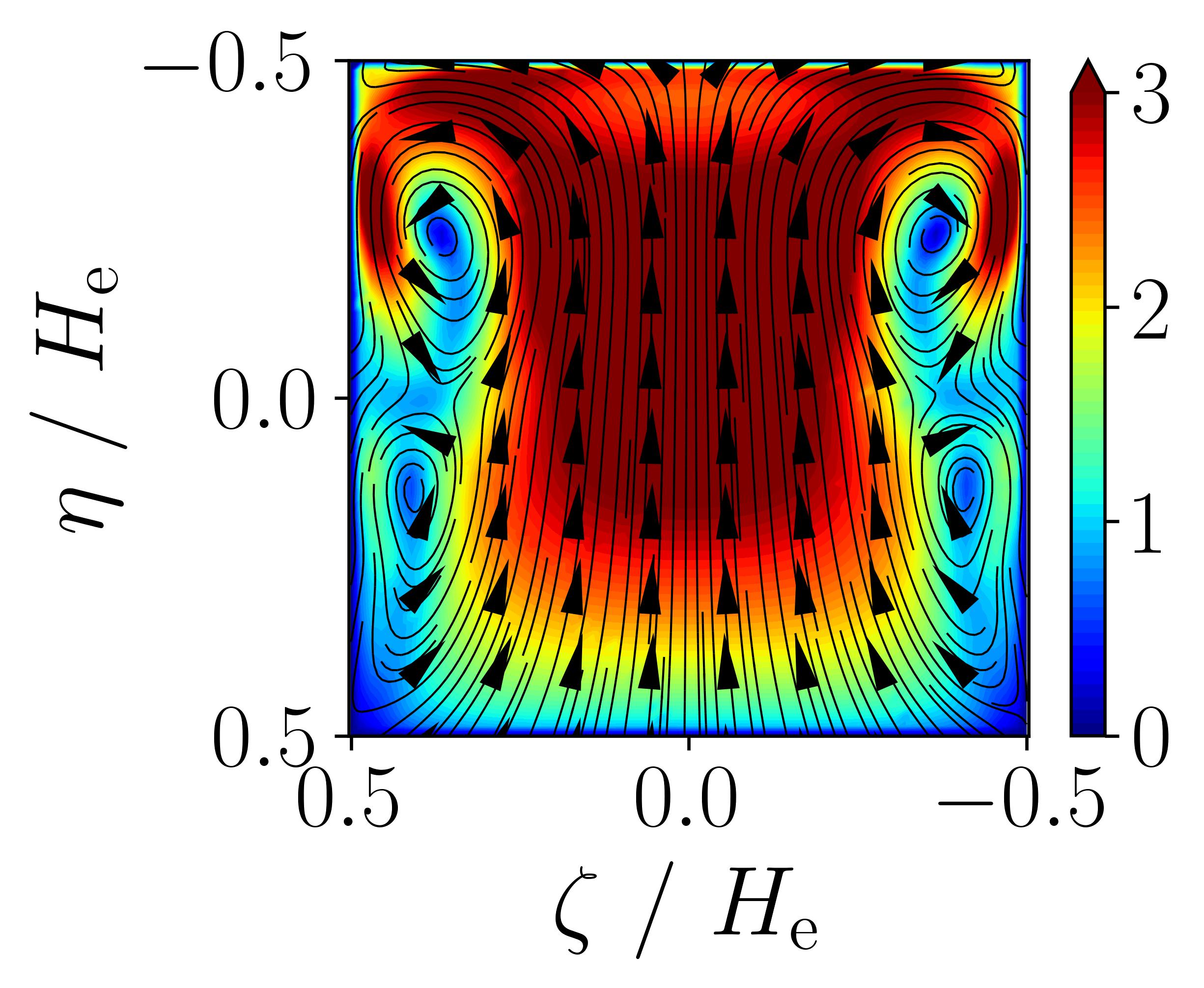}}} \\
		%
		\parbox[t]{2mm}{\multirow{1}{*}[-1em]{\rotatebox[origin=c]{90}{ $\langle u_\mathrm{\xi} \rangle\ /\ U_\mathrm{e}$ }}}   &
		{\rlap{\subcaptionmark\label{fig:Y_elec_inlet_Y1_xi_av}}\adjustbox{trim=-4mm 0mm 0mm 0mm, clip, valign=t}{
				\includegraphics[height=\FigHeightElec,trim=0mm 0mm 0mm 0mm,clip]{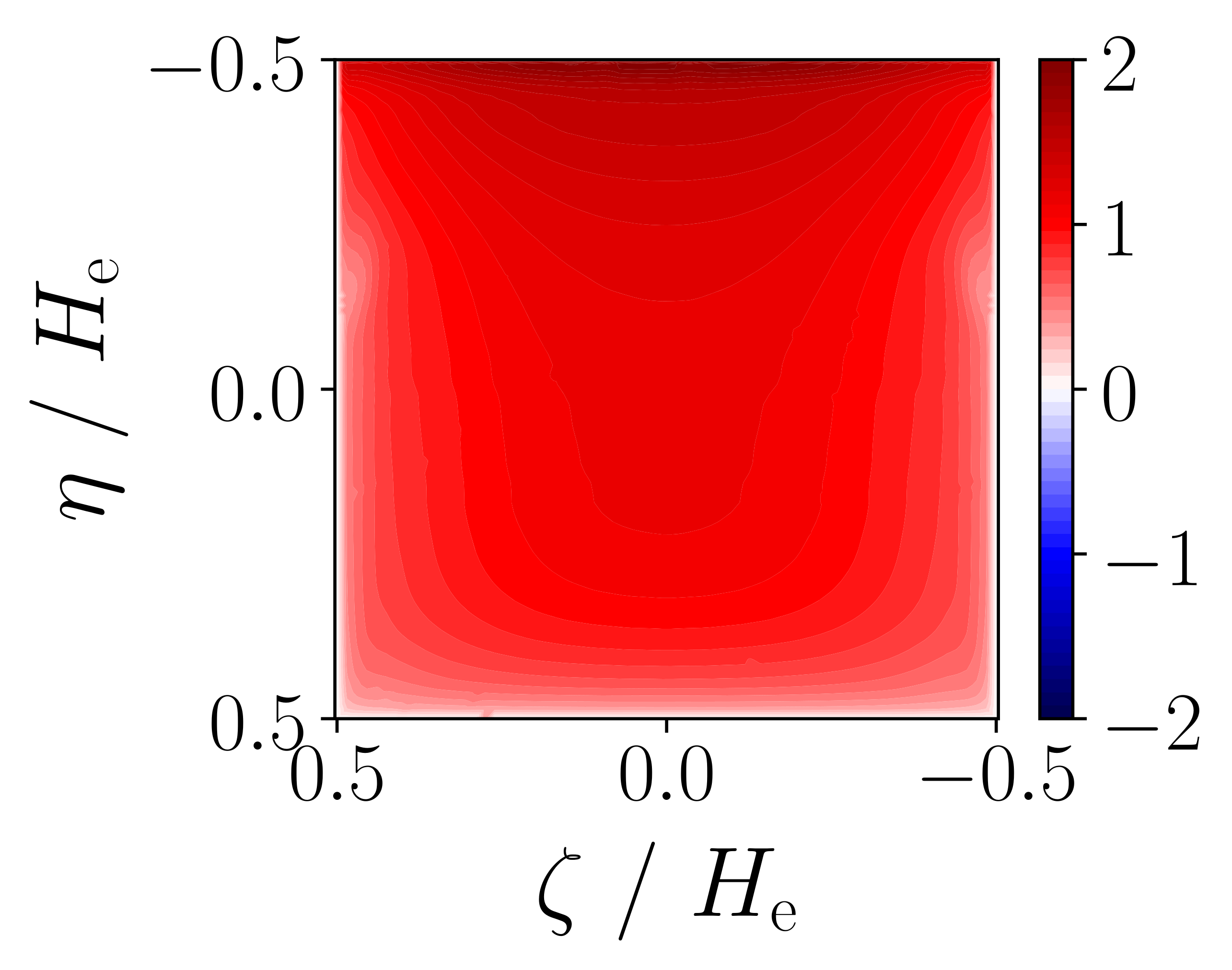}}} &
		{\rlap{\subcaptionmark\label{fig:Y_elec_inlet_Y5_xi_av}}\adjustbox{trim=-5mm 0mm 0mm 0mm, clip, valign=t}{
				\includegraphics[height=\FigHeightElec,trim=0mm 0mm 0mm 0mm,clip]{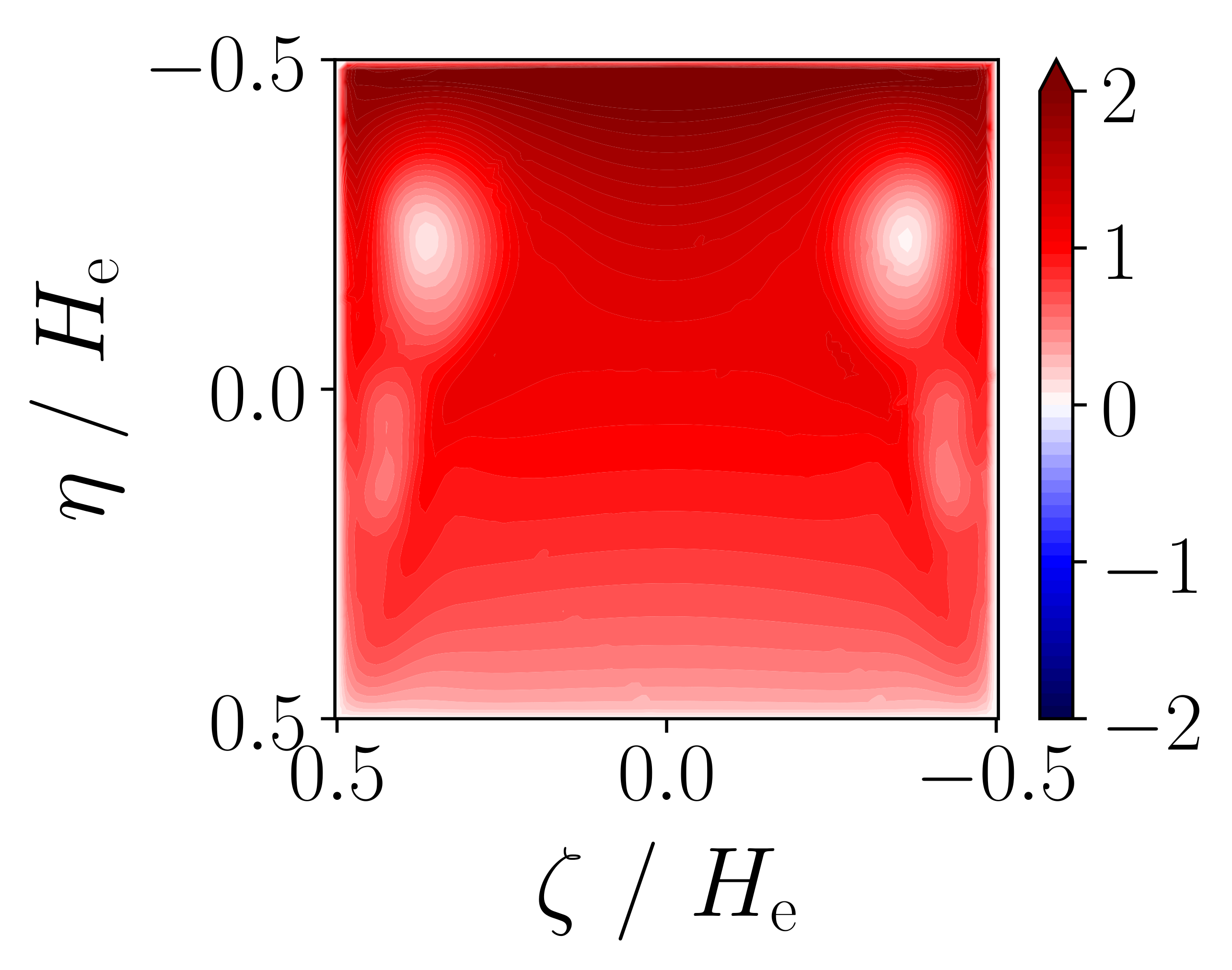}}} \\
		%
		\parbox[t]{2mm}{\multirow{1}{*}[-1em]{\rotatebox[origin=c]{90}{ $ u_\mathrm{\xi}\  /\ U_\mathrm{e}$ }}}   &
		{\rlap{\subcaptionmark\label{fig:Y_elec_inlet_Y1_xi}}\adjustbox{trim=-4mm 0mm 0mm 0mm, clip, valign=t}{
				\includegraphics[height=\FigHeightElec,trim=0mm 0mm 0mm 0mm,clip]{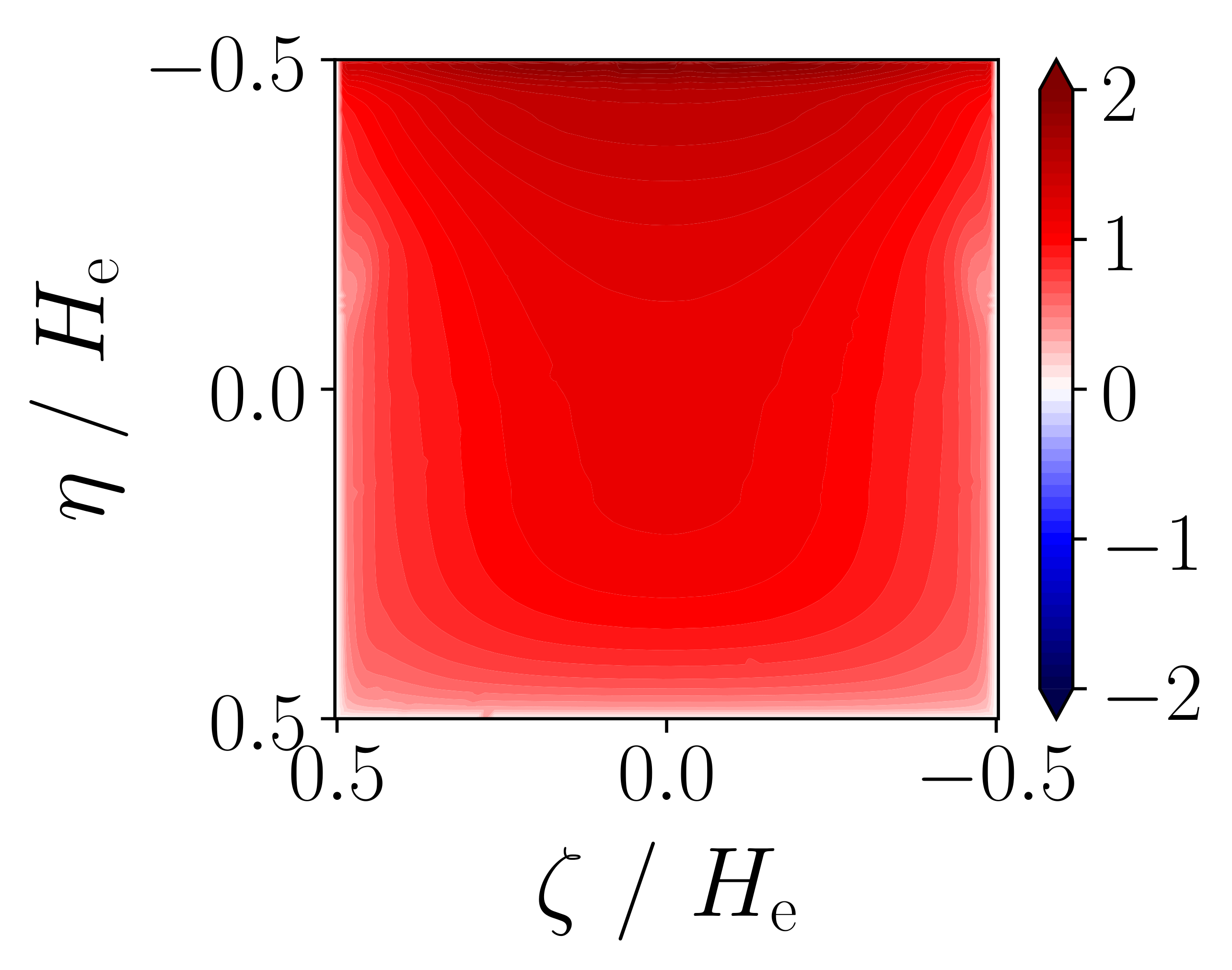}}} &
		{\rlap{\subcaptionmark\label{fig:Y_elec_inlet_Y5_xi}}\adjustbox{trim=-5mm 0mm 0mm 0mm, clip, valign=t}{
				\includegraphics[height=\FigHeightElec,trim=0mm 0mm 0mm 0mm,clip]{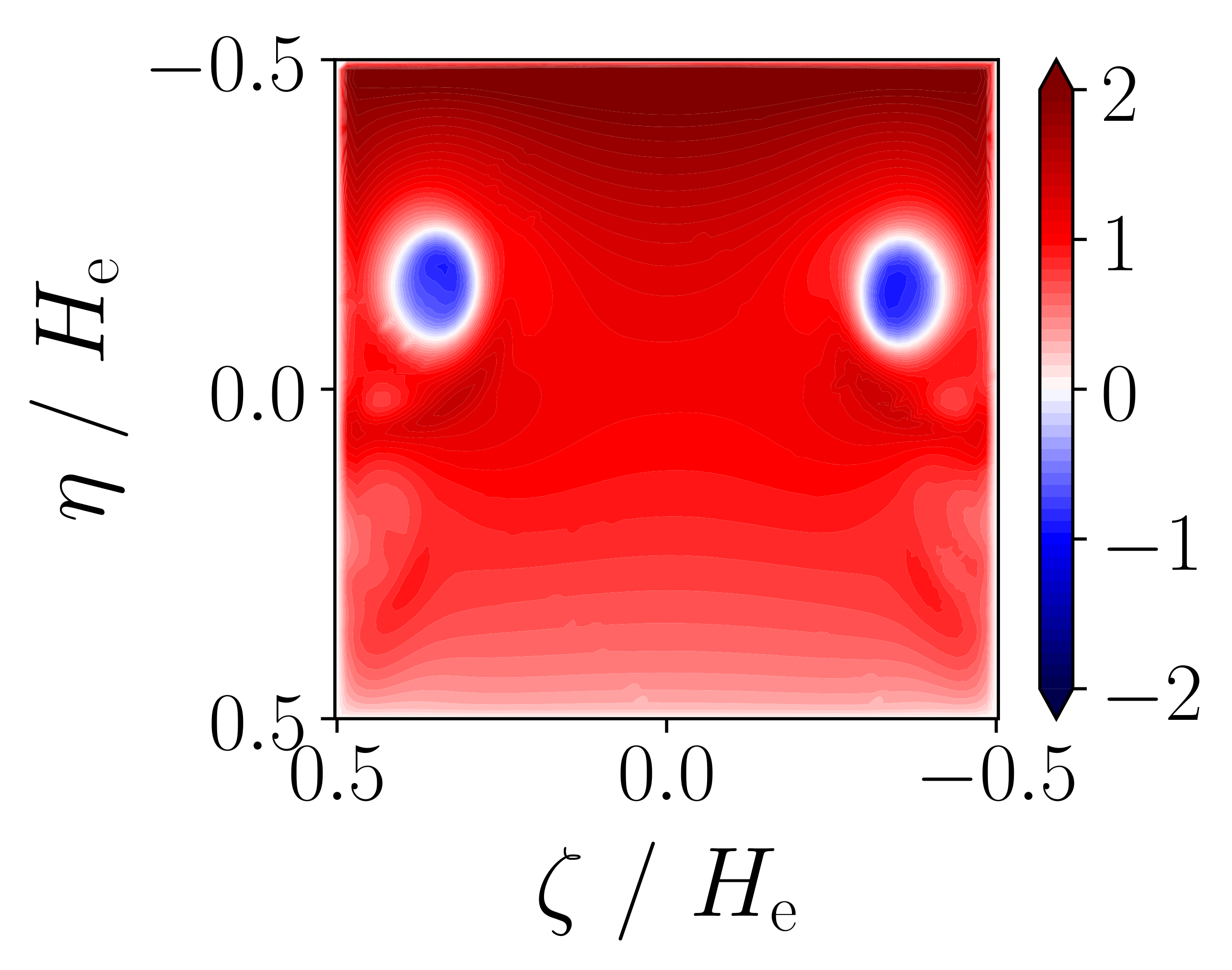}}} \\
		%
		\parbox[t]{2mm}{\multirow{1}{*}[-1em]{\rotatebox[origin=c]{90}{ $ \sqrt{\langle u_\mathrm{\xi}' u_\mathrm{\xi}' \rangle }\ /\ U_\mathrm{e}$ }}}   &
		{\rlap{\subcaptionmark\label{fig:Y_elec_inlet_Y1_xi_rms}}\adjustbox{trim=-4mm 0mm 0mm 0mm, clip, valign=t}{
				\includegraphics[height=\FigHeightElec,trim=0mm 0mm 0mm 0mm,clip]{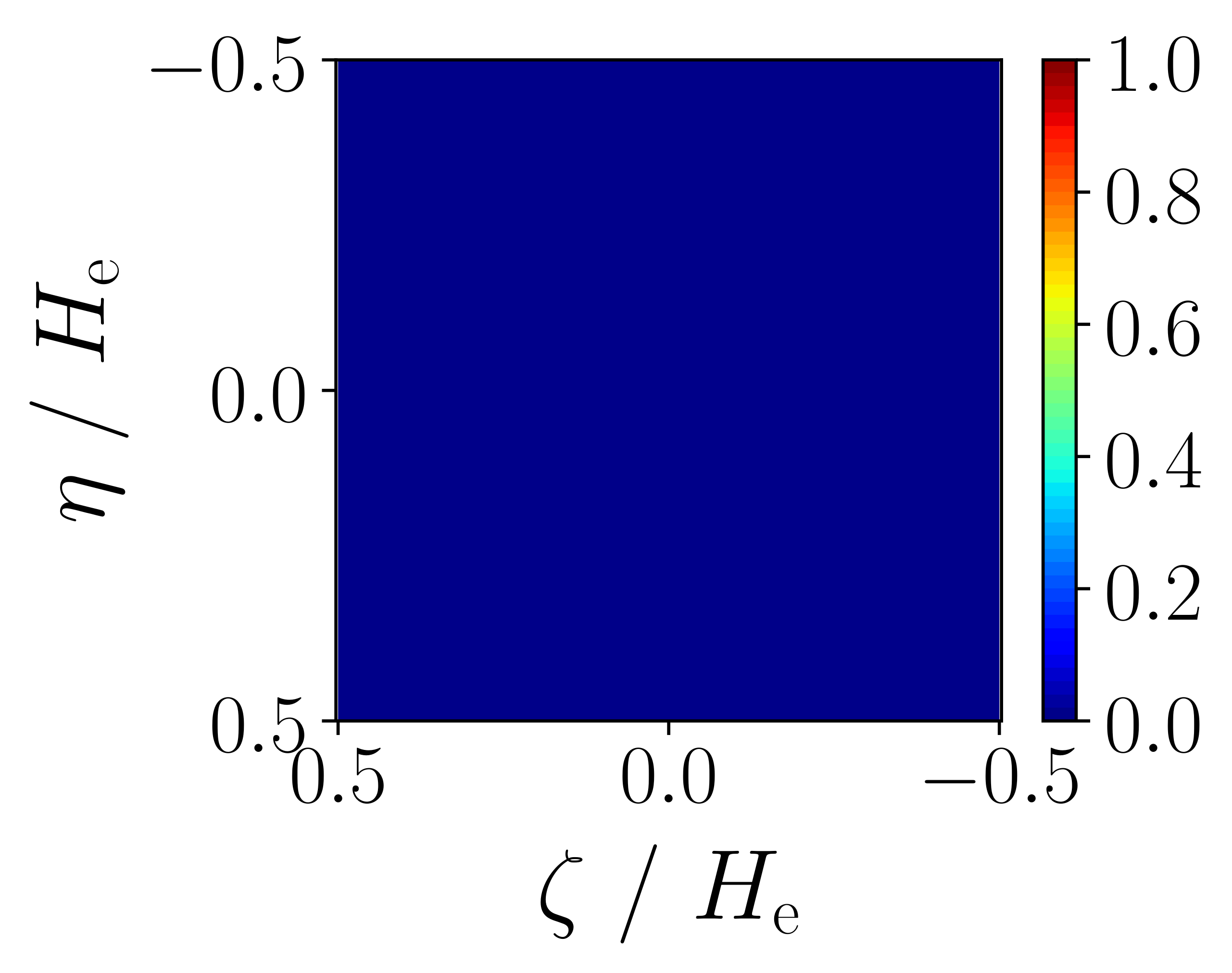}}} &
		{\rlap{\subcaptionmark\label{fig:Y_elec_inlet_Y5_xi_rms}}\adjustbox{trim=-5mm 0mm 0mm 0mm, clip, valign=t}{
				\includegraphics[height=\FigHeightElec,trim=0mm 0mm 0mm 0mm,clip]{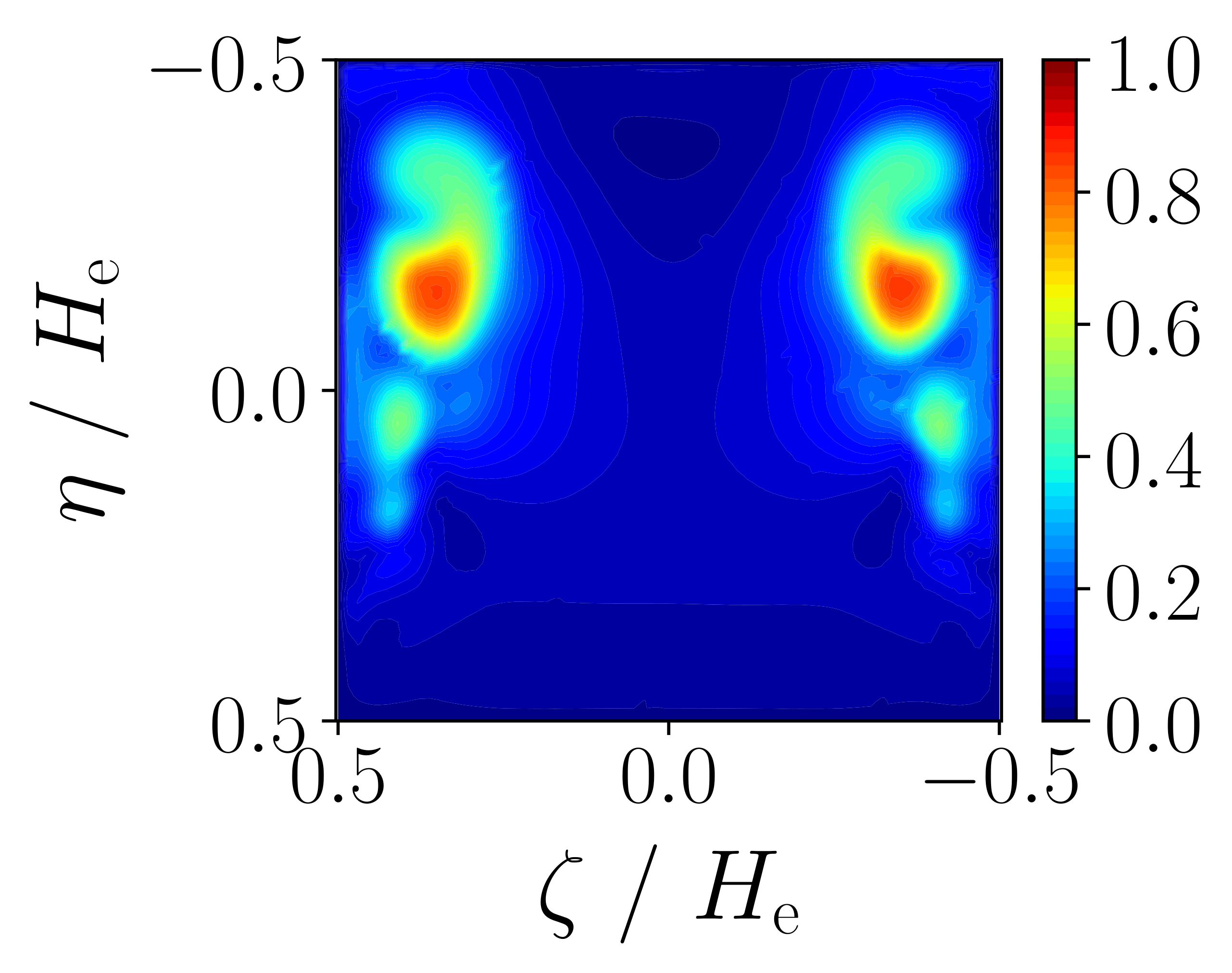}}} \\
	\end{tabular}
	\caption{ 
		Flow in $\zeta\eta$-plane at $\xi = -T/2$ (upstream side of electrode)
		for two variants of the Y-electrolyzer: 
		Y1 and Y5 in left and right column, respectively.
		From upper to lower row:
		Magnitude of time-averaged velocity 
		$\lVert \langle \boldsymbol{u} \rangle \rVert$ 
		and streamlines,
		time-averaged normal velocity component $\langle u_\xi \rangle$,
		instantaneous normal velocity component $u_\xi$ at an instant with the strongest reverse flow observed over the simulated time
		and temporal standard deviation of $u_\xi$.
	} \label{fig:appendix_Y_elec_inlet}
\end{figure}

\section{Further evaluation of the design criteria: non-parametric skew $S_{\zeta\eta}$ of ratios ${R}^{(1)}$ and ${R}^{(2)}$}
\label{sec:skew}

Ideally, however, the standard deviation $\sigma_{\zeta\eta}$ should vanish, indicating a constant design ratio. This is not the case, even for Y2.5, so that there still might be a small risk of hotspots where $R$ would be too small, resulting in possibility of gas crossover.
To investigate this further, the non-parametric skew is evaluated, which is defined for a quantity $R$ by 
\begin{equation}
	S_{\zeta\eta}(R) = \frac{ \langle R \rangle_{\zeta\eta} - M_{\zeta\eta}(R) }{\sigma_{\zeta\eta}(R)} ,
\end{equation}
where $M_{\zeta\eta}$ is the median over all points $(\zeta, \eta)$. In the context of the present problem, a negative skew means that there is more area on the electrode with $R > 1$, i.e.\ stronger bubble removal compared to bubble generation, and less area with $R < 1$. Consequently, a negative skew is preferable because it implies that there is less area of outliers with strong gas production and low flow velocity, hence lower risk of gas crossover. For most values of the non-parametric skew listed in Tab.~\ref{tab:std_skew_zeta_eta_plane} applies $S_{\zeta\eta} < 0$. The Y-Electrolyzer has the strongest negative skew for both $R^{(1)}$ and $R^{(2)}$ so that it represents the most conservative design with respect to outliers.

\begin{table}[h]
	\caption{Standard deviation $\sigma_{\zeta\eta}$ and non-parametric skew $S_{\zeta\eta}$ of ratios ${R}^{(1)}$ and ${R}^{(2)}$ for their change in upstream electrode plane ($\zeta\eta$-plane at $\xi=-T/2$).}
	\label{tab:std_skew_zeta_eta_plane}
	\centering
	\begin{tabular}{ l c c c c}
		\toprule
		Electrolyzer & $\sigma_{\zeta\eta}\left( R^{(1)} \right) $ & $S_{\zeta\eta} \left( R^{(1)} \right) $ & $\sigma_{\zeta\eta}\left( R^{(2)} \right) $  & $S_{\zeta\eta}     \left( R^{(2)} \right) $ \\
		\hline
		I1 & $\num{0.39}$ & $\num{-0.32}$ & \num{0.86} & \num{-0.02}\\
		T7 & $\num{0.45}$ & $\num{-0.05}$ & \num{0.88} & \phantom{-}\num{0.15}\\
		Y2.5 & $\num{0.30}$ & $\num{-0.36}$ & \num{0.55} & \num{-0.09}  \\
		\bottomrule
	\end{tabular}
\end{table}


\section{Experimental setup of Y-shaped FT-ME}
\label{sec:appendix_experimental}
\begin{figure}[ht]
	\centering
	\includegraphics[width=0.7\textwidth]{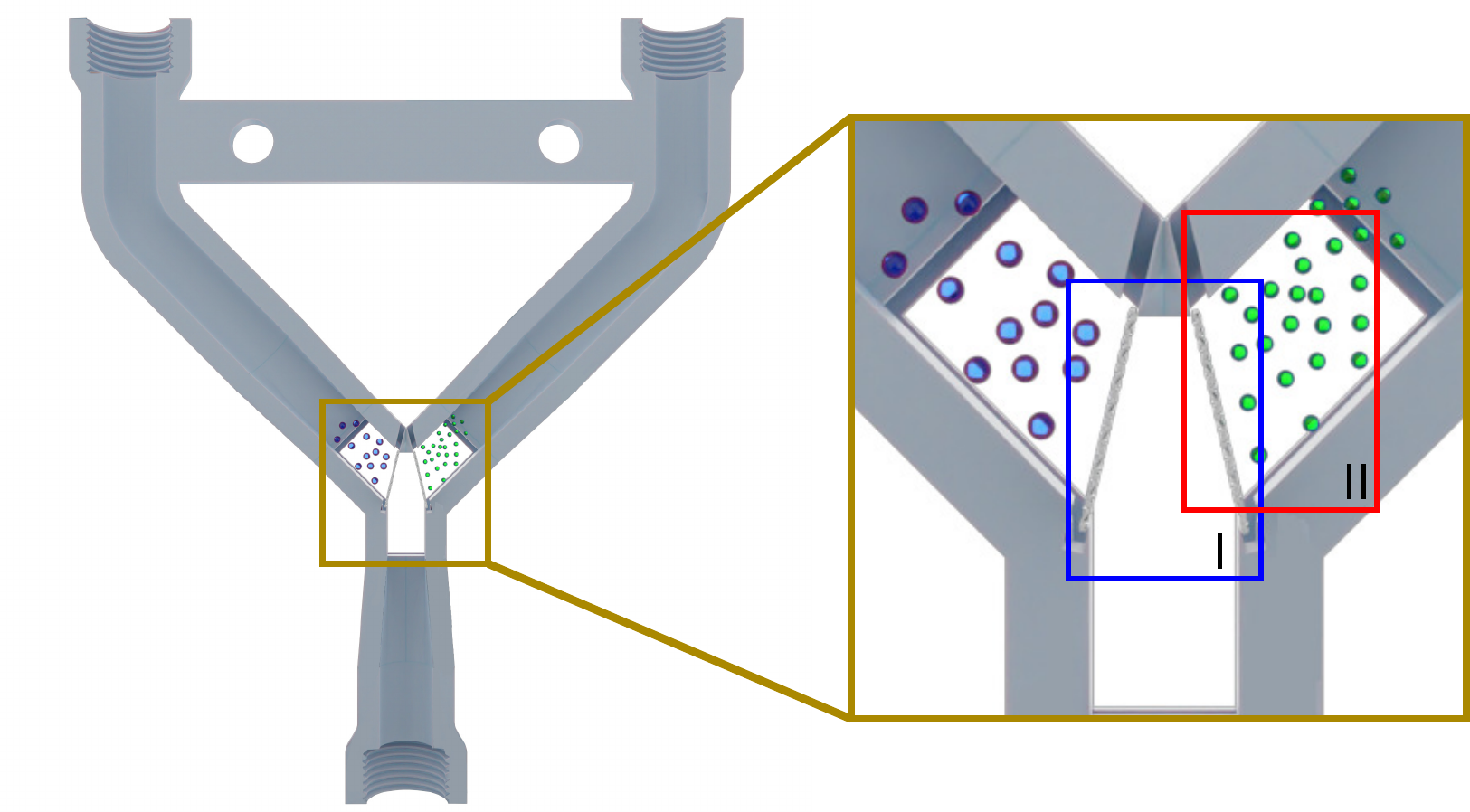}
	\caption{Geometry of the Y-shaped electrolyzer Y2.5 investigated experimentally. The cathode is located on the right-hand side, the anode on left-hand side of the drawing.  Two regions of interest are highlighted for later use: region~I representing the electrode gap, and region~II downstream of cathode.}
	\label{fig:exp_setup}
\end{figure}
\section{PIV image processing}
\label{sec:appendix_piv_processing}
Common preprocessing steps were applied to remove image and background noise, including background image subtraction, enhancement of particle contrast and sharpening of particle borders. A time-resolved PIV algorithm with a multipass vector calculation and an initial window size of \SI{256}{px}~$\times$~\SI{256}{px} was employed to determine the vector field on a grid of \SI{32}{px}~$\times$~\SI{32}{px}. Vector outliers were removed using a median filter and all data averaged in time. All image processing steps of the PIV images were performed in DaVis (DaVis 10.2.1, LaVision, Germany).

\newpage
\section{Gray value distributions for remaining $Re$}
\label{sec:appendix_experimental_result}

\begin{figure}[ht]
	\centering
	\hspace*{0pt}{\rlap{\subcaptionmark\label{fig:void_fraction_a}}\adjustbox{trim=-5mm 1mm 0mm 0mm, valign=t}
		{\includegraphics[width=0.7\textwidth]{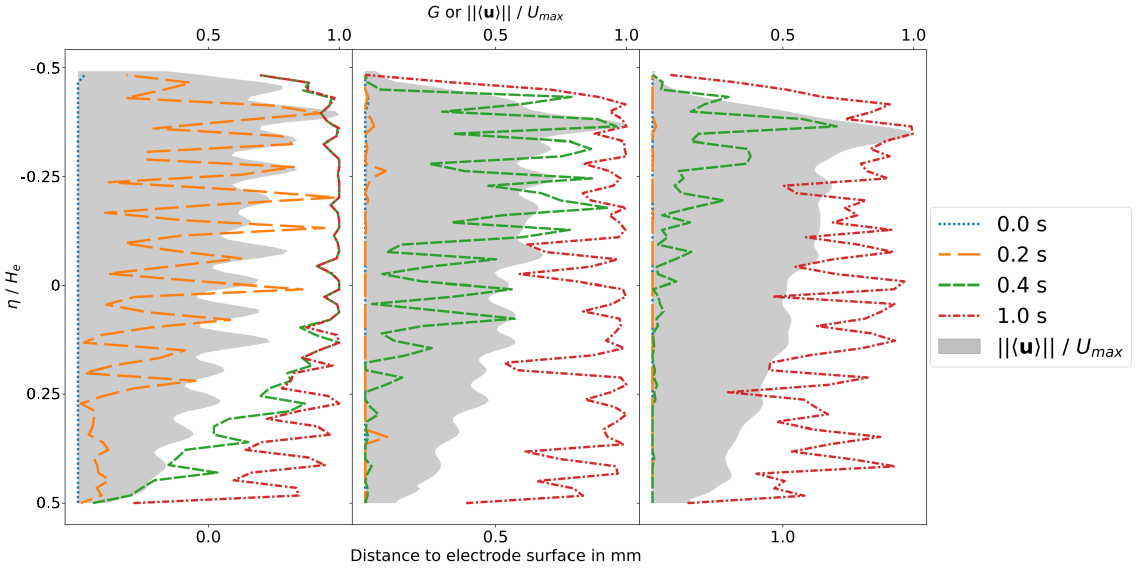}}} 
	
	\vspace*{2mm}
	\hspace*{0pt}{\rlap{\subcaptionmark\label{fig:void_fraction_c}}\adjustbox{trim=-5mm 1mm 0mm 0mm, valign=t}
		{\includegraphics[width=0.7\textwidth]{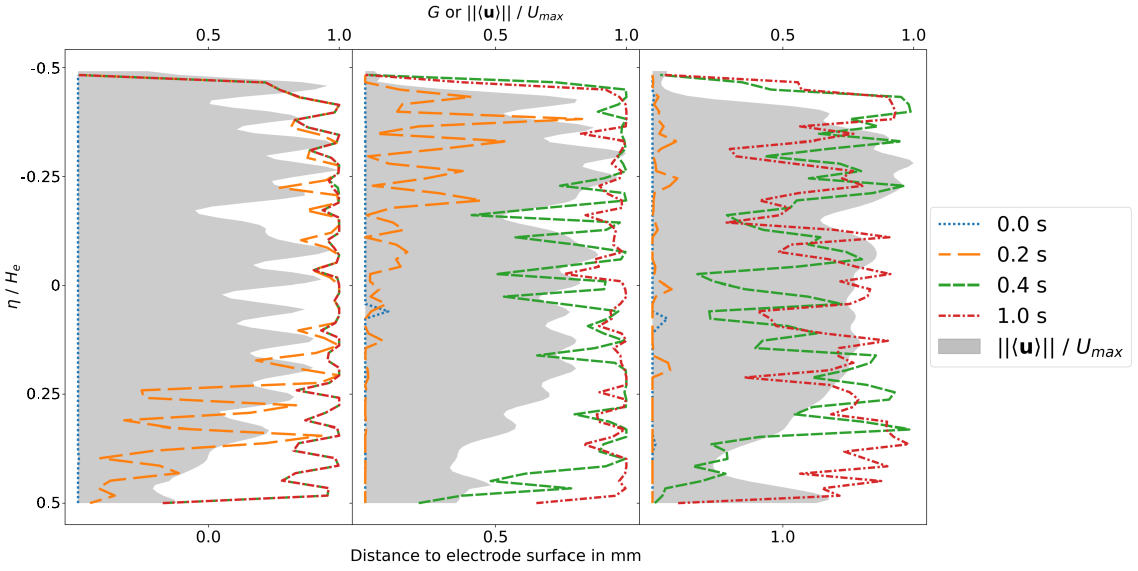}}} 
	
	\caption{
		Presence of bubbles in terms of the GVD for different Reynolds numbers, different times and different positions behind the cathode.
		\subref{fig:void_fraction_a}~$Re = 1000$, 
		\subref{fig:void_fraction_c}~$Re = 3000$.
		Each subfigures assembles plots at different positions behind the cathode, 
		$\SI{0}{mm}$, $\SI{0.5}{mm}$, $\SI{1}{mm}$ from left to right, 
		as indicated below the respective graphs.
		The lines show $G$ 
		along the electrode obtained from the measurements at different times after starting the current,
		$\SI{0}{s}$, $\SI{0.2}{s}$, $\SI{0.4}{s}$, $\SI{1.0}{s}$.
		The shaded area shows the fluid velocity of the single-phase flow.
	}
	\label{fig:appendix_void_fraction}
\end{figure}

\end{document}